\documentclass[10pt,twocolumn,twoside]{IEEEtran}
\usepackage[ruled]{algorithm2e}
\usepackage{threeparttable}
\usepackage{dcolumn}
\usepackage{booktabs}
\usepackage{amsmath,graphicx}
\usepackage{url}
\usepackage{cite}
\usepackage{psfrag}
\usepackage{amsfonts,amssymb}
\usepackage{amsbsy}
\usepackage{graphicx}
\usepackage{array}
\usepackage{multirow}
\usepackage{bm}
\usepackage{subfigure}
\usepackage{verbatim}
\usepackage{colortbl}
\usepackage{stfloats}
\usepackage{bm}
\usepackage{bigstrut}
\usepackage{color}

\newcommand{\comments}[1]{}
\newcommand{\etal}{\textit{et al}. }
\newcommand{\ie}{\textit{i}.\textit{e}., }
\newcommand{\eg}{\textit{e}.\textit{g}., }

\ifCLASSINFOpdf
\else
\fi

\begin{document}
	\title{Surface Light Field Compression \\ using a Point Cloud Codec}
	
	\author{Xiang~Zhang,
		~Philip~A.~Chou,
		~Ming-Ting~Sun,
		~Maolong~Tang,
		~Shanshe~Wang,
		~Siwei~Ma,
		and~Wen~Gao
		
		\thanks{X. Zhang, S. Wang, S. Ma and W. Gao are with the Institute of Digital Media, School of Electronics Engineering and Computer Science, Peking University, Beijing, China. (e-mail: x\_zhang@pku.edu.cn; sswang@pku.edu.cn; swma@pku.edu.cn; wgao@pku.edu.cn)}
		\thanks{P. A. Chou is with the 8i Labs, Inc., Seattle, WA, USA (e-mail: pachou@ieee.org)}
		\thanks{M. Sun and M. Tang are with the Department of Electrical Engineering, University of Washington, Seattle, WA, USA. (e-mail: mts@uw.edu; mltang@uw.edu)}
	}
	
	
	{}
	\maketitle
	
	\begin{abstract}
		Light field (LF) representations aim to provide photo-realistic, free-viewpoint viewing experiences. However, the most popular LF representations are images from multiple views.  Multi-view image-based representations generally need to restrict the range or degrees of freedom of the viewing experience to what can be interpolated in the image domain, essentially because they lack explicit geometry information. 
		We present a new surface light field (SLF) representation based on explicit geometry, and a method for SLF compression.  First, we map the multi-view images of a scene onto a 3D geometric point cloud.  The color of each point in the point cloud is a function of viewing direction known as a view map. We represent each view map efficiently in a B-Spline wavelet basis. This representation is capable of modeling diverse surface materials and complex lighting conditions in a highly scalable and adaptive manner. The coefficients of the B-Spline wavelet representation are then compressed spatially. To increase the spatial correlation and thus improve compression efficiency, we introduce a smoothing term to make the coefficients more similar across the 3D space. We compress the coefficients spatially using existing point cloud compression (PCC) methods. On the decoder side, the scene is rendered efficiently from any viewing direction by reconstructing the view map at each point. In contrast to multi-view image-based LF approaches, our method supports photo-realistic rendering of real-world scenes from arbitrary viewpoints, \ie with an unlimited six degrees of freedom (6DOF). In terms of rate and distortion, experimental results show that our method achieves superior performance with lighter decoder complexity compared with a reference image-plus-geometry compression (IGC) scheme, indicating its potential in practical virtual and augmented reality applications.
	\end{abstract}
	\section{Introduction and Related Works}
	\label{sec:intro}
	In emerging virtual reality (VR) and augmented reality (AR) applications, it is important to be able to render a scene from arbitrary points of view, allowing free-viewpoint navigation for example.  While conventional computer graphics (CG) allow synthesis of CG-modeled scenes from arbitrary points of view, the photorealism of natural scenes using CG models is elusive, at least without extreme computation, especially in the presence of complex material and lighting phenomena such as reflection, refraction, and scattering.
	
\begin{figure}
		\centering
		\subfigure[Multi-view representation]{\includegraphics[width=0.49\linewidth]{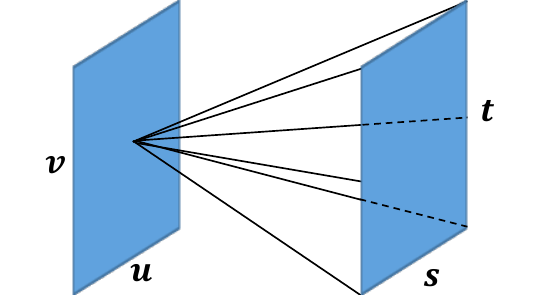}}
		\subfigure[Lenslet representation]{\includegraphics[width=0.49\linewidth]{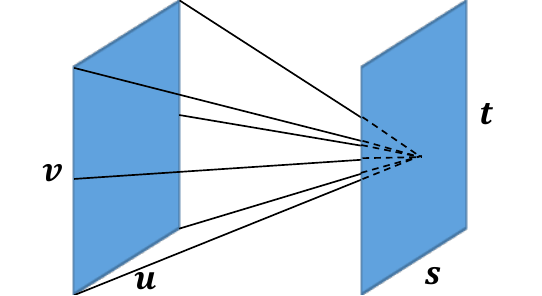}}
		\caption{Two-plane light field representations \cite{levoy1996light}, where $(u,v)$ and $(s,t)$ indicate the camera plane and the focal plane, respectively. (a) Multi-view (or sub-aperture) representation: an image of the scene is captured from each camera position $(u,v)$. (b) Lenslet representation: an image (or {\em view map}) of the color of a point as seen from different directions is captured for each point $(s,t)$.}
		\label{fig:light_field}
	\end{figure}

	Light field (LF) representations aim to provide photo-realistic renderings of 3D scenes from a range of viewpoints even in the presence of such complex material and lighting phenomena, to enable rich and immersive viewing experiences. An LF is most frequently represented as a 4D function of a light ray in which each ray is parameterized by its intersections with two parallel planes \cite{levoy1996light}. There are two types of such an LF representation in common use: a multi-view (or sub-aperture) representation and a lenslet representation, as shown in Fig.~\ref{fig:light_field}. A multi-view representation is essentially a collection of images captured from different viewpoints, while a lenslet representation is essentially a collection of images (or view maps) of the color each point on a plane as the point appears from different directions.
	
	An LF may be captured directly into a multi-view representation using a dense array of cameras \cite{lf_camera_array}, or it may be captured directly into a lenslet representation using a large aperture lens focused onto a compact micro-lens sensor \cite{ng2005light}. Either representation may then be converted to the other.  Because an LF records the color of many rays of light through a scene, its representation requires a large amount of information, hindering transmission, storage, and application development. Therefore, LF compression has attracted extensive attention recently, including standardization efforts in JPEG \cite{jpeg-pleno} and MPEG \cite{mpeg-i}. An efficient, robust, and flexible LF compression approach is highly needed.
	
	Recently emerging LF compression methods are based on either the multi-view representation or the lenslet representation.  Methods based on the multi-view representation use video coding techniques to remove redundancy among the multi-view images by inter-view prediction, as first proposed by Magnor and Girod \cite{magnor2000data}. With advances in video coding, multi-view based algorithms have achieved remarkable improvements by advanced inter-image prediction techniques \cite{liu2016pseudo,li2017pseudo,zhao2016light,jia2017light,jia2017optimized} and transform based algorithms \cite{chen2017light,jiang2017light,vagharshakyan2017light}. In \cite{volino2014optimal}, the authors propose to align multi-view images to multi-layer texture maps according to their 3D mesh representation.
	LF compression methods that are alternatively based on the lenslet representation use image coding techniques to directly compress the lenslet image, as captured by a micro-lenses camera, by exploiting intra-image similarities \cite{li2014efficient,li2016intra,li2016intra2,li2016intra3,conti2016hevc,conti2016hevc2}. For example, Li \etal \cite{li2014efficient,li2016intra,li2016intra2,li2016intra3} propose a macro-pixel prediction mode, which resembles the intra-block copy technique in video compression. Conti \etal \cite{conti2016hevc,conti2016hevc2} present a self-similarity compensated prediction to further exploit spatial correlations among lenslet images.
	
	However, these LF compression methods barely use the geometric information, and as a result have significant limitations. For one, since the images are captured by cameras at a discrete set of positions, high quality view interpolation requires a dense camera array to avoid occlusion artifacts. Even more significantly, extrapolation of views outside a narrow range of view angles close to the original camera positions is generally not feasible. For static scenes, it may be barely feasible to scan the scene from a very large set of camera positions, thereby increasing the range in which views can be synthesized. But for dynamic scenes, such scanning is completely infeasible. This all but eliminates conventional LF approaches for VR and AR applications in which arbitrary viewpoints of dynamic scenes must be generated.
	
	A more efficient and flexible approach is the surface light field (SLF) representation, which was introduced by Miller \etal \cite{miller1998lazy}. The SLF enables synthesis from an arbitrary viewpoint, interactive rendering, and rudimentary editing of the LF. 
	Essentially, the SLF defines the light rays emanating from each point on each surface of the 3D scene. As shown in Fig.~\ref{fig:slf}, the SLF can be regarded as a function $f(\bm{\omega}|\bm{p})$, where $\bm{p}$ is the location in 3D of a surface point, and $\bm{\omega}$ is the direction of a ray emanating from the point. The SLF can be viewed as a generalization of the lenslet representation, if one considers the surface point at $\bm{p}$ as a point on the $(s, t)$ plane in Fig.~\ref{fig:light_field}(b). Since the point $\bm p$ is in 3D coordinates, the SLF generalizes the lenslet representation by generalizing the $(s, t)$ plane to a 2D manifold embedded in 3D.  This not only solves occlusion problems but also enables free-viewpoint rendering.  Since an SLF is a generalization of an LF, it can represent anything that an LF can represent.  Moreover, it has the potential to be a more efficient representation. Indeed, for Lambertian or near-Lambertian objects, the view map at each point $\bm{p}$ is a constant or near-constant image, reducing $f(\bm{\omega}|\bm{p})$ essentially to a function only of $\bm{p}$, like a 2D CG texture map.  An alternative view of an SLF is as a CG texture map whose value at every point is an image, which can be arbitrarily complex yet is frequently near-constant. Thus an SLF can also be viewed as a generalization of a CG texture map.  In a sense, SLFs combine the best of LFs and CG modeling, allowing photo-realistic rendering from arbitrary points of view.
	
	For SLF compression, Miller \etal \cite{miller1998lazy} compress view maps as images by first partitioning each view map into blocks of pixels and then applying the discrete cosine transform (DCT) to each block. Wood \etal \cite{wood2000surface} represent each view map as a linear combination of a set of prototypes, which are determined by either vector quantization or principle component analysis (PCA). Chen \etal \cite{chen2002light}, in addition to PCA, apply non-negative matrix factorization (NMF) to project the view maps to a subspace with a non-negativity constraint. However, these methods have the following limitations.  First, they discretize each view map of SLF into either a grid or octahedral representation to enable discrete transform/projection. However, these schemes inherently introduce errors during discretization, and a coarse granularity may decrease the reconstruction quality significantly. Second, the view maps are compressed independently for each point and the spatial redundancies across the 3D space are not considered.
	
	In this work, we propose a new SLF representation and method for its compression.
	In a nutshell, we propose to represent and compress the SLF function $f(\bm{\omega},\bm{p})$ as a separable linear transform $F(i,j)$, where $i$ is an image frequency index and $j$ is a spatial frequency index.  Specifically, first, for every surface point $\bm{p}$, we apply a linear transform to transform the view map $f(\bm{\omega}|\bm{p})$ into a sequence of image transform coefficients $\alpha_0(\bm{p}),\alpha_1(\bm{p}),\ldots$, and second, for every image transform coefficient $\alpha_i(\bm{p})$, we use a spatial transform (\ie a transform across the surface) to transform $\alpha_i(\bm{p})$ as a function of $\bm{p}$ into a sequence of spatial transform coefficients $F(i,0),F(i,1),\ldots$.
	Our contributions include:
	\begin{itemize}
		\item A new LF representation method based on a SLF representation by mapping multi-view images to a point cloud. We propose to use B-Spline wavelets to represent view maps for each point on the point cloud. This representation is highly efficient, robust, and scalable. It is able to handle complicated surface radiance patterns and different numbers of observations.  It supports true free-viewpoint rendering and immersive experiences for virtual reality applications.
		\item A new SLF compression framework in which the view map coefficients are spatially transformed and compressed by using a point cloud codec. To improve coding efficiency, a smoothness term is introduced when solving the coefficients of the view maps, to increase their spatial correlation across 3D space.
		\item A comparison with an image-plus-geometry compression (IGC) approach, against which the proposed method can significantly relieve the burden of rendering on the decoder side with competitive rate-distortion (RD) performance.
	\end{itemize}
	
	Section~\ref{sec:representation} proposes the SLF representation by B-Spline wavelets. Section~\ref{sec:compression} presents the SLF compression framework and algorithms in detail. Section~\ref{sec:results} shows experimental results. Section~\ref{sec:conclusion} concludes.
	
	\section{Surface Light Field Representation\\ by B-Spline Wavelets}
	\label{sec:representation}
	We aim to determine an SLF representation in an efficient, robust, and scalable manner. Efficiency means that the representation is compact and friendly to compression. Robustness means that the representation is capable of approximating views from various directions of various surface materials under various lighting conditions. Scalability means that the representation can approximate simple view maps (as is the case with near-Lambertian surfaces) through arbitrarily complex view maps (as is the case with reflective surfaces) using a bitrate commensurate with its complexity.
	
	To this end, we propose to approximate the view map at each point $\bm{p}$ as a linear combination of basis functions,
	\begin{equation}\label{eq:slf}
	f(\bm{\omega}|\bm{p})\approx\sum_{i=0}^{N-1}{\alpha_i(\bm{p})\cdot g_i(\bm{\omega})},
	\end{equation}
	where $g_i(\bm{\omega})$ is the $i^{th}$ basis function, which is a function of the viewing direction $\bm\omega$; $\alpha_i(\bm{p})$ is the corresponding coefficient at point $\bm{p}$; and $N$ is the number of basis functions.
	
	\begin{figure}
		\centering
		\includegraphics[width=0.6\linewidth]{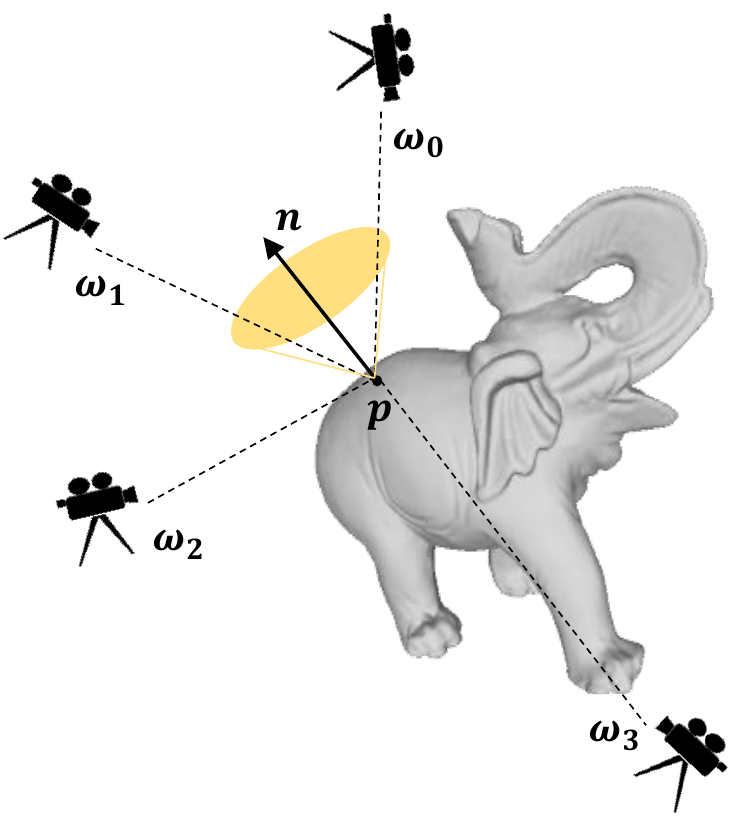}\\
		\caption{Illustration of capturing an SLF. SLF defines every light ray emanating from a surface point $\bm{p}$. One can capture an SLF by a number of cameras in different positions. We define a cone area as shown in the figure, within which the camera observations will be regarded as valid.}\label{fig:slf}
	\end{figure}
	
	The number of cameras to capture the SLF is always limited in practice. As shown in Fig.~\ref{fig:slf}, the SLF of the object is captured by several cameras from different viewpoints. For a surface point $\bm p$, we denote its pixel values from different camera directions as a vector $\bm{c}=\left\{ c_0, c_1, \ldots, c_{M-1} \right\}$, where $M$ is the number of the valid observations. We eliminate invalid observations caused by being occluded or out of the camera field of view. The corresponding camera directions $\bm{\omega}_m$ are parameterized by spherical coordinates with azimuth $\theta \in [-\pi,\pi]$ and elevation $\phi \in [-\pi/2,\pi/2]$. Further, we re-parameterize $\phi$ as $\gamma=\sin(\phi)\in[-1,1]$, which ensures that equal areas in the $(\theta,\gamma)$ plane are equal areas on the sphere.  In turn this ensures that any basis that is orthogonal on the $(\theta,\gamma)$ plane is orthogonal on the sphere.
	
	Since we have only a limited number of cameras to measure the SLF, the number of valid observations $M$ is usually smaller than the number of basis functions $N$, making the problem under-determined. Thus we regularize the solution.  To be precise, let each element in matrix $\bm{G} \in \mathbb{R}^{M\times N}$ be $\bm{G}_{j,i}=g_i\left( \bm{\omega}_j \right)$.  Then, given the observation vector $\bm{c}$ at point $\bm{p}$, we determine $\bm{\alpha}(\bm{p})$ as
	\begin{equation}\label{eq:problem}
	\begin{aligned}
	\bm{\alpha}(\bm{p}) &= \arg\min_{\bm{\alpha}}{\|\bm{c}-\bm{G}\bm{\alpha}\|_2^2+\lambda\|\bm{\alpha}\|_2^2}\\
	&=\left(\bm{G}^T\bm{G}+\lambda\bm{I}\right)^{-1}\left( \bm{G}^T\bm{c} \right),
	\end{aligned}
	\end{equation}
	where $\lambda$ is a regularization factor. $\lambda$ is significant for two reasons: First, when the problem is under-determined, \ie $M<N$, $\lambda$ avoids over-fitting and yield compressible coefficients with reasonable range. Second, $\lambda$ makes the solution robust to outliers due to camera noise and other imprecisions.
	
	Since the representation coefficients $\bm{\alpha}(\bm{p})$ would then be compressed over the surface by spatial transforms or predictions, increasing their spatial coherence leads to more efficient compression. Therefore, we further introduce a smoothing term into the optimization problem in \eqref{eq:problem} as follows,
	\begin{equation}\label{eq:problem_spatial}
	\begin{aligned}
	\bm{\alpha}(\bm{p}) &= \arg\min_{\bm{\alpha}}{\|\bm{c}-\bm{G}\bm{\alpha}\|_2^2+\lambda\|\bm{\alpha}\|_2^2+\beta\|\bm{\alpha}-\bar{\bm{\alpha}}\|_2^2}\\
	&=\left(\bm{G}^T\bm{G}+\lambda\bm{I}+\beta\bm{I}\right)^{-1}\left( \bm{G}^T\bm{c} + \beta \bar{\bm \alpha} \right),
	\end{aligned}
	\end{equation}
	where $\bar{\bm{\alpha}}$ indicates the averaged coefficients over the nearest surface points to $\bm{p}$. $\beta$ is the corresponding regularization factor, which is significant in improving the overall compression efficiency, as a larger $\beta$ will increase the spatial similarity and reduce the coding bitrates.
	
	To solve \eqref{eq:problem_spatial}, we first obtain a good initialization by \eqref{eq:problem} as the coefficients of neighboring points are unavailable at the first iteration. Then we iteratively solve \eqref{eq:problem_spatial} until convergence. The detailed algorithm is shown in  Algorithm~\ref{al:coef}.
	
	\begin{algorithm}
		\label{al:coef}
		\KwIn{
			\begin{itemize}
				\item Pixel values from different observation directions for each point $\bm p$: $\bm{c}(\bm{p})$
				\item Observation directions for each point $\bm p$: $\bm{\omega}(\bm p)$
				\item Regularization factors $\lambda>0$ and $\beta>0$
				\item Maximum iteration number $T$
			\end{itemize}
		}
		\KwOut{
			\begin{itemize}
				\item SLF representation coefficients for each point $\bm p$: $\bm{\alpha}(\bm p)$
			\end{itemize}
		}
		\textbf{Initialization}:\\
		\For {each point $\bm p$}{
			Calculate the basis function: $\bm{G}(\bm \omega)$\;
			Calculate the SLF coefficient $\bm{\alpha}^0$ by \eqref{eq:problem}\;
		}
		\For {$k \gets 1$ to $T$}{
			\For {each point $\bm p$}{
				Calculate the average SLF coefficient $\bar{\bm{\alpha}}$ of $\bm p$'s neighboring points\;
				Calculate the SLF coefficient $\bm{\alpha}^k$ by \eqref{eq:problem_spatial}\;
			}
		}
		\caption{Calculating SLF representation coefficients.}
	\end{algorithm}
	
	The design of basis functions is significant, since a good basis compacts the energy in the coefficients and make them easier to compress. In this work, we use 2D separable B-Spline wavelets for the basis, as they can describe the local variation of view maps more efficiently. The 2D B-Spline wavelets can be formulated as follows,
	\begin{equation}\label{splinew2d}
	g_i(\theta,\gamma)=w_{i_0}\left( \frac{\theta}{2\pi} \right) w_{i_1}\left( \frac{\gamma}{2} \right),
	\end{equation}
	where $w_i$ is the $i^{th}$ offset of the periodicized 1D B-Spline wavelet function
	\begin{equation}\label{splinew}
	w_i\left( x \right)=\sum_{m\in \mathbb{Z}}{\psi_o\left( 2^s x-i+m 2^s \right)},
	\end{equation}
	and $\psi_o$ is the basic B-Spline wavelet function with order $o$ and scale $s$, which can be defined by the sum of cardinal B-Spline functions,
	\begin{equation}\label{splinepsi}
	\begin{matrix}
	{\psi_o}(x) = \sum\limits_{n = 0}^{3o-2} {{q_n}{N_o}(2x - n)},\\
	{q_n} = \frac{{( - 1)}^n}{2^{o - 1}}\sum\limits_{j = 0}^o {\left( {\begin{matrix}
			m \\
			j
			\end{matrix} } \right)} {N_{2o}}(n - j + 1),
	\end{matrix}
	\end{equation}
	where $N_o$ is the cardinal B-Spline function. To ensure $\theta$ and $\gamma$ have roughly equal resolution, we make the scales of $\theta$ and $\gamma$, \ie $s_0$ and $s_1$, satisfy $s_0=s_1+1$. Therefore $i_0=0,1,\ldots, 2^{s_0-1}$, $i_1=0,1,\ldots, 2^{s_1-1}$, and $i=0,1,\ldots, 2^{s_0+s_1-1}$; the total number of basis functions equals $N=2^{s_0+s_1}$.
	
\begin{figure}
		\centering
		\includegraphics[width=1\linewidth]{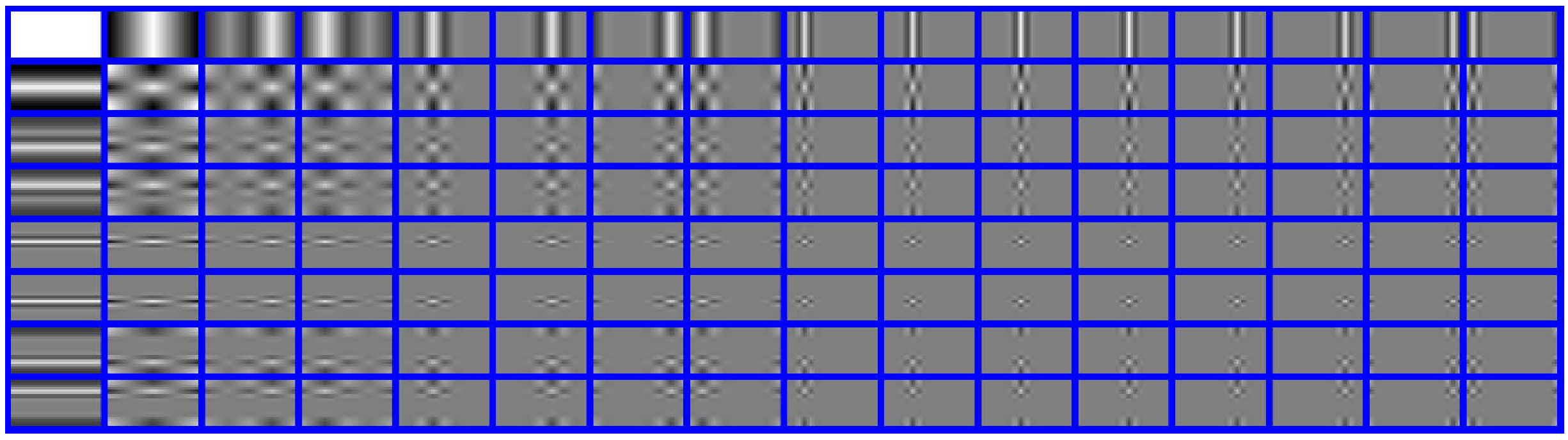}\\
		\caption{Illustration of B-Spline Wavelet basis functions.}\label{fig:spline}
	\end{figure}

	The first 128 B-Spline wavelet basis functions are shown in Fig.~\ref{fig:spline}. There are 16 basis functions per row ($\theta$ direction) and 8 basis functions per column ($\gamma$ direction). The top-left basis function is constant, so the corresponding coefficient is termed the DC coefficient as it represents the mean value. From the top-left to the bottom-right corner, the basis functions are able to describe more high-frequency signals.
	
\begin{figure}
	\centering
	\subfigure{\includegraphics[width=0.48\linewidth]{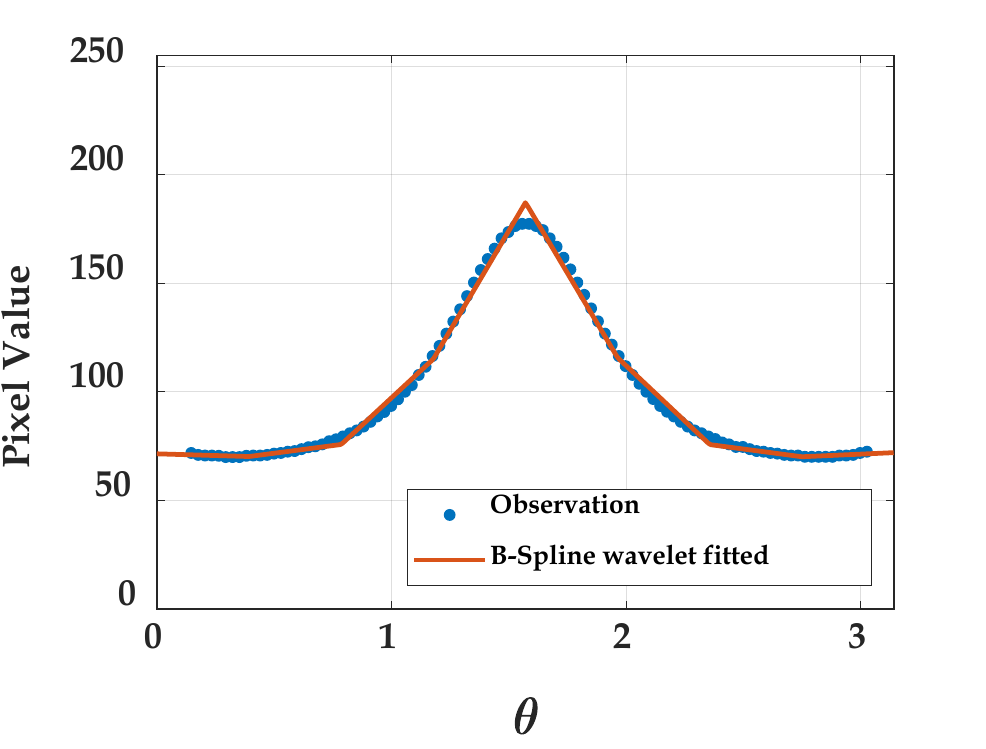}}
	\subfigure{\includegraphics[width=0.48\linewidth]{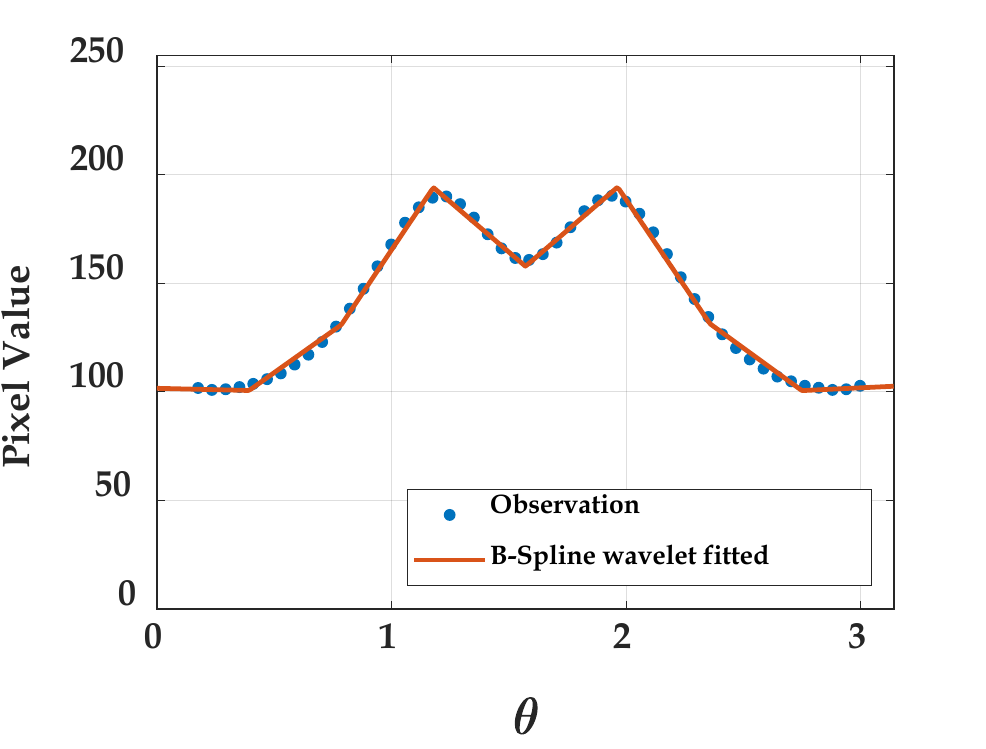}}
	\subfigure{\includegraphics[width=0.48\linewidth]{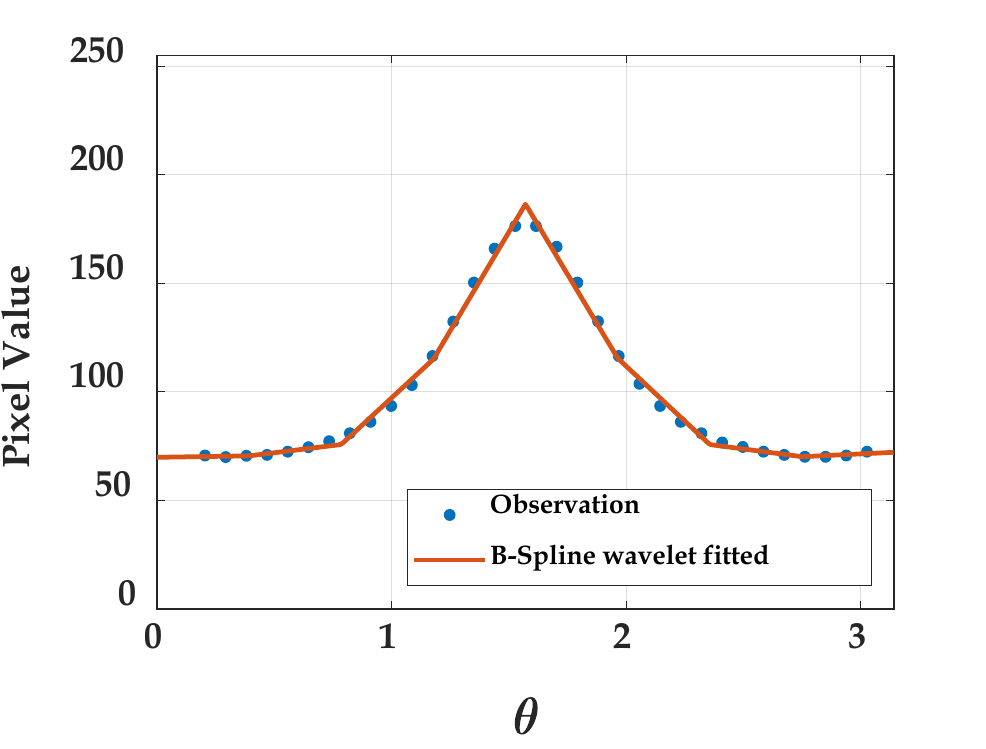}}
	\subfigure{\includegraphics[width=0.48\linewidth]{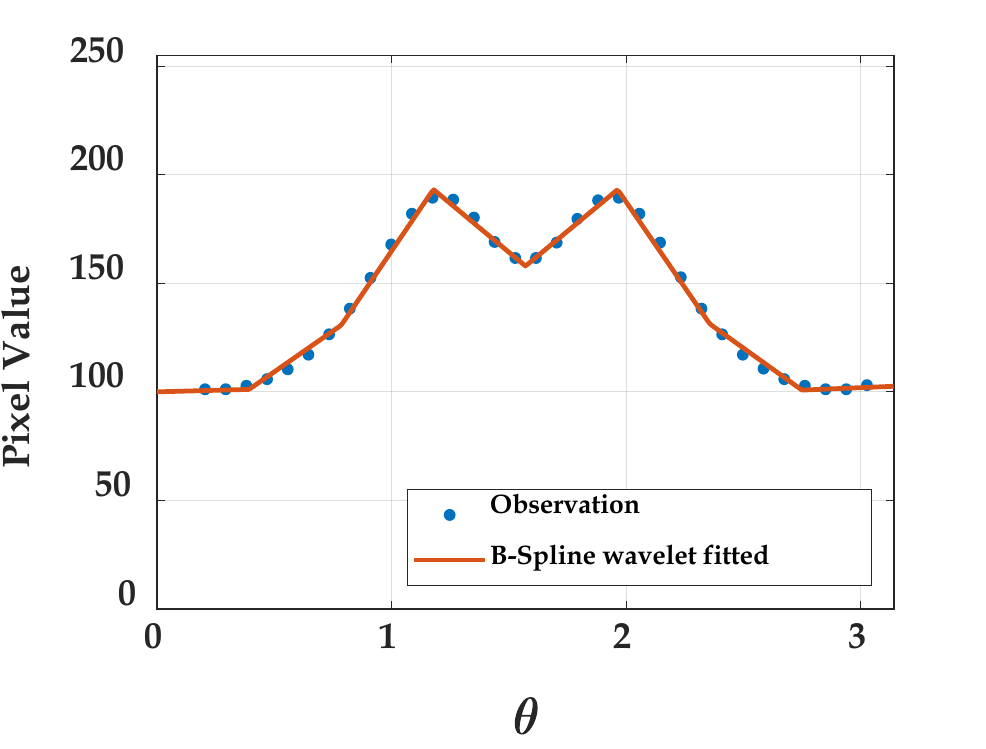}}
	\subfigure{\includegraphics[width=0.48\linewidth]{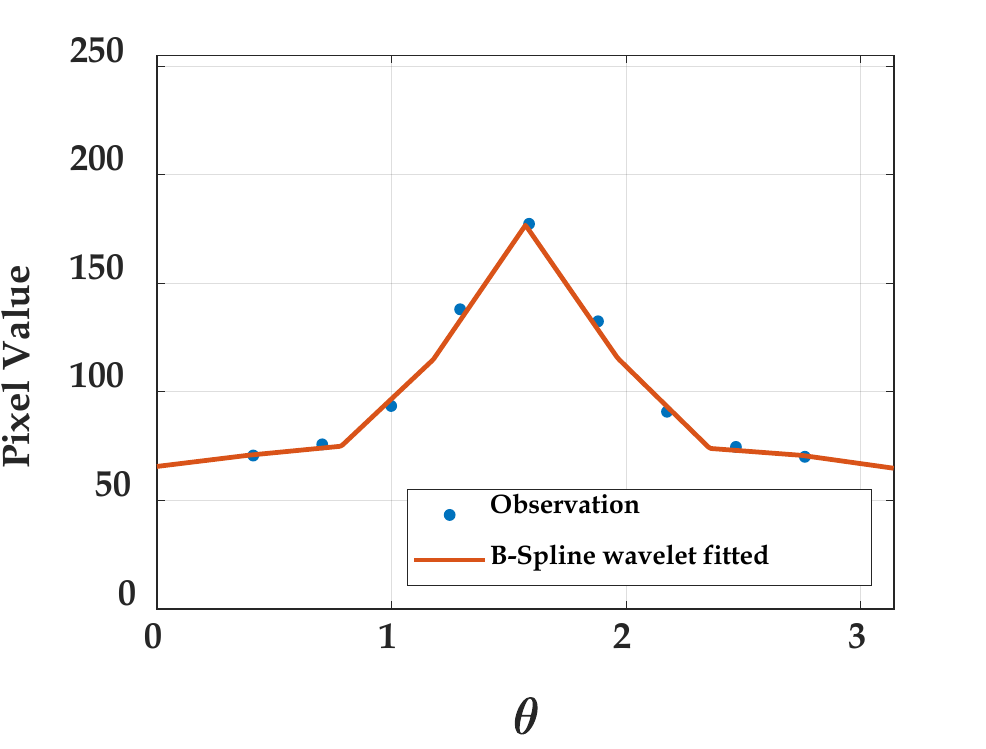}}
	\subfigure{\includegraphics[width=0.48\linewidth]{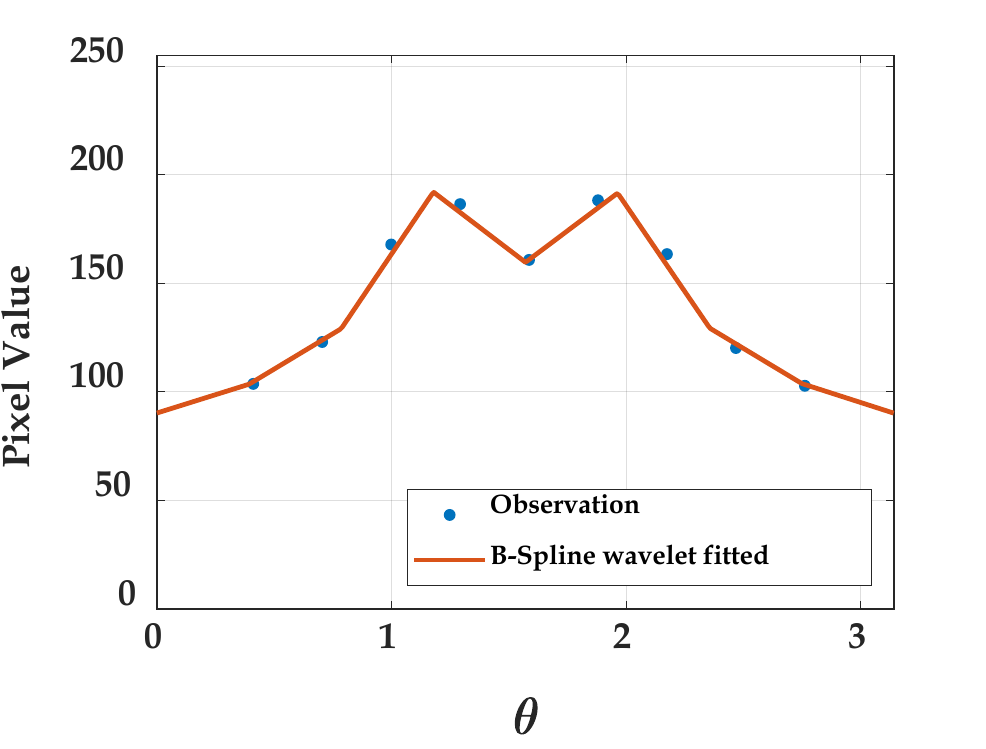}}
	\caption{1D B-Spline wavelet fitting, where blue dots indicate the observed luminance of a surface point observed from different discrete angles and the red solid lines indicate continuous functions reconstructed by fitting B-Spline wavelet basis functions to the observations. The left subfigures show the case of one light source (one peak), and the right subfigures show the case of two light sources (two peaks). From top to bottom subfigures, the density of observations is decreasing.}
	\label{fig:1dfit}
\end{figure}

	To demonstrate the effectiveness and robustness, we show the reconstruction of a 1D B-Spline wavelet fit to image data captured from a synthetic scene (in terms of $\theta$ only, for simplicity). The results are shown in Fig.~\ref{fig:1dfit}, where the blue dots in each subfigure represent pixel values of a surface point observed from different view angles. The left three subfigures show the case of one light source (one peak in the figures), and the right three subfigures show the case of two light sources (two peaks in the figures). From the top row to bottom row, the density of observations is decreasing. From the results, one can see that the 1D B-Spline wavelet fits the observed data well and obtains a good inference of unobserved data even for sparser observation density, indicating the B-Spline wavelet basis can accurately and robustly represent the SLF.
	
\begin{figure}
		\centering
		\includegraphics[width=0.8\linewidth]{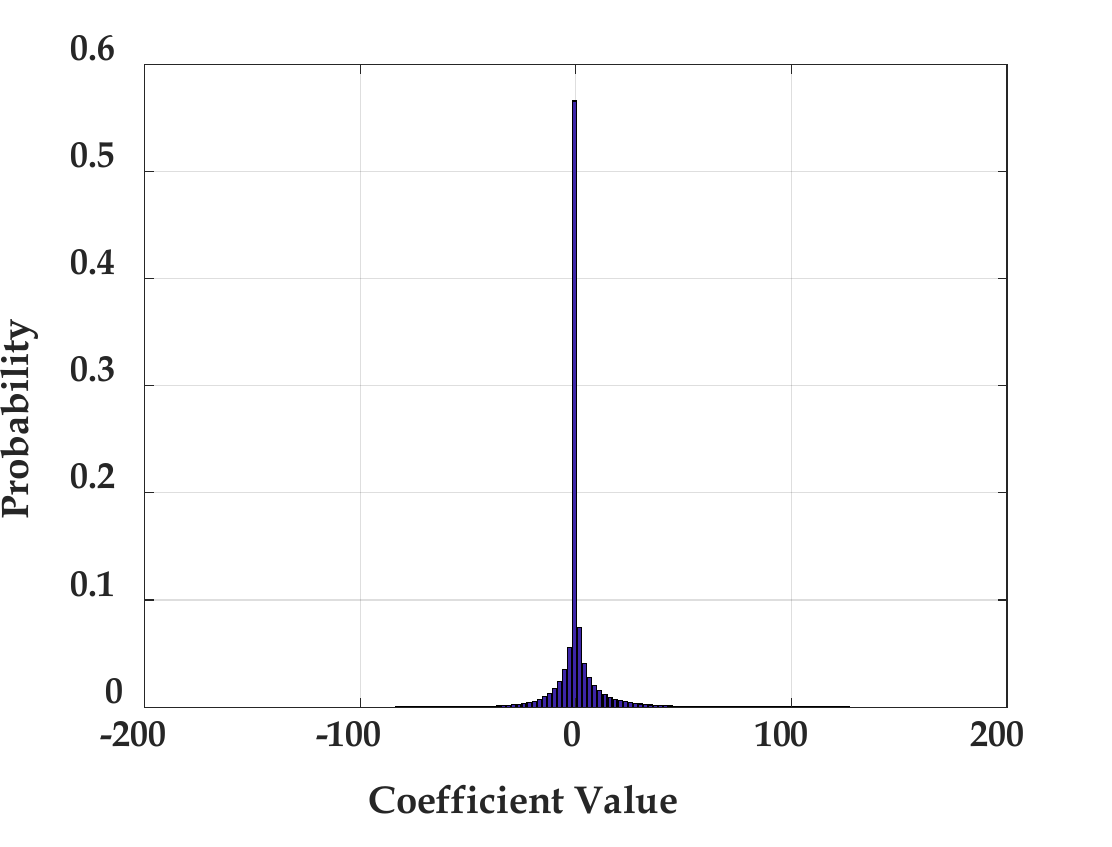}\\
		\caption{Coefficient distribution of B-Spline wavelets for representing SLF.}\label{fig:coef_dist}
	\end{figure}

	The distribution of the SLF coefficients is shown in Fig.~\ref{fig:coef_dist}.  Most coefficients are close to zero, indicating that the SLF representation is highly compressible.
	
	\section{Surface Light Field Compression}
	\label{sec:compression}
	
	\begin{figure}
		\centering
		\includegraphics[width=1\linewidth]{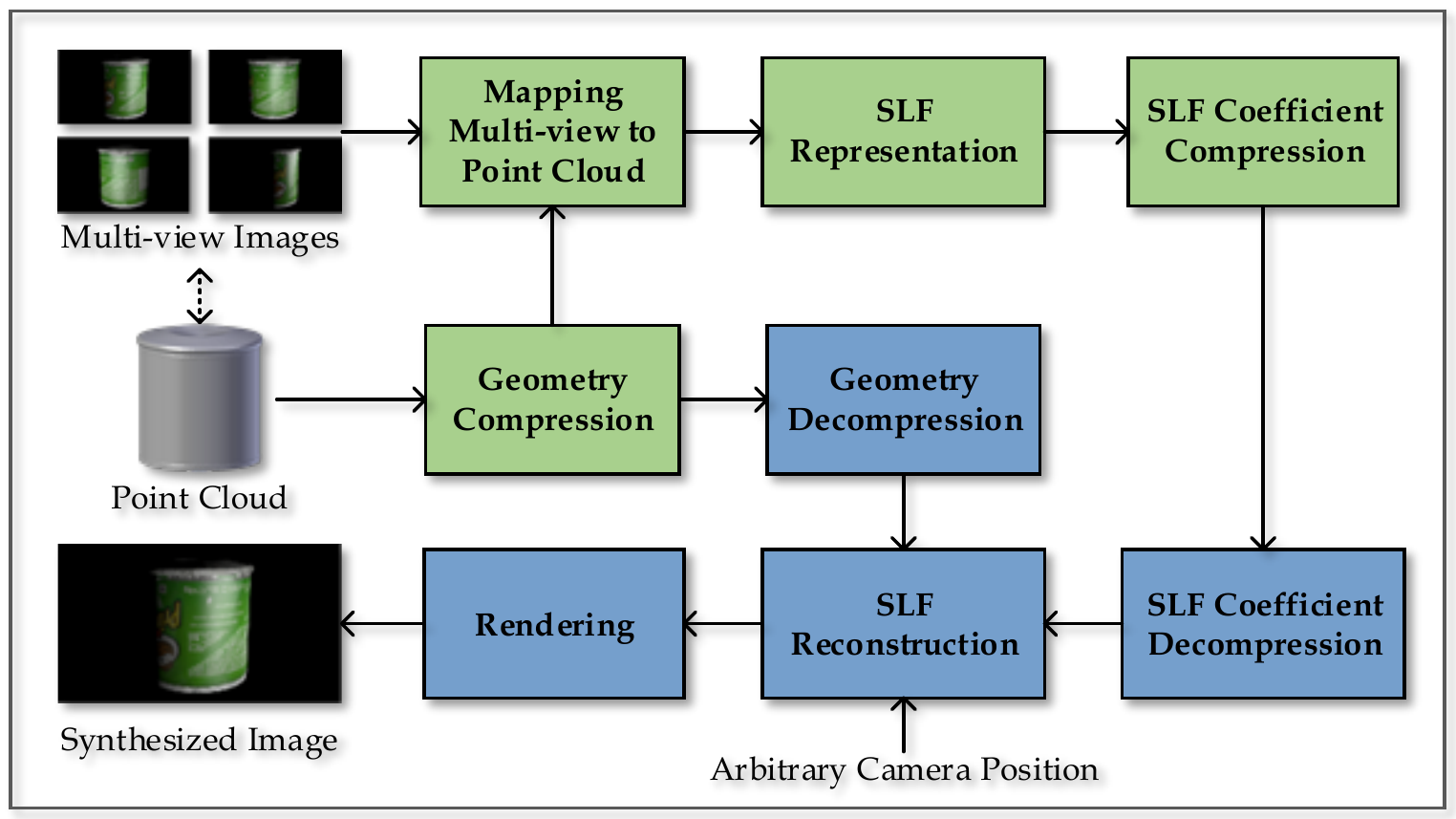}\\
		\caption{The proposed framework of SLF compression. The input data include the geometry and multi-view images. The output is synthesized images from arbitrary view angles. Green and blue boxes indicate the processes on encoder and decoder sides, respectively.}\label{fig:framework}
	\end{figure}
	
	Fig.~\ref{fig:framework} illustrates the pipeline of the proposed SLF compression. The input data include a point cloud that represents object geometry and a number of images captured from different points of view. It is worth noting that the obtaining of multi-view images and point cloud are usually dependent on each other in practice. For real scenes, the geometry of 3D scenes can be estimated from multi-view images by structure from motion (SfM) techniques \cite{book_computer_vision}. For synthetic scenes, usually the multi-views are obtained from CG models. The geometry of the point cloud, \ie the 3D position of each point $\bm{p}$, needs to be compressed, since the geometry information will be used at both the encoder and decoder for representing and rendering the LF. The remainder of our pipeline for SLF compression, where our major contribution lies, consists of the following steps: 1) mapping multi-view images to the point cloud, by collecting observations $\{f(\bm{\omega}_m|\bm{p})\}$ of the view map $f(\bm{\omega}|\bm{p})$ at each point $\bm{p}$, 2) representation of the view map $f(\bm{\omega}|\bm{p})$ at each point $\bm{p}$ by a linear combination of B-Spline wavelet basis functions with coefficients $\alpha_i(\bm{p})$, $i=0,\ldots,N-1$, 3) independent compression of each wavelet coefficient $\alpha_i(\bm{p})$ by utilizing its spatial coherence across $\bm{p}$, and 4) decompression, reconstruction, and rendering of the SLF from arbitrary points of view. Technical details of each component are discussed in the following subsections.
	
	\subsection{Mapping Multi-view Images to Point Cloud}
	\label{sec:framework_preprocessing}
	
	The purpose of mapping is to find for each surface point $\bm{p}$ a set of valid observations $\{f(\bm{\omega}_m|\bm{p})\}$ from the captured images. Obviously, a surface point may not be observed by all cameras for several possible reasons.  For example, it might be out of the camera's field of view, occluded, or self-occluded (back-facing).
	
	To determine whether a point $\bm{p}$ is in the field of view of camera $m$, we project $\bm{p}$ to the 2D camera plane as
	\begin{equation}\label{eq:color}
	\bm{p}'=\bm{K}\left[ \bm{R}|\bm{t} \right] \bm{p},
	\end{equation}
	where $\bm{K}$ and $\left[ \bm{R}|\bm{t} \right]$ are the intrinsic and extrinsic parameters of camera $m$, respectively. From this, one can easily determine whether $\bm{p}$ is located within the field of view of the camera.
	
	To determine whether a point is occluded, we project all the points to an image plane for each camera. If multiple points are projected to a same image position, the one with minimal depth (whose distance to the camera is nearest) is kept and the others are determined to be occluded. Changing the resolution of the image plane will determine the granularity of the occlusions.
	
\begin{figure}
		\centering
		\includegraphics[width=0.8\linewidth]{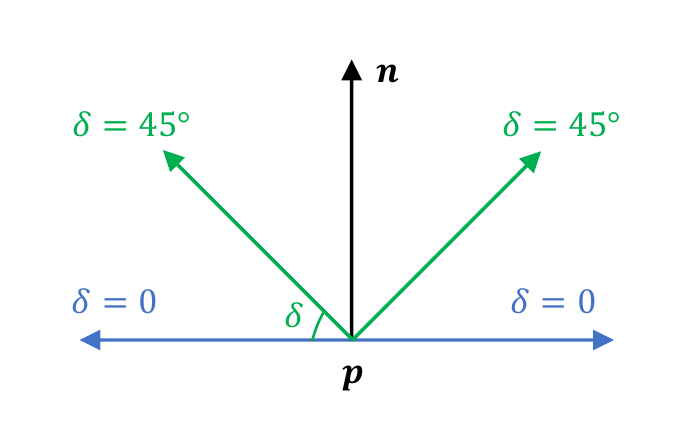}\\
		\caption{Illustration of $\delta$ in 2D, where $\bm{p}$ indicates a surface point and $\bm{n}$ is the normal vector at the point. $\delta$ is the complement of the maximum valid observation angle between the normal direction and the viewing direction.}\label{fig:dn}
	\end{figure}

	To determine whether a point is back-facing (or nearly back-facing), we define a cone centered around the normal at $\bm{p}$ as a valid observation region, within which the observations are regarded as valid while others are not, ruling out back-facing and extreme angles of view of the surface point. In Fig.~\ref{fig:slf}, the valid observation region is illustrated by a yellow cone. In this case, cameras from directions $\bm{\omega}_0$ and $\bm{\omega}_1$ are valid because they are located inside the cone. We parameterize the cone by the angle $\delta$, which is the complement of the maximum angle between the normal direction and a valid camera (or observation) direction. A 2D illustration of $\delta$ is shown in Fig.~\ref{fig:dn}.
	
	After checking the validity of each camera for each point, the pixel value at the 2D point $\bm{p}'$ in the valid camera's image is derived by linear interpolation and it is a potentially valid observation for $f(\bm{\omega}_m|\bm{p})$, where $\bm{\omega}_m$ is the direction between point $\bm{p}$ and the center of projection of camera $m$.
	
	\subsection{SLF Coefficient Compression and Reconstruction}
	\label{sec:framework_rec}
	
	Section~\ref{sec:representation} explained how to obtain for each point $\bm{p}$ on the surface a compact representation of the view map $f(\bm{\omega}|\bm{p})$ as an $N$-dimensional vector $\bm{\alpha}(\bm{p})$ of coefficients of B-spline wavelets. In this subsection, we describe how to compress $\bm{\alpha}(\bm{p})$ efficiently across the surface.
	
\begin{figure}
		\centering
		\subfigure[$\bm{\alpha}_0$]{\includegraphics[height=0.23\linewidth]{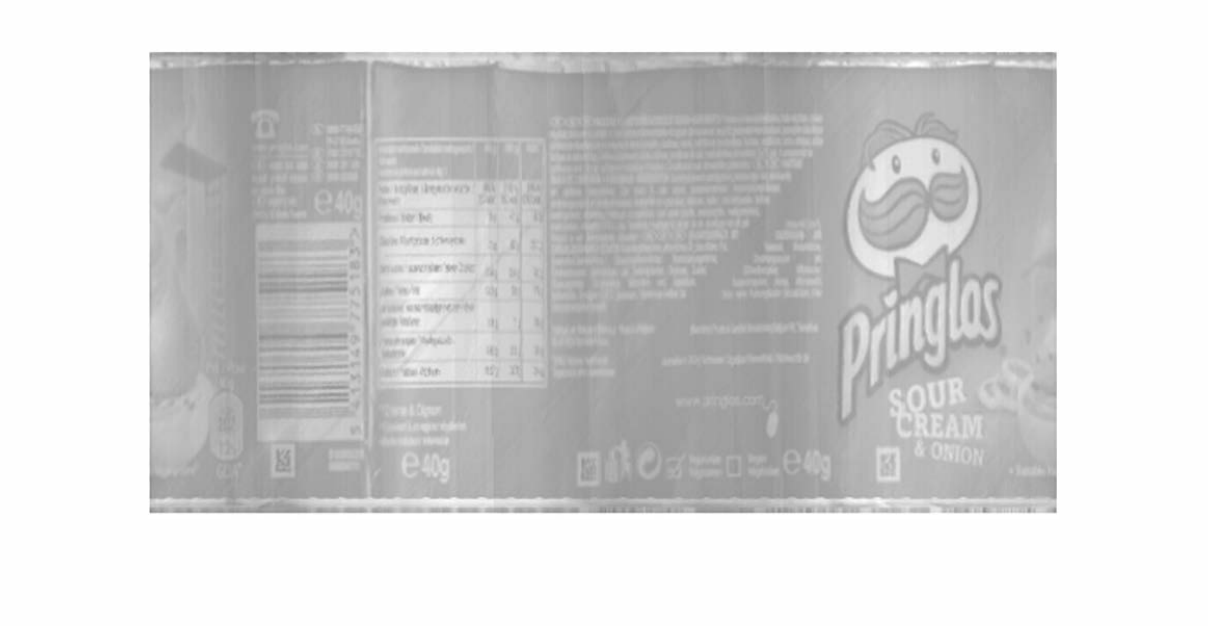}}
		\subfigure[$\bm{\alpha}_1$]{\includegraphics[height=0.23\linewidth]{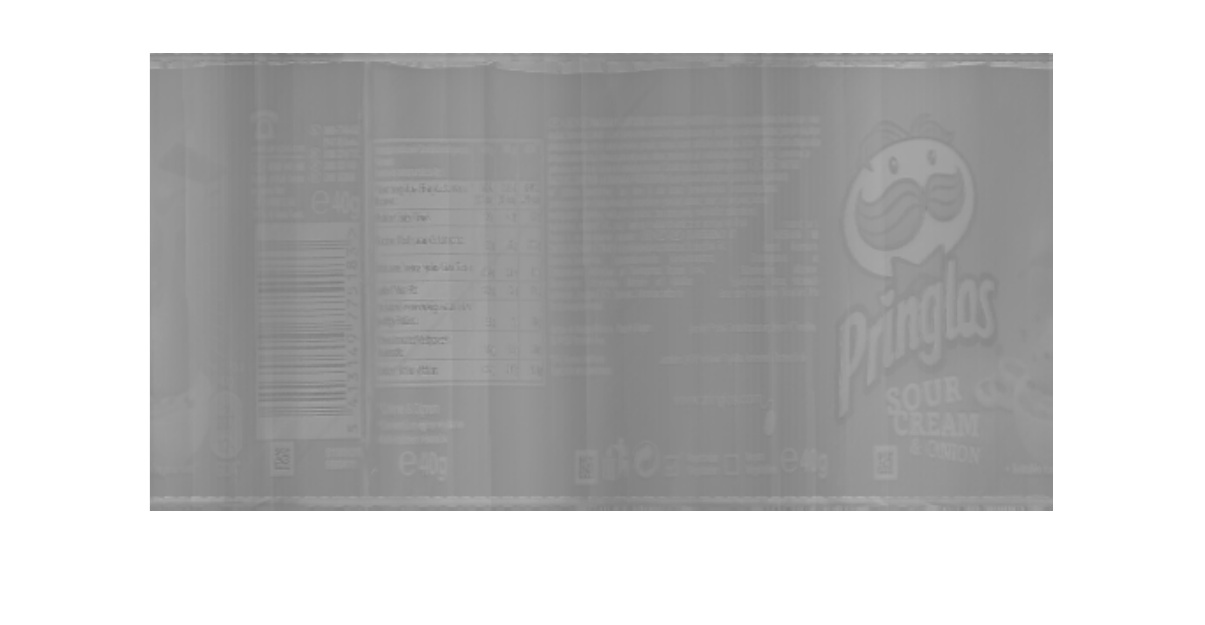}}
		\subfigure[$\bm{\alpha}_2$]{\includegraphics[height=0.23\linewidth]{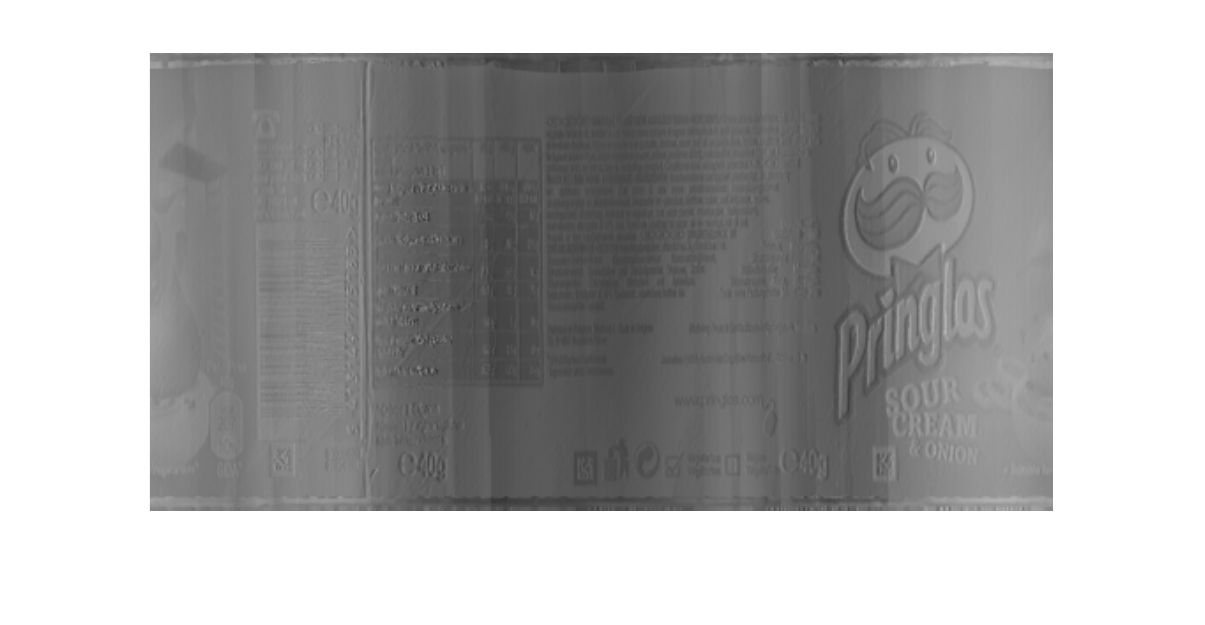}}
		\subfigure[$\bm{\alpha}_3$]{\includegraphics[height=0.23\linewidth]{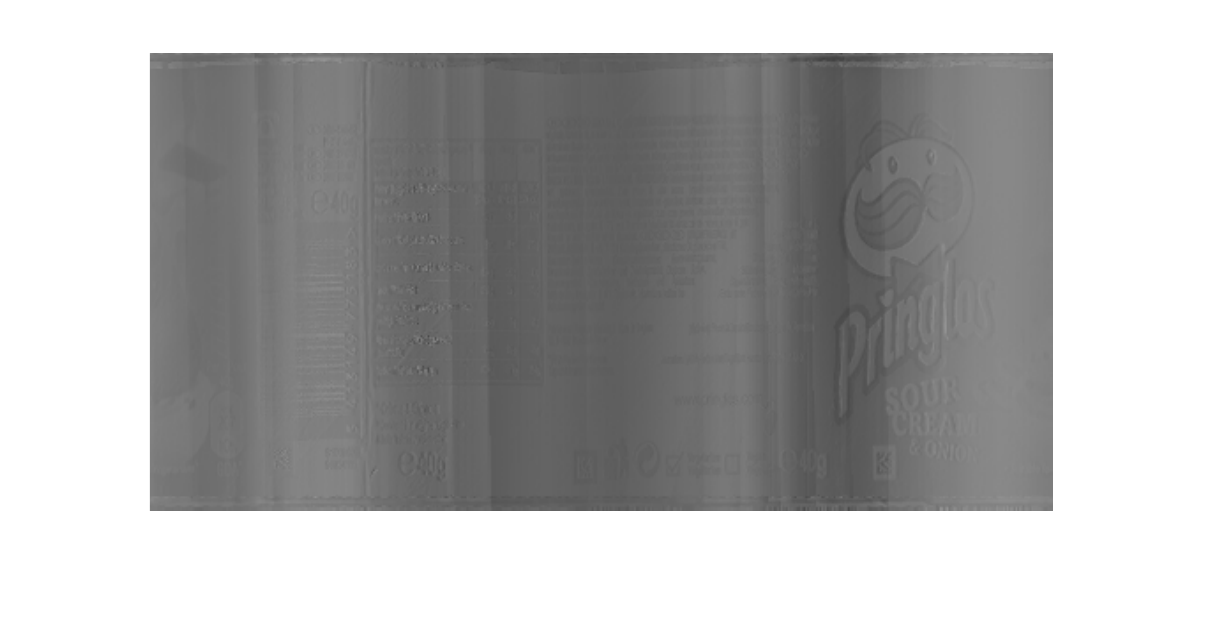}}
		\caption{Visualization of the first four SLF coefficients after projecting them into a 2D image plane. An affine transform is applied to the coefficients in order to constrain their values in the range of [0, 255], with brighter pixels indicating larger coefficients and vice versa.}
		\label{fig:texturemap}
	\end{figure}

	The motivation is based on the fact that the SLF coefficients are strongly correlated across the surface. As an example, Fig.~\ref{fig:texturemap} shows the first four SLF coefficients (of the surface of the {\em Can} dataset shown in Fig.~\ref{fig:camera}) visualized by projecting them into an image plane, where in this case two neighboring positions in the 2D plane are also neighbors in 3D. One can observe high correlations among the SLF coefficient maps, motivating us to remove the spatial redundancy between neighboring points. Since the B-Spline wavelet functions are orthogonal to each other, each coefficient plane can be compressed independently. By viewing the coefficients as additional attributes of the point cloud, any point cloud attribute coding method can be applied to compress them.
	
	We use two different methods for point cloud attribute coding.  The first is called region adaptive hierarchical transform (RAHT) coding \cite{raht}, which is at the core of the MPEG Point Cloud Codec (PCC) Test Model Category 1 (TMC1) \cite{tmc1}. Since the distribution of points on a discretized surface, or point cloud, can be rather sparse compared to the whole space, RAHT adaptively applies the 2-point Haar transform to two spatially neighboring points and progressively groups them. After transformation, scalar quantization is applied to each coefficient given a quantization step-size $Q$,
	\begin{equation}\label{quant}
	\hat{\bm{F}}_i = {Round}\left(\frac{\bm{F}_i}{Q}\right)Q,
	\end{equation}
	where $\bm{F}_i$ denotes the coefficients transformed by RAHT and $\hat{\bm{F}}_i$ the quantized transformed coefficients. Generally, larger values of $Q$ yield lower bitrate but also an inferior reconstruction quality. Then, the quantized coefficients $\hat{\bm{F}}_i$ are entropy encoded. On the decoder side, the coefficients can be recovered by entropy decoding, inverse quantization, and inverse RAHT.
	
	We also use a more classical method, which we call texture map coding (TMC) in this paper. Texture mapping is done by mapping each point on the 3D surface to a point in 2D, such that the colors on the discretized surface, or point cloud, can be viewed as an image. One example of a texture map is already shown in Fig.~\ref{fig:texturemap}, where the cylinder surface is unwrapped to a rectangle. By texture mapping, the $N$-dimensional vectors $\bm{\alpha}(\bm{p})$ can be mapped onto $N$ images. Since the B-Spline wavelet basis functions are orthogonal to each other, we can compress each SLF coefficient $\alpha_i(\bm p)$ independently.  Therefore each image can be compressed by any image codec. Before image coding, the coefficients are uniformly rescaled and rounded to integers in the range of [0, 255] for an 8 bit representation. It is worth noting that the mapping from 3D to 2D must also be encoded for the decoder reconstruction. In this work, we use the reference software for the MPEG PCC Test Model Category 2 (version 0.0 TMC2v0) \cite{tmc2}, where an adaptive texture mapping is applied and the texture maps along with the geometry are then encoded by High Efficiency Video Coding (HEVC) \cite{hevc}.
	
	
	Both RAHT and TMC are lossy, in that errors are introduced by quantization. We denote the recovered SLF coefficients on the decoder side as $\hat{\bm{\alpha}}$. Given $\hat{\bm{\alpha}}$, the view maps can be easily reconstructed at the decoder side as
	\begin{equation}\label{eq:slf_rec}
	\hat f(\bm{\omega}|\bm{p})=\sum_{i=0}^{N-1}{\hat{\alpha}_i(\bm{p}) \cdot g_i\left( \bm{\omega} \right)}.
	\end{equation}
	Accordingly, the radiance of any point $\bm{p}$ from an arbitrary viewpoint can then be obtained as $\hat f(\bm{\omega}|\bm{p})$, where $\bm{\omega}=(\theta,\phi)$ indicates the direction from $\bm{p}$ to the viewpoint. Thus, an efficient free viewpoint rendering can be achieved feasibly at the decoder side.
	
	\section{Experimental Results}
	\label{sec:results}
	
	In this section, the datasets and evaluation methodology are introduced in Subsections~\ref{sec:results_datasets} and \ref{sec:results_evaluation_method}. We then evaluate our method from four perspectives. First, we show the robustness and adaptiveness of our method by investigating the impact of super-parameters in Subsections~\ref{sec:results_impact_para} and \ref{sec:results_impact_camera_density}. Second, we analyze the scalability of our method in Subsection~\ref{sec:results_scalability}. Third, we demonstrate the subjective quality of our method for free viewpoint rendering in Subsection~\ref{sec:results_recon}. Finally, we compare the RD performance and complexity of our method with an image based method in Subsection~\ref{sec:results_rd_compare}.
	
	\subsection{Datasets}
	\label{sec:results_datasets}
\begin{figure*}
		\centering
		\includegraphics[width=1\linewidth]{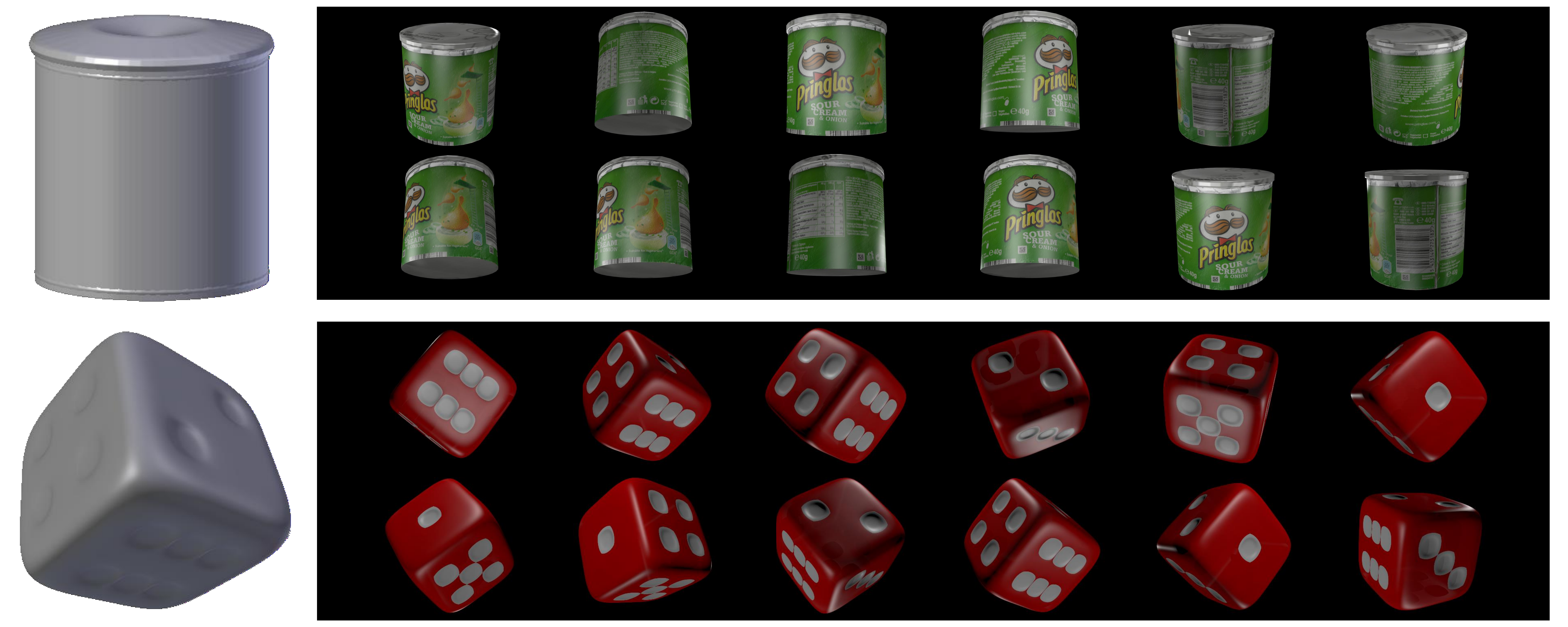}\\
		\caption{Two synthetic datasets: {\em Can} and {\em Die}. (left) 3D geometry, (right) Images from various viewpoints.}\label{fig:database}
	\end{figure*}

\begin{figure}
		\centering
		\includegraphics[width=0.8\linewidth]{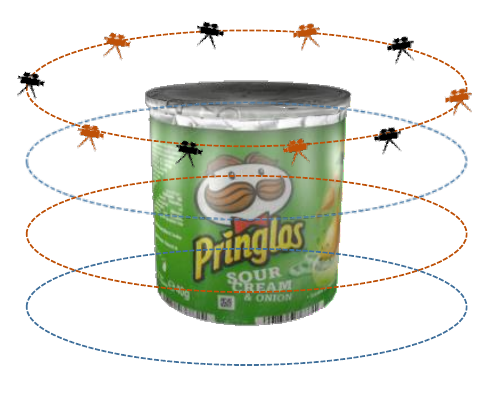}\\
		\caption{Illustration of camera distribution. Cameras are located on several circles around the object. They are used for capturing images of the object from different angles, some serving as inputs to the system and others as ground truth for evaluation.}\label{fig:camera}
	\end{figure}

	For evaluation, we use two synthetic and two real datasets.  The synthetic datasets, {\em Can} and {\em Die}, are derived from CG objects \cite{free3d}, whose 3D geometries are shown in Fig.~\ref{fig:database} on the left. Blender v2.78c \cite{blender} is used to render images of the objects from various viewpoints, some of which are shown in Fig.~\ref{fig:database} on the right. The {\em Can} has a complex texture and specular surface. Four light sources illuminate the {\em Can}. One can observe from the right images in Fig.~\ref{fig:database} how the surface of the {\em Can} reflects light differently from different directions. The {\em Die} dataset illustrates specularity as well as transparency. Additional details are given in Table~\ref{tab:dataset}.
	For each synthetic object, 550 images are captured from virtual cameras positioned around the object.  As illustrated in Fig.~\ref{fig:camera}, the cameras are located in several circles parallel to the $x$-$y$ plane. In our experiments, there are 11 circles spaced evenly along the $z$ axis and 50 cameras spaced evenly around each circle, for a total of 550 cameras.
	
	\begin{table}
		\centering
		\caption{Dataset specifications, including the number of vertices, surface material properties, and number of light sources. Material properties include diffuse, specular, and transparent properties range from 0 to 1.}
		\begin{tabular}{|c|c|c|c|c|c|}
			\hline
			\textbf{} & \textbf{\# Vertices} & \textbf{Diffuse} & \textbf{Specular} & \textbf{Transparent} & \textbf{\# Lights} \bigstrut\\
			\hline
			\textbf{\emph{Can}} & 91,462 & 0.8   & 0.5   & 0.0     & 4 \bigstrut\\
			\hline
			\textbf{\emph{Die}} & 59,906 & 0.8   & 0.5   & 0.2   & 3 \bigstrut\\
			\hline
		\end{tabular}%
		\label{tab:dataset}%
	\end{table}%
	
	For the real datasets, we use the dataset proposed in \cite{wood2000surface}, containing two objects, {\em Elephant} and {\em Fish}. They have rich texture, specular surfaces, and complex lighting. {\em Elephant} has 316 images captured by calibrated cameras from different viewpoints and {\em Fish} has 582 images. The corresponding camera parameters for each image are included in the dataset. The 3D geometries for each object are also provided, where {\em Elephant} and {\em Fish} have 155,688 and 64,982 vertices on the object surface, respectively.
	
	\subsection{Evaluation Methodology}
	\label{sec:results_evaluation_method}
	
	We evaluate the RD performance of the proposed method.  The overall bitrates reported in our experiments include the bits required to represent the geometry, as well as the SLF coefficients.  The reconstruction error at a particular point $\bm{p}$ for direction $\bm{\omega}$ is calculated as the difference between the reconstruction and the ground truth, \ie $\|\hat f(\bm{\omega_e}|\bm{p})-\bm{c_e}\|_2^2$, where $\bm{c_e}$ and $\bm{\omega_e}$ indicate the pixel values and directions of valid evaluation cameras, respectively.  Here the valid evaluation camera is determined by a parameter $\delta'$, which is similar to $\delta$ as described in Section~\ref{sec:framework_preprocessing}. $\delta'$ specifies the range of valid angles for the evaluation of each surface point. A $\delta'$ closer to $0$ indicates that the evaluation is performed over a wider viewing range, including glancing surface angles.  In the experiments, the reconstruction error is evaluated for different values of $\delta'$ in order to assess quality from various angles of view.
	
	For each dataset, we separate the images into two sets, an input set and an evaluation set, where the input set is used for SLF representation and the evaluation set for evaluation. If one calculates the reconstruction errors on the input set, the result reflects how accurate the SLF model fits the input SLF data. If one evaluates on the evaluation set, the result shows the generalization ability of the SLF model and how well it can estimate virtual views from arbitrary directions.  We also compare the performances under different input camera densities, from dense to sparse. We evaluate on three test configurations as detailed in Table~\ref{tab:cameras}. The sparse case may be more practical since the number of input cameras is always limited in real-life applications.  Regarding two synthetic objects {\em Can} and {\em Die}, in the dense case, for example, there are 125 input cameras arranged in five circles of 25 cameras each, while the remaining 425 cameras are used for evaluation.  For two real objects {\em Elephant} and {\em Fish}, the input images are selected proportionally, where 1/2, 1/4, and 1/8 of the images serve as input for dense, intermediate and sparse cases, respectively, and the remaining serves as evaluation.
	
	\begin{table}
		\centering
		\caption{Input (and evaluation) camera numbers of three test cases.}
		\begin{tabular}{|c|c|c|c|c|}
			\hline
			\textbf{} & \textit{\textbf{Can / Die}} & \textit{\textbf{Elephant}} & \textit{\textbf{Fish}} \bigstrut\\
			\hline
			\textbf{Dense} & $5\times25=125 (425)$ & 158 (158) & 291 (291) \bigstrut\\
			\hline
			\textbf{Intermediate} & $3\times13=39 (511)$ & 79 (237) & 146 (436) \bigstrut\\
			\hline
			\textbf{Sparse} & $2\times7=14 (536)$  & 40 (276) & 73 (509) \bigstrut\\
			\hline
		\end{tabular}%
		\label{tab:cameras}%
	\end{table}%
	
	Regarding the point cloud codecs for compressing the coefficients spatially, we use RAHT from MPEG PCC Test Model Category 1 to evaluate robustness and scalability in Subsections from \ref{sec:results_impact_para} to \ref{sec:results_recon} and we use texture map coding (TMC) from MPEG PCC Test Model Category 2 to compare RD performance in Subsection~\ref{sec:results_rd_compare}, because RAHT is much more computationally efficient and TMC achieves higher RD performance.
	
	\subsection{Impact of parameters}
	\label{sec:results_impact_para}
	
\begin{figure*}
		\centering
		\subfigure[$\delta'=0^{\circ}$]{\includegraphics[width=0.24\linewidth]{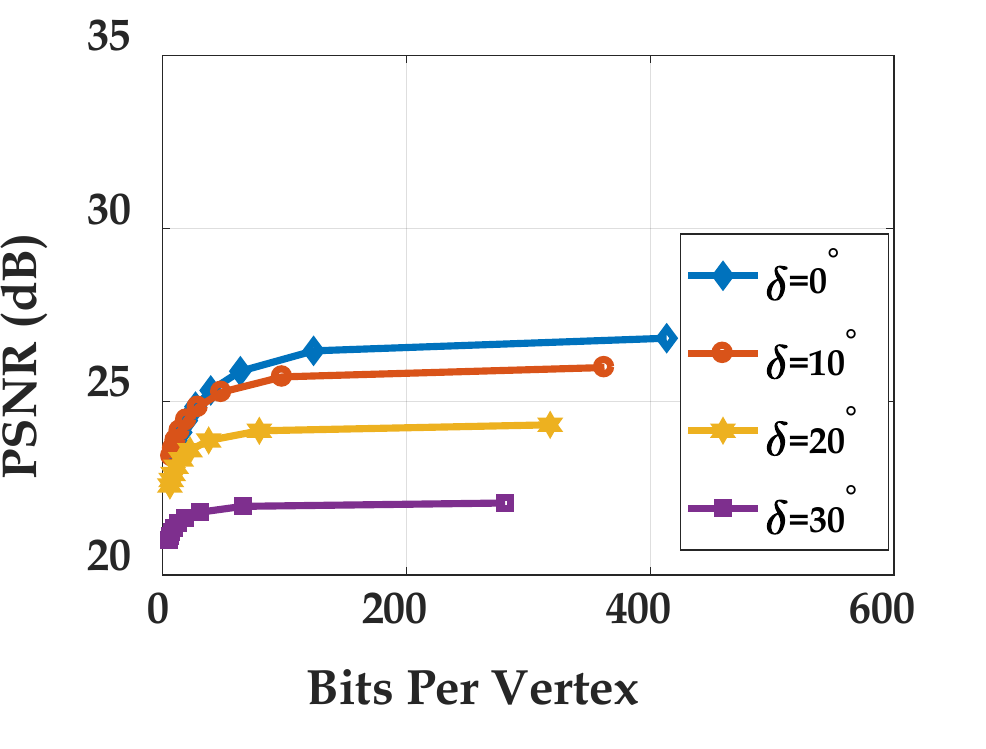}}
		\subfigure[$\delta'=10^{\circ}$]{\includegraphics[width=0.24\linewidth]{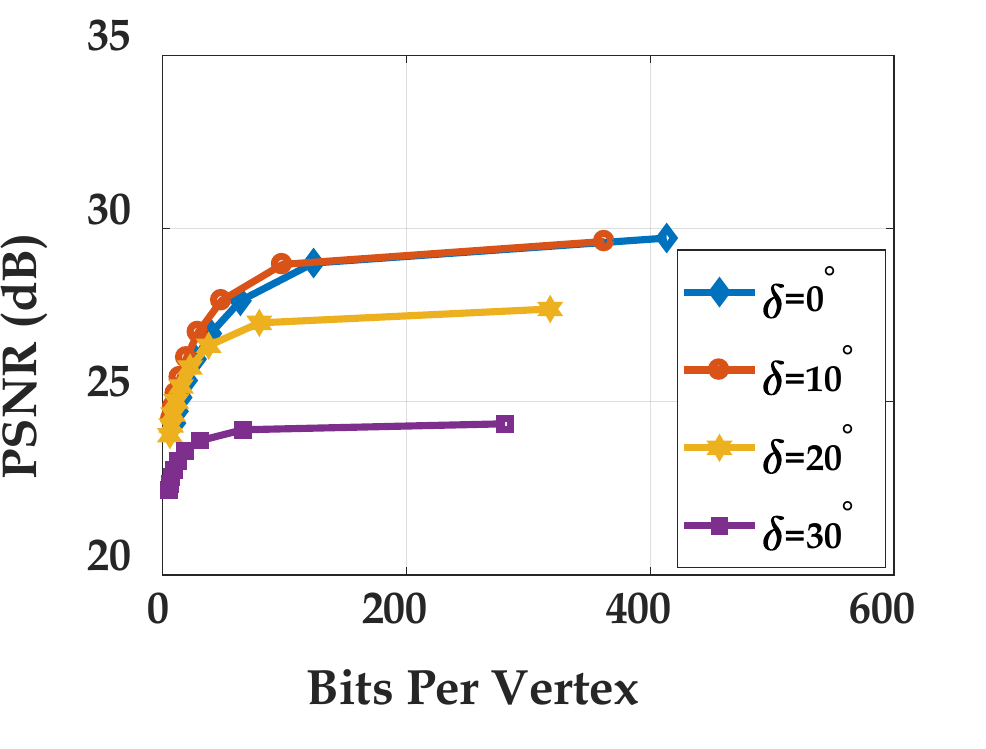}}
		\subfigure[$\delta'=20^{\circ}$]{\includegraphics[width=0.24\linewidth]{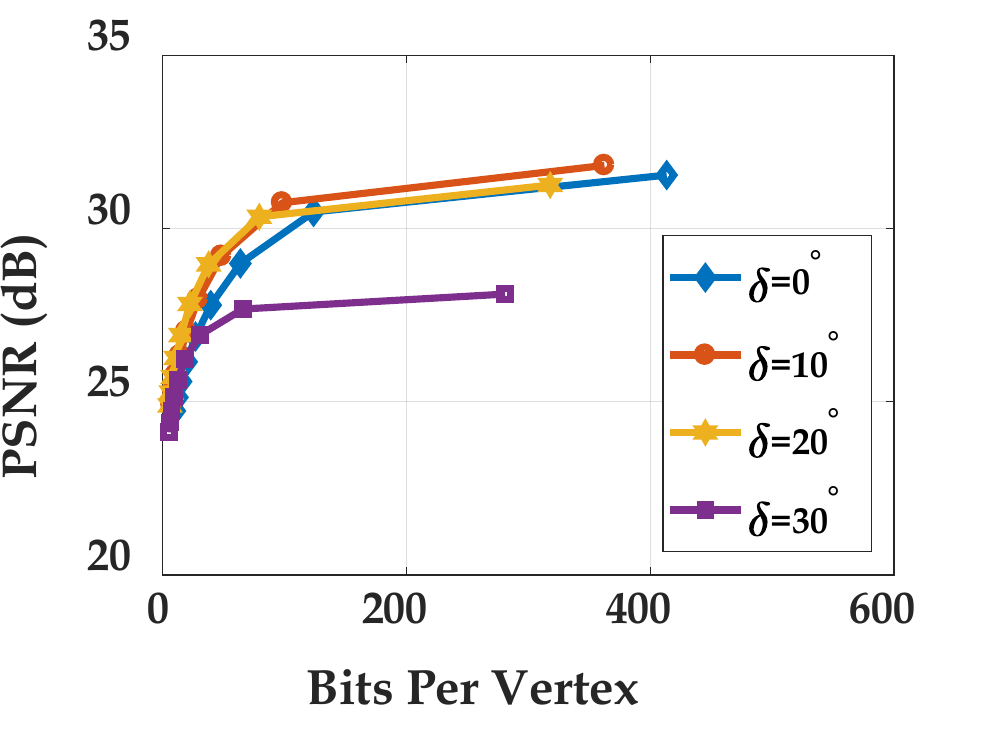}}
		\subfigure[$\delta'=30^{\circ}$]{\includegraphics[width=0.24\linewidth]{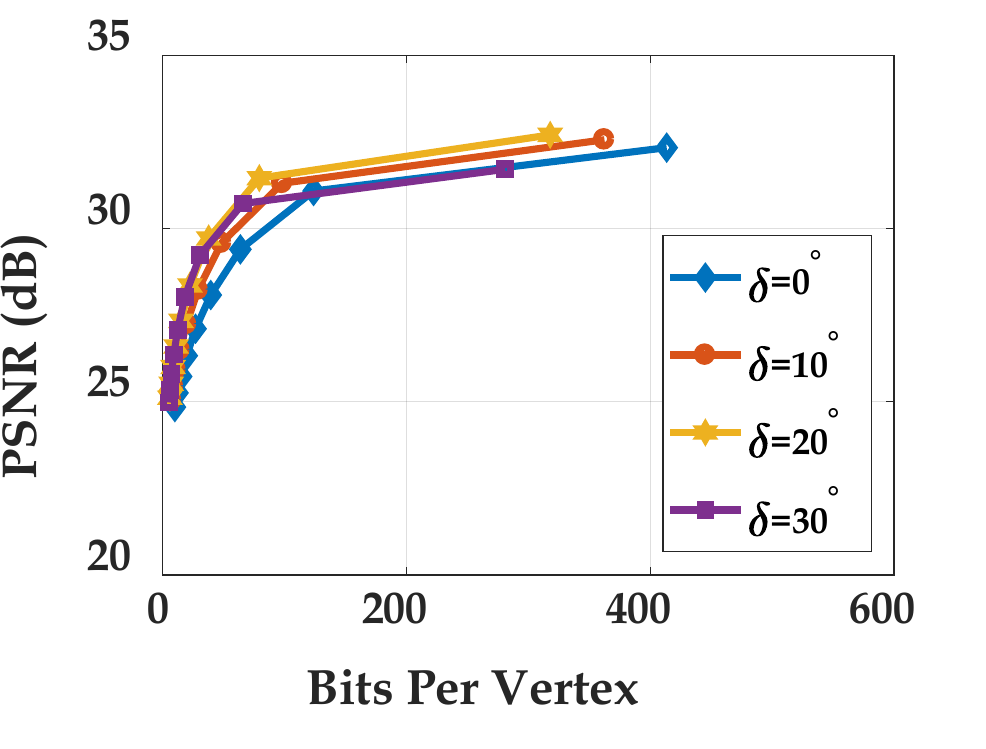}}
		\caption{RD performance comparisons of different values of $\delta$ and $\delta'$.}
		\label{fig:delta_c}
	\end{figure*}

	First, we investigate how $\delta$ impacts RD performance. Fig.~\ref{fig:delta_c} shows the RD performance for different input and evaluation angles $\delta$ and $\delta'$. One can see that $\delta$ significantly influences RD performance across all evaluation angles. On the one hand, small $\delta$ includes observations from glancing angles, which may introduce noise and outliers, thus resulting in degraded fitting accuracy.  On the other hand, large $\delta$ can suppress noise but may also reduce the ability to render accurately surfaces at glancing angles because of a lack of valid observations there.  A good balance is achieved when $\delta=10^{\circ}$, since RD performance is always best or close to best for different evaluation angles (determined by $\delta'$). Therefore, we choose $\delta=10^{\circ}$ for the remainder of our experiments.
	One can also see that overall quality improves as $\delta'$ increases, indicating viewers may observe better reconstruction quality from frontal views than side views. This makes sense since from extreme angles one can barely observe a surface because of self-occlusion.
	
\begin{figure}
		\centering
		\subfigure[Input set]{\includegraphics[width=0.48\linewidth]{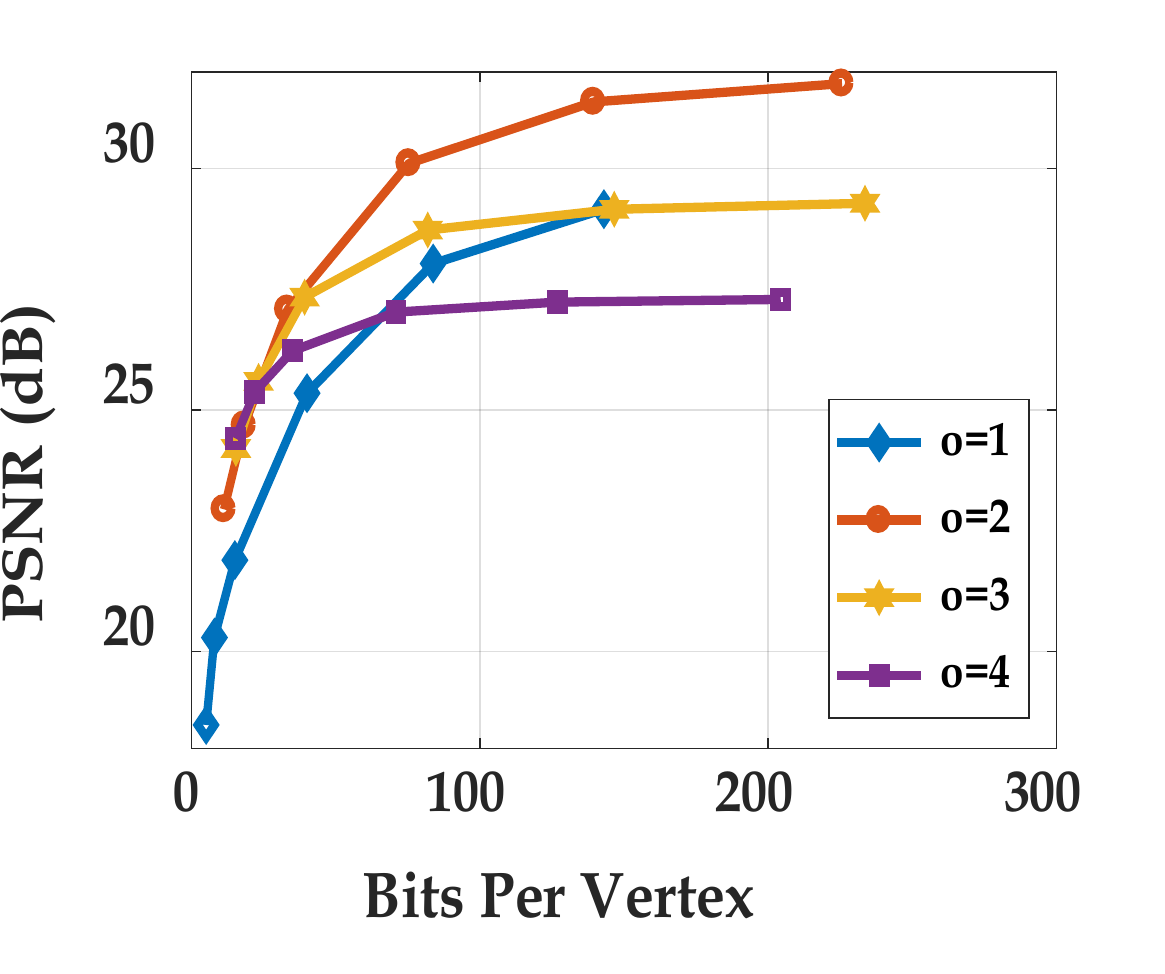}}
		\subfigure[Evaluation set]{\includegraphics[width=0.48\linewidth]{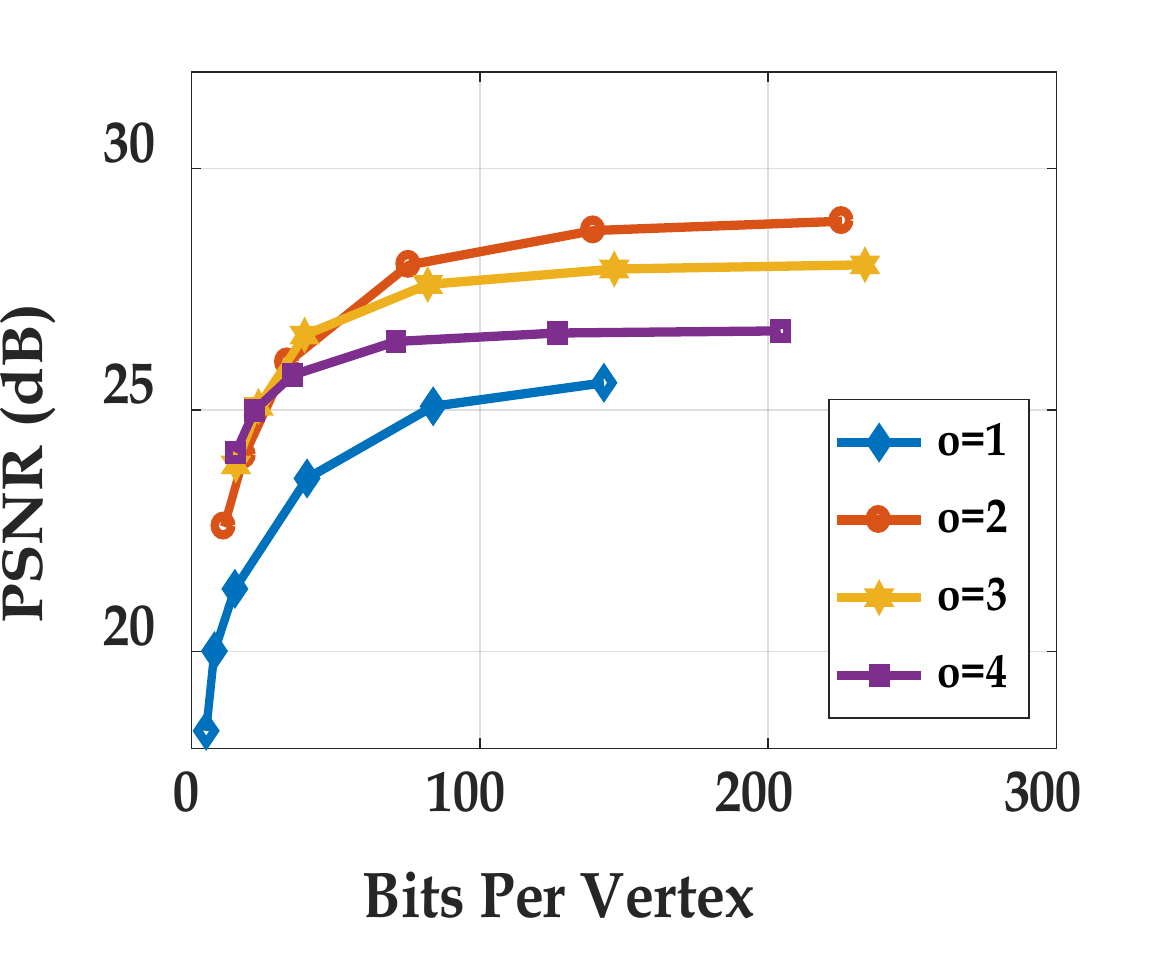}}
		\caption{RD performance comparisons of different B-Spline wavelet orders $o$, where the PSNR is calculated on (a) input image set and (b) evaluation image set, respectively.}
		\label{fig:order}
	\end{figure}

	Second, we investigate how the order $o$ of the B-Spline wavelet function impacts RD performance. Fig.~\ref{fig:order} compares constant, linear, quadratic, and cubic B-Spline wavelet basis functions, for which $o=1,2,3,4$, respectively. When $o=1$, the functions degrade to the well-known Haar wavelets. One can observe that the order dramatically influences the overall RD performance and the linear B-Spline wavelet ($o=2$) outperforms the others especially at high bitrate, indicating that piece-wise linear functions fit the SLF well. Thus, we use the linear B-Spline wavelets as the basis functions in the remainder of this work.
	
\begin{figure}
		\centering
		\subfigure[Input set]{\includegraphics[width=0.48\linewidth]{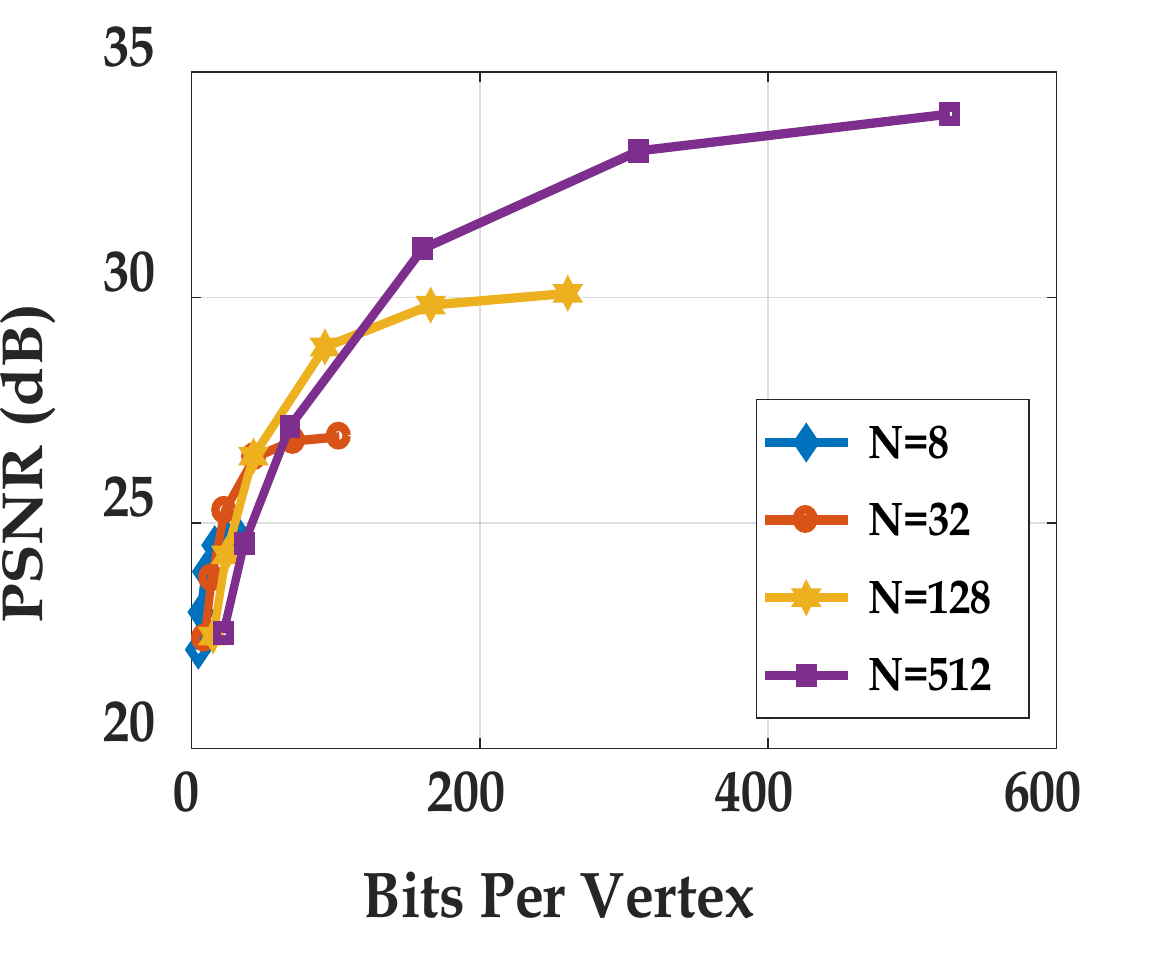}}
		\subfigure[Evaluation set]{\includegraphics[width=0.48\linewidth]{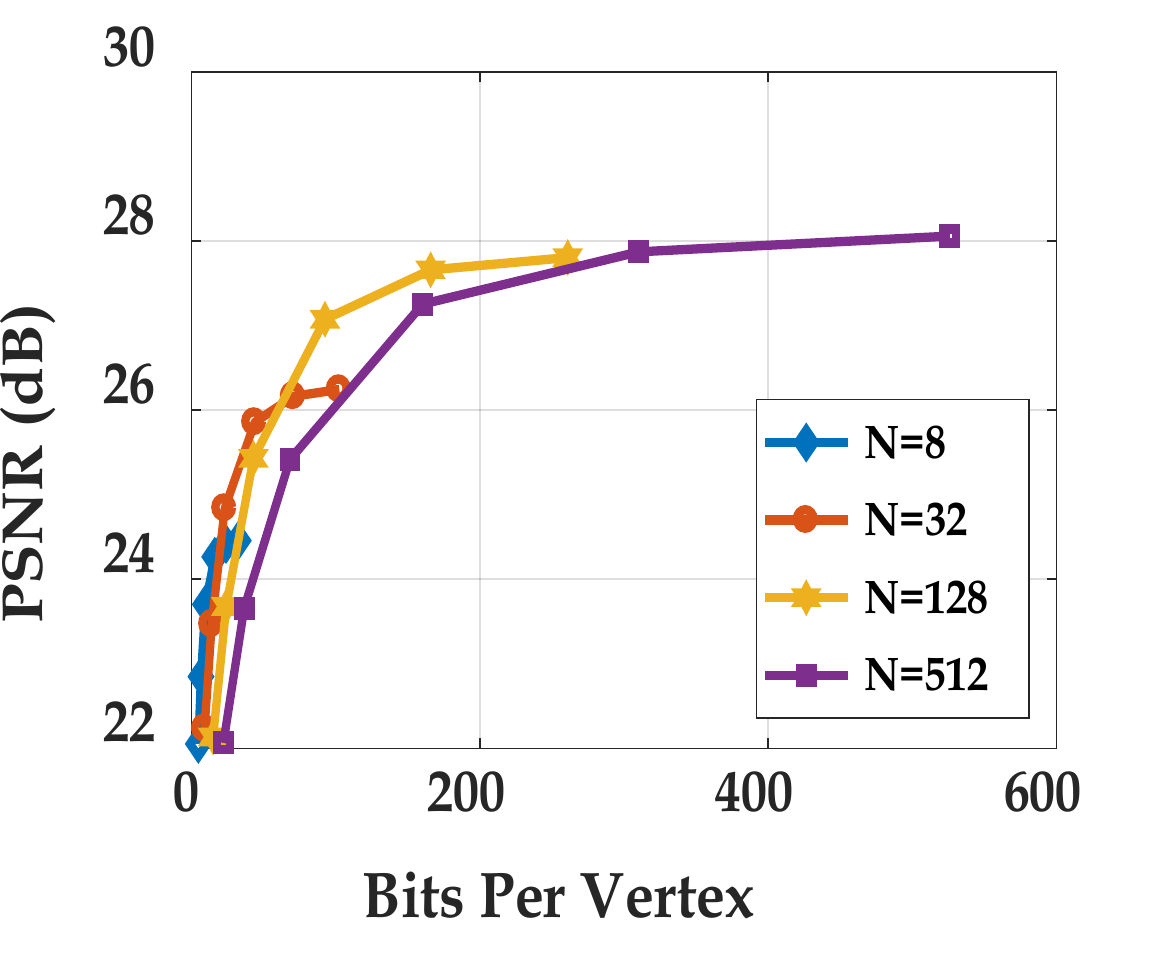}}
		\caption{RD performance comparisons of different numbers of basis functions $N$, where the PSNR is calculated on (a) input image set and (b) evaluation image set, respectively.}
		\label{fig:n_basis}
	\end{figure}

	Third, we investigate how the number of basis functions, $N$, impacts RD performance. Fig.~\ref{fig:n_basis} shows that on the input set, higher PSNR can be achieved with higher $N$, though at the cost of higher bitrate. On the evaluation set, however, there is evidence of over-fitting, as at high bitrate, the PSNR appears to saturate as $N$ increases, while at low bitrates, the PSNR peaks and then deteriorates as $N$ increases. Therefore, we set $N=128$ in this work to strike a balance at the bitrates of interest.
	
\begin{figure}
		\centering
		\subfigure[Input set]{\includegraphics[width=0.48\linewidth]{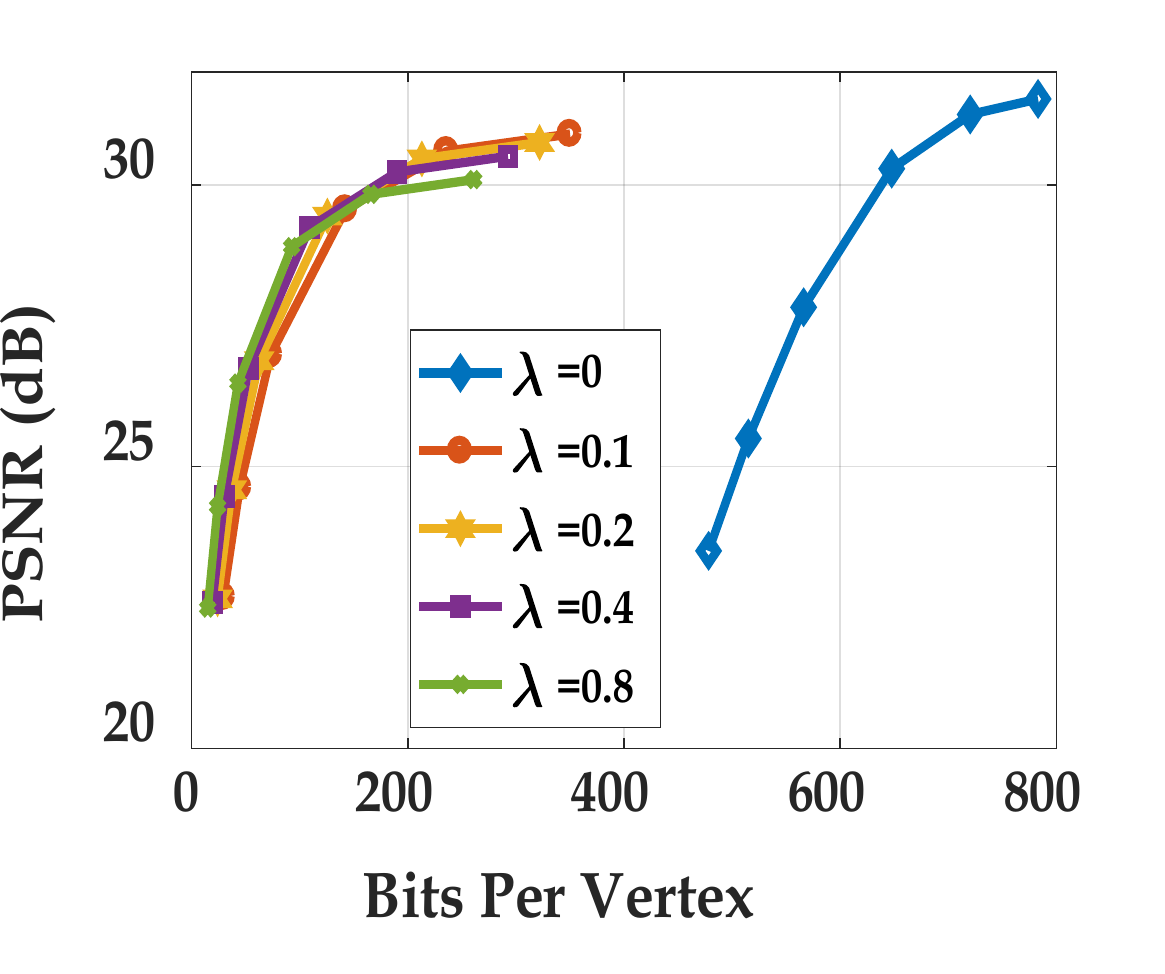}}
		\subfigure[Evaluation set]{\includegraphics[width=0.48\linewidth]{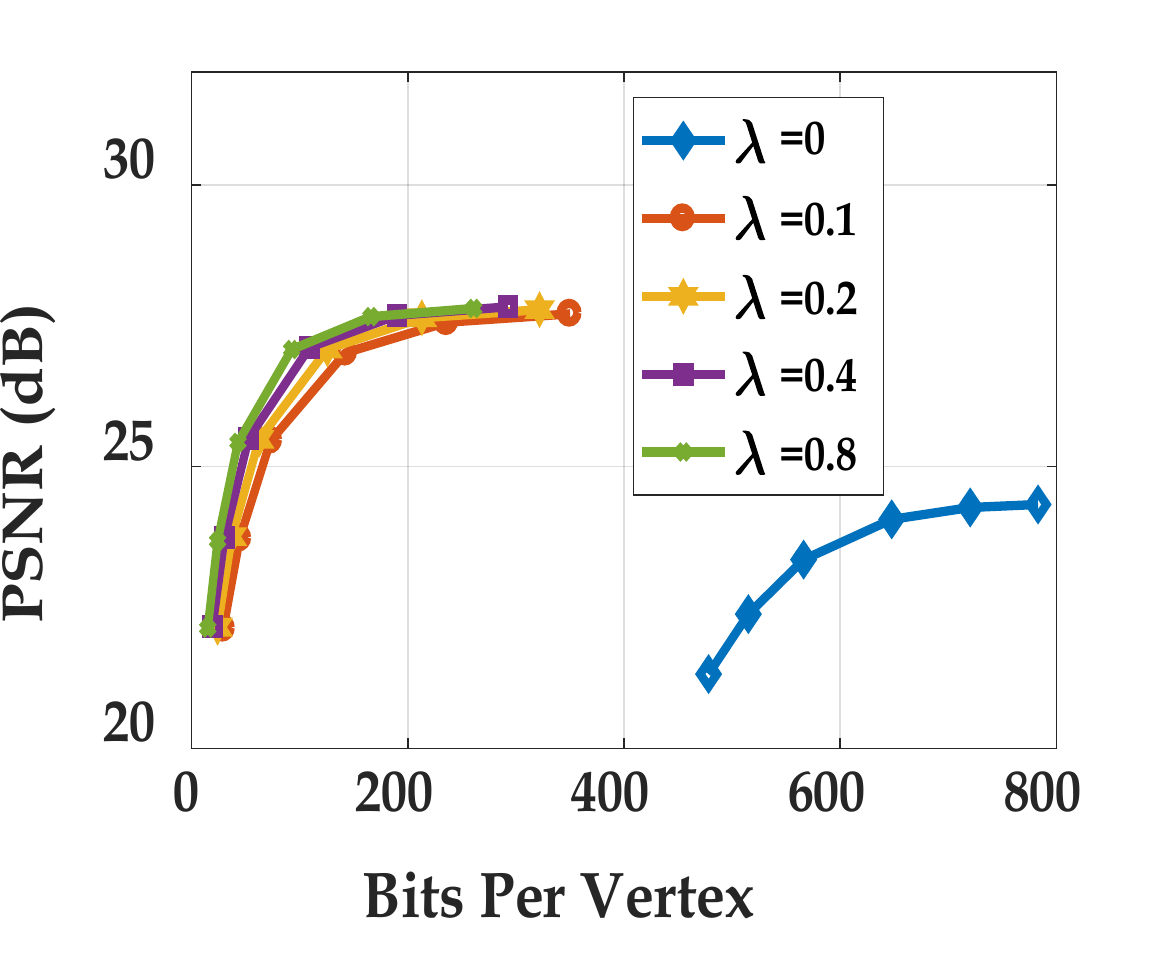}}
		\caption{RD performance comparisons of different values of $\lambda$, where the PSNR is calculated on (a) input image set and (b) evaluation image set, respectively.}
		\label{fig:lambda}
	\end{figure}

	Fourth, we evaluate the impact of $\lambda$ on RD performance. Fig.~\ref{fig:lambda} shows that any non-zero value of $\lambda$ is far superior to $\lambda=0$.  This is because if the number of capture cameras is lower than the number of basis functions, the problem is under-determined.  A small non-zero value of $\lambda$ immediately causes a regularized solution to be found.  Beyond that, though, Fig.~\ref{fig:lambda}(a) shows that a larger $\lambda$ slightly decreases the PSNR on the input set, while Fig.~\ref{fig:lambda}(b) shows that a larger $\lambda$ may slightly increase the PSNR on the evaluation set. This is because a positive $\lambda$ makes the view map smoother and reduces the influence of outliers and noise. Since we target rendering images from arbitrary viewpoints, the performance on evaluation set is more important. Therefore, we set $\lambda=0.8$.
	
\begin{figure}
		\centering
		\subfigure[Input set]{\includegraphics[width=0.48\linewidth]{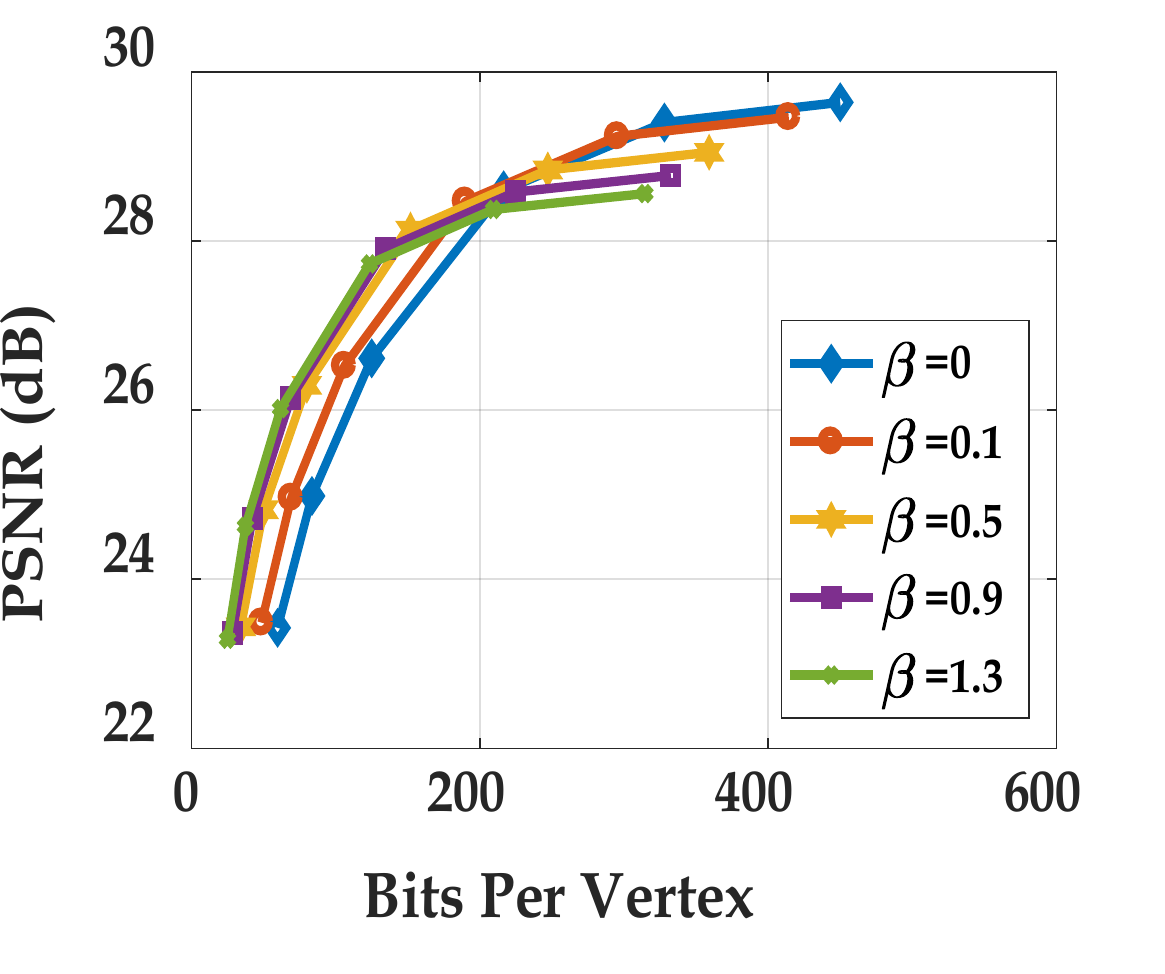}}
		\subfigure[Evaluation set]{\includegraphics[width=0.48\linewidth]{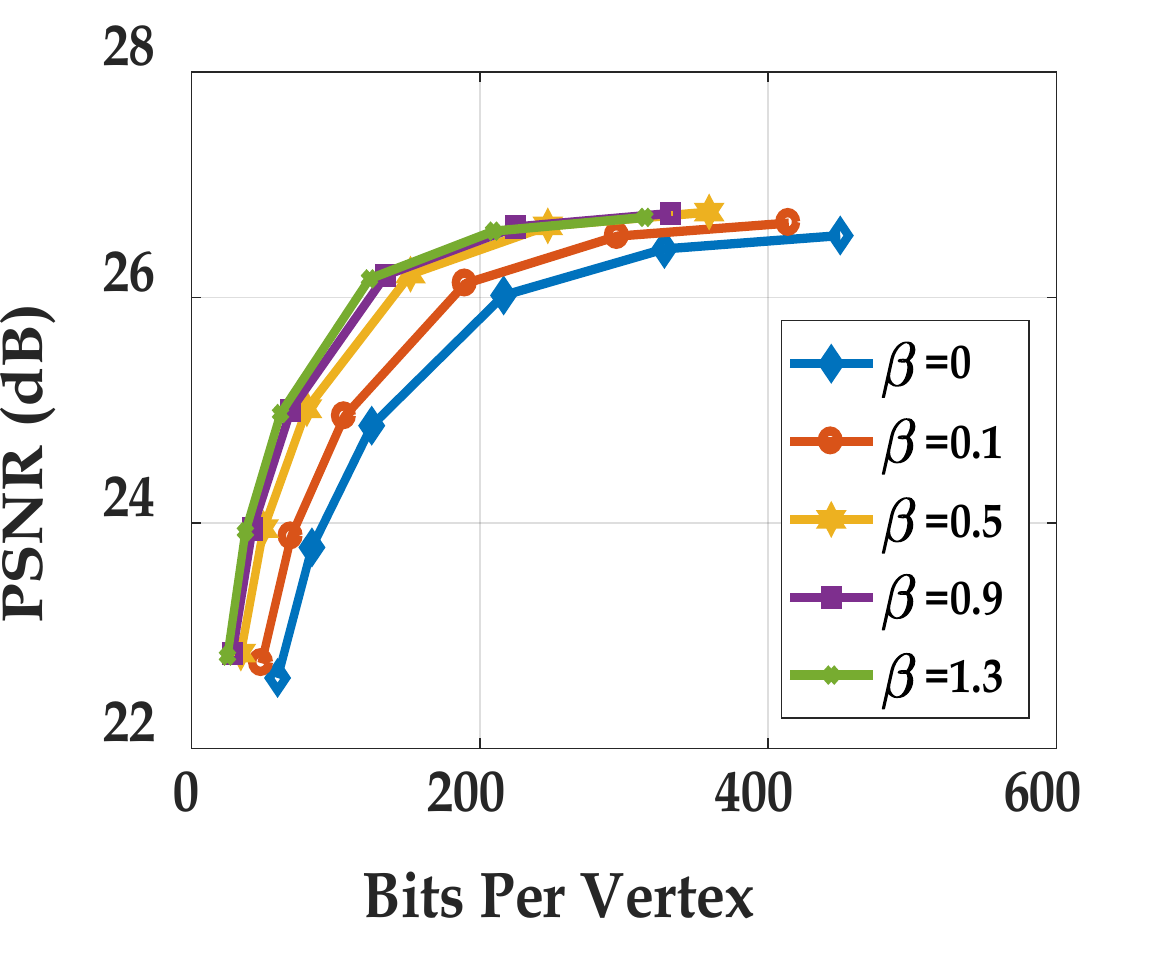}}
		\caption{RD performance comparisons of different values of $\beta$, where the PSNR is calculated on (a) input image set and (b) evaluation image set, respectively.}
		\label{fig:beta}
	\end{figure}

	Lastly, we evaluate RD performance as a function of $\beta$ in Fig.~\ref{fig:beta}. It can be observed that $\beta$ has a similar impact as $\lambda$. Increasing $\beta$ decreases the reconstruction quality for original viewpoints but can increase the PSNR for virtual viewpoints, or alternatively can decrease the bitrate. Considering the overall RD performance on evaluation set, we choose $\beta$ to be $1.3$ in this work.
	
	\subsection{Impact of Camera Density}
	\label{sec:results_impact_camera_density}
	
\begin{figure*}
		\centering
		\subfigure[Groundtruth]{\includegraphics[width=0.24\linewidth]{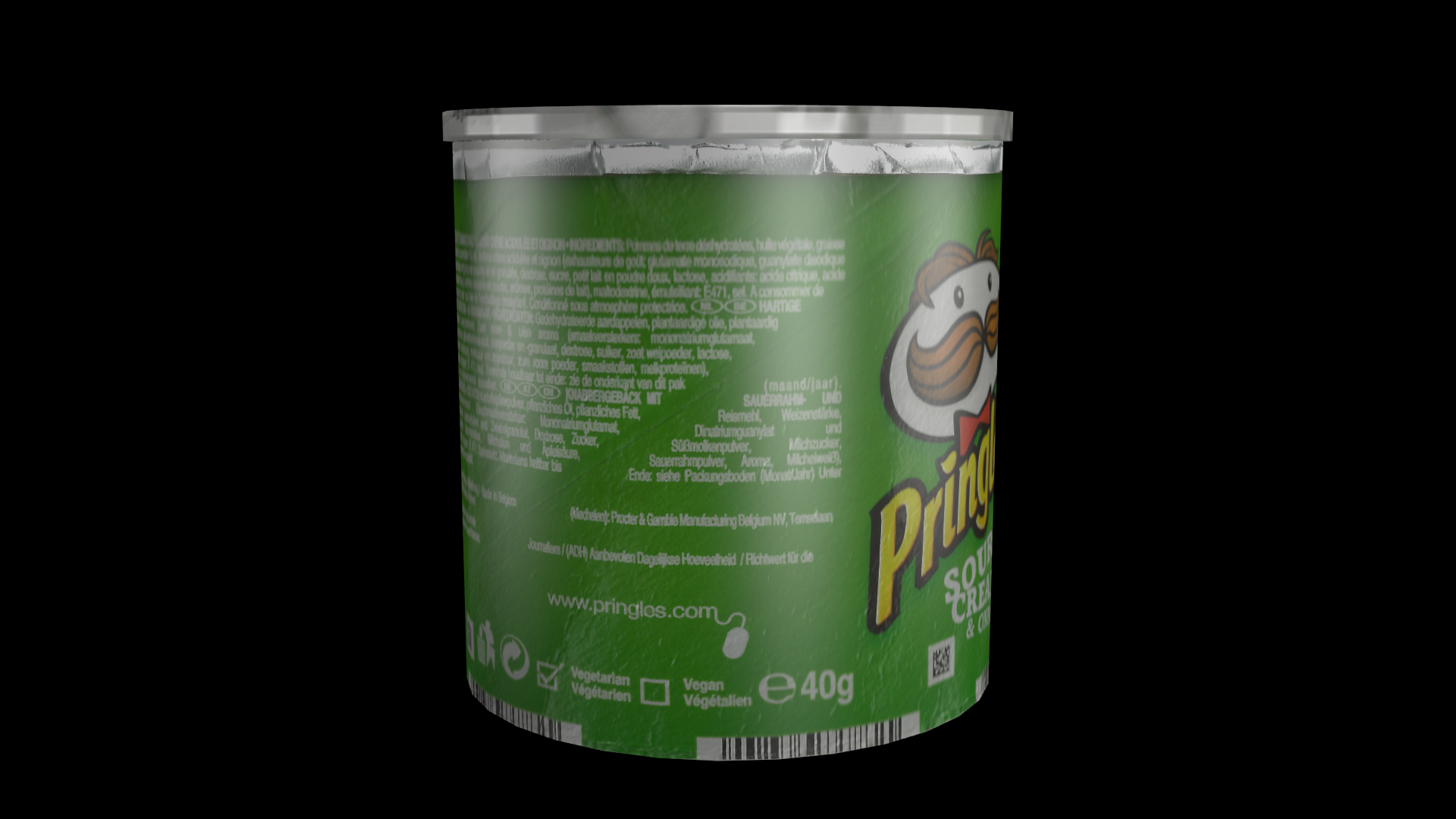}}
		\subfigure[Dense case]{\includegraphics[width=0.24\linewidth]{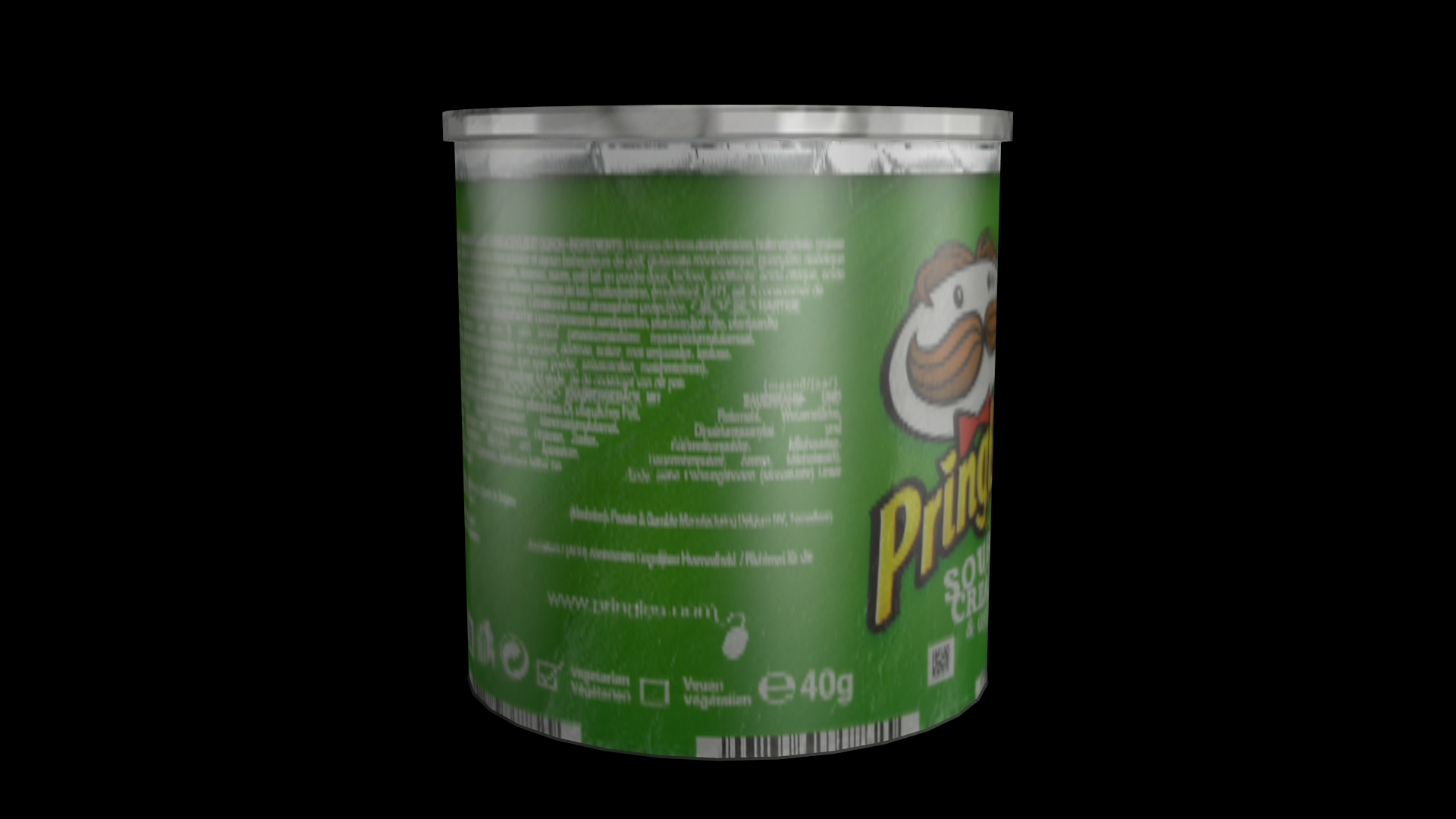}}
		\subfigure[Intermediate case]{\includegraphics[width=0.24\linewidth]{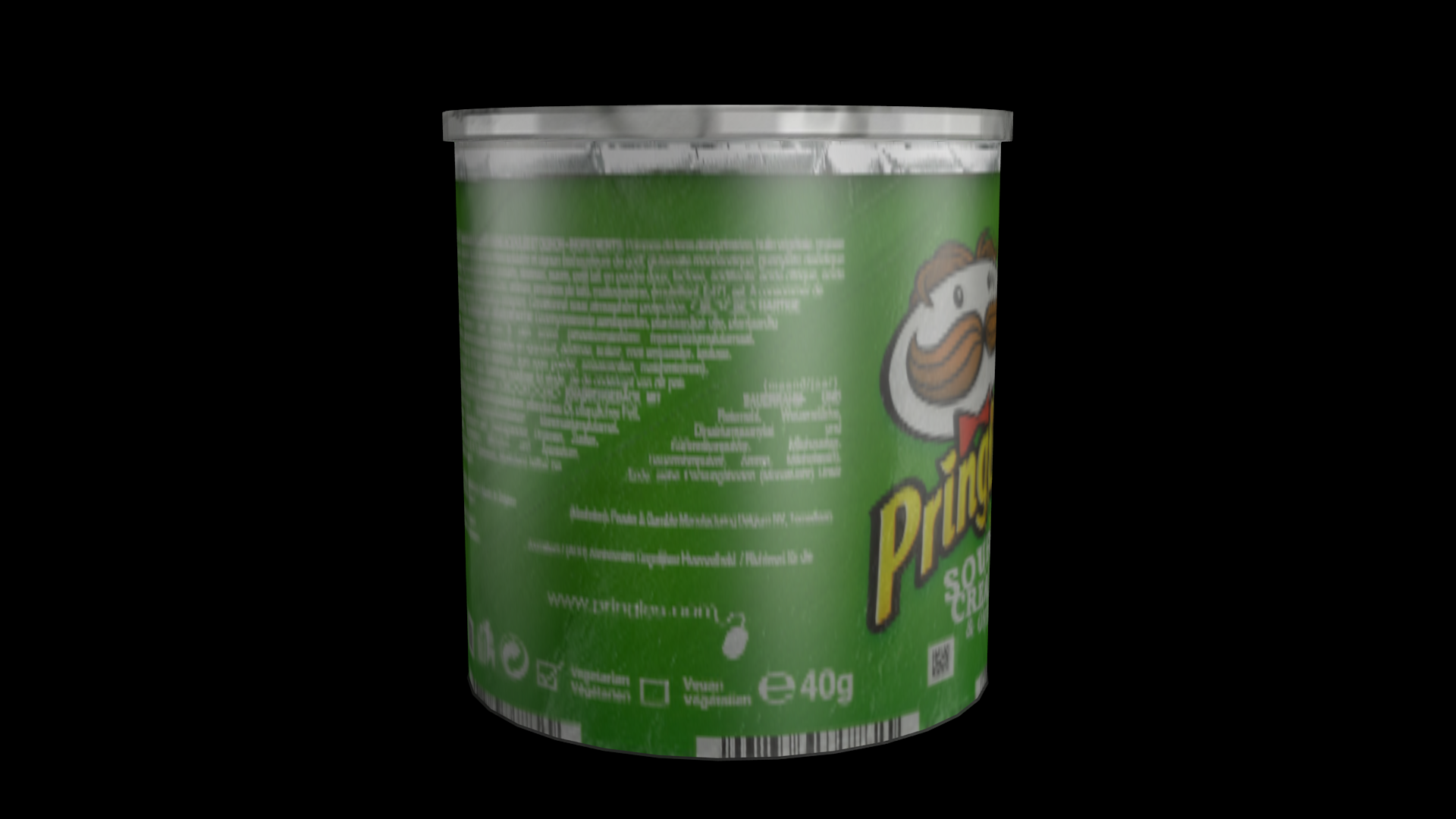}}
		\subfigure[Sparse case]{\includegraphics[width=0.24\linewidth]{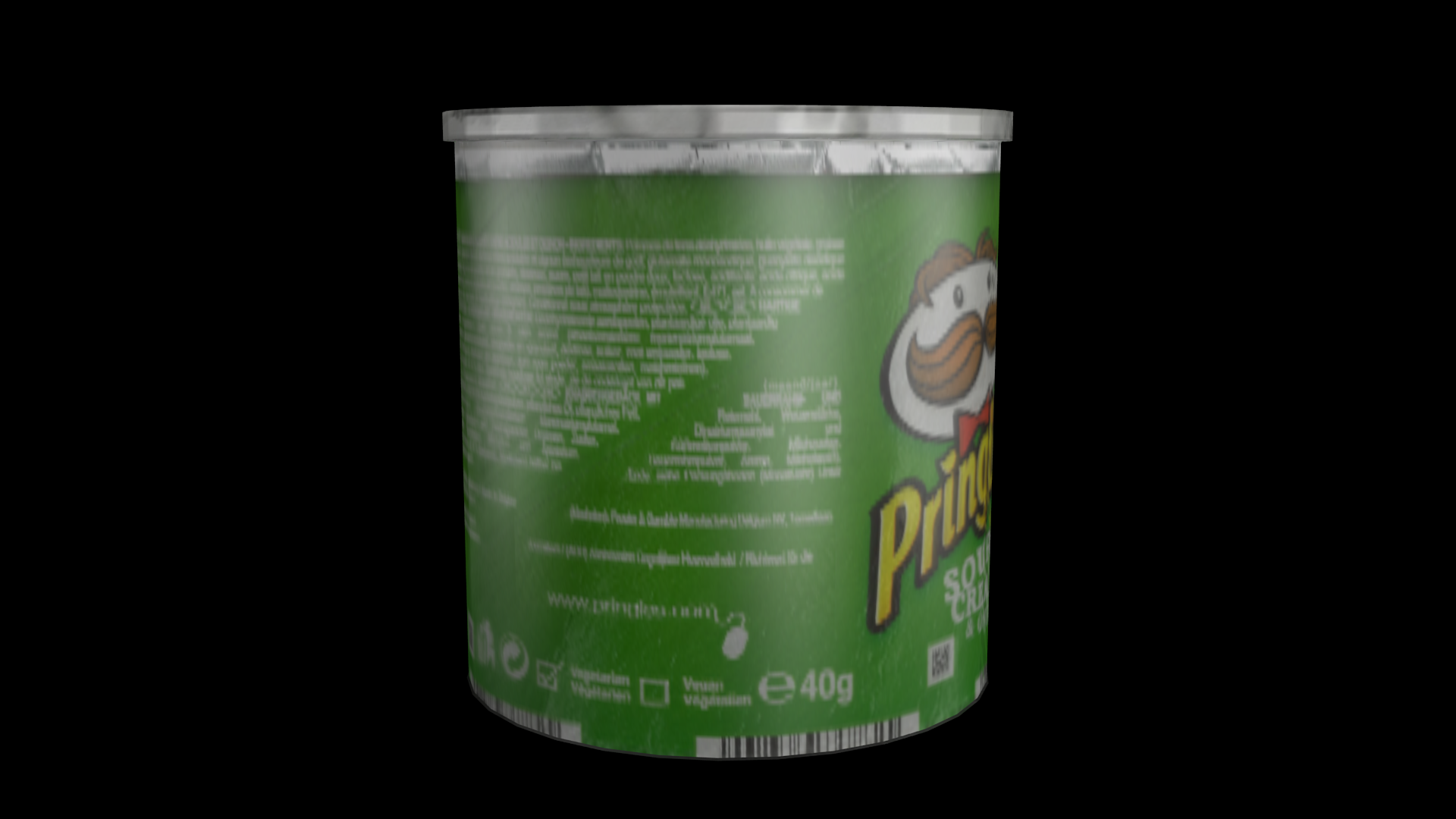}}
		\caption{Reconstruction of {\em Can} from a specific viewpoint as a function of camera density, where the number and placement of cameras in dense, intermediate, and sparse cases are specified in Table~\ref{tab:cameras}. The groundtruth image is shown in (a).}
		\label{fig:recon_density_can}
	\end{figure*}

\begin{figure*}
		\centering
		\subfigure[Groundtruth]{\includegraphics[width=0.24\linewidth]{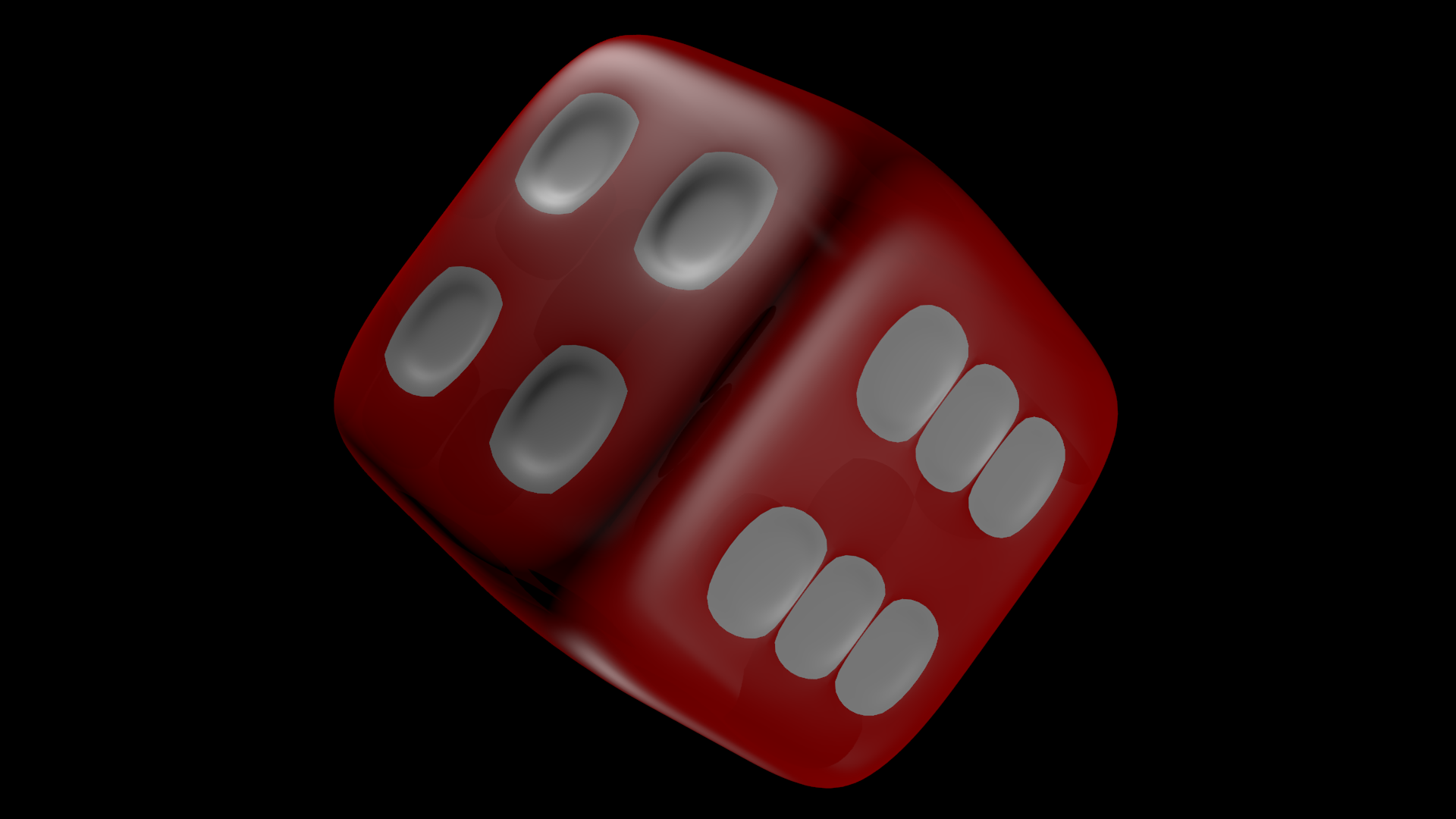}}
		\subfigure[Dense case]{\includegraphics[width=0.24\linewidth]{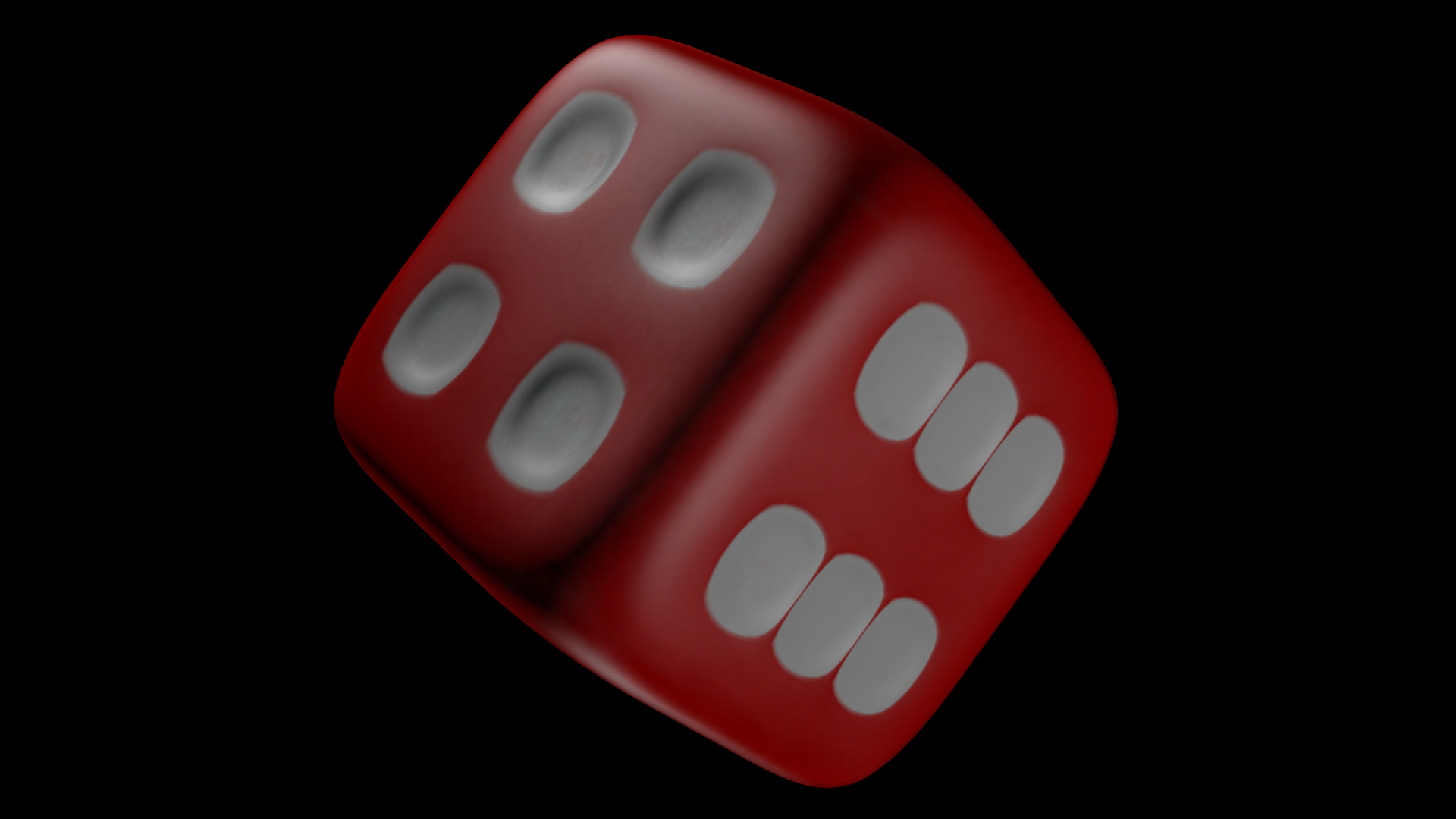}}
		\subfigure[Intermediate case]{\includegraphics[width=0.24\linewidth]{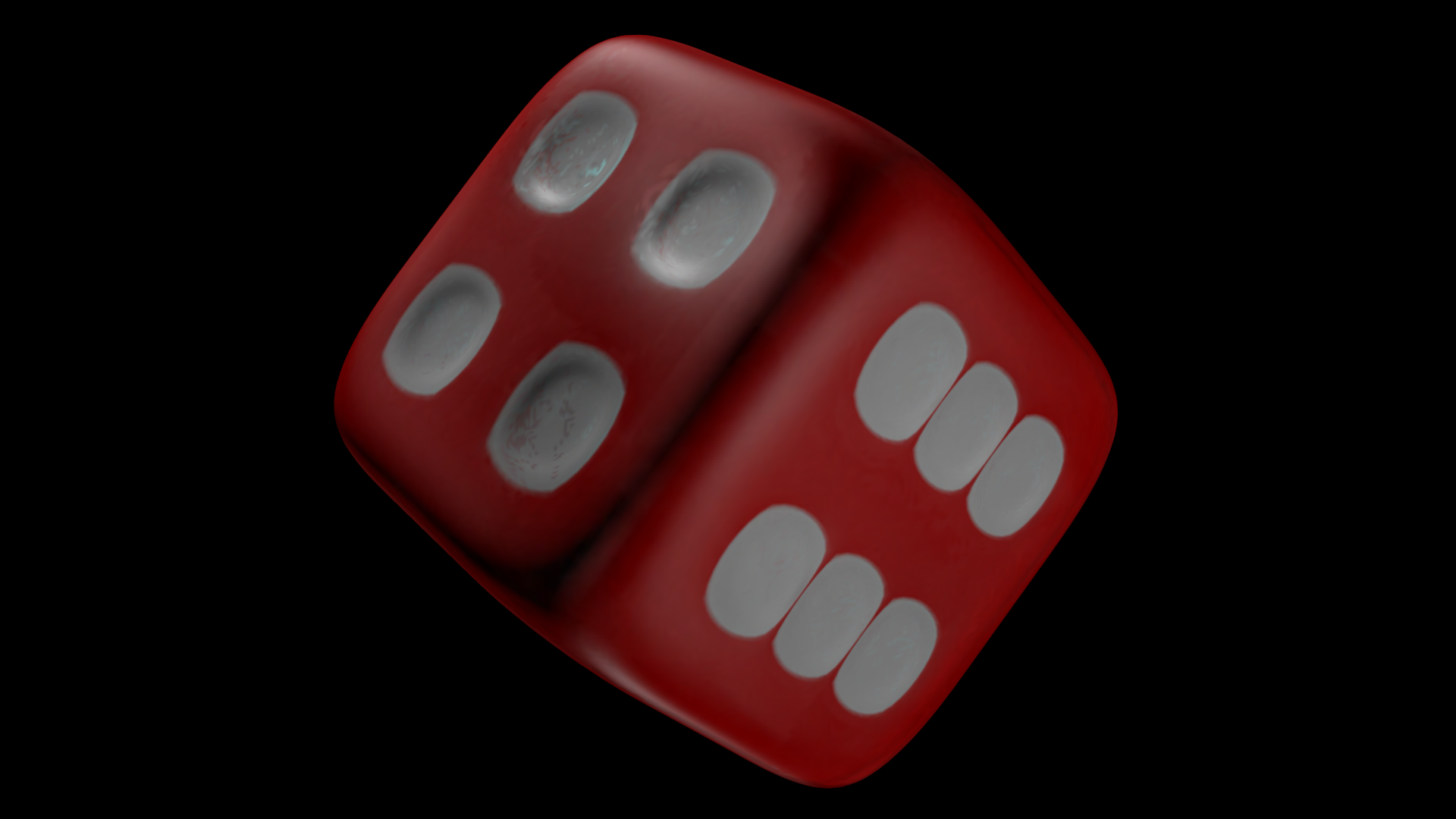}}
		\subfigure[Sparse case]{\includegraphics[width=0.24\linewidth]{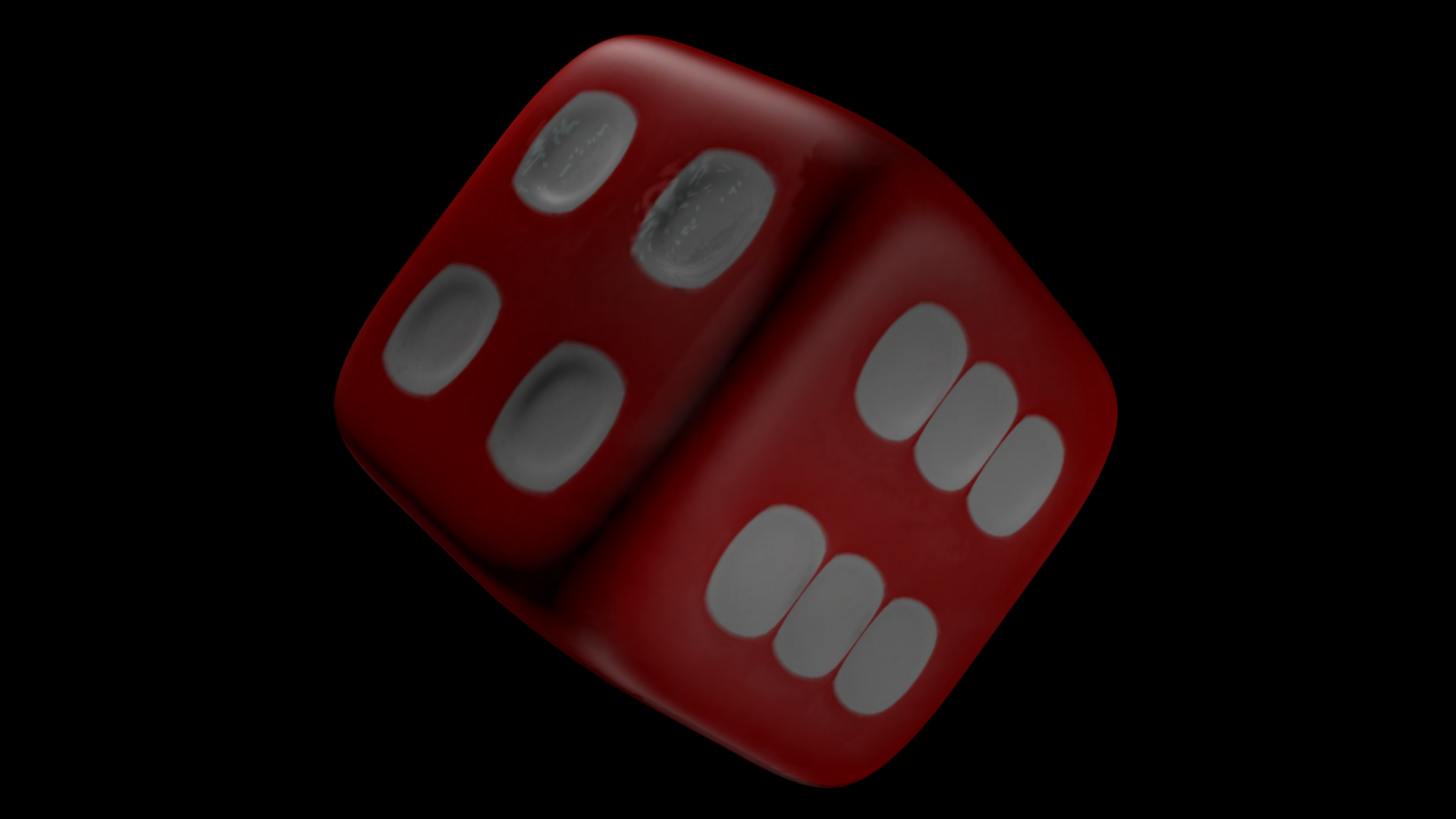}}
		\caption{Reconstruction of {\em Die} from a specific viewpoint as a function of camera density, where the number and placement of cameras in dense, intermediate, and sparse cases are specified in Table~\ref{tab:cameras}. The groundtruth image is shown in (a).}
		\label{fig:recon_density_dice}
	\end{figure*}

	Figs.~\ref{fig:recon_density_can} and \ref{fig:recon_density_dice} show reconstructions of {\em Can} and {\em Die} as a function of camera density, where the configurations for different densities are described in Table~\ref{tab:cameras}.  Comparing the two figures, one can conclude that the impact of camera density depends on the complexity of the SLF.  For simple SLFs, \eg {\em Can}, one can barely tell the differences between a dense and a sparse camera density. For more complex SLFs, \eg {\em Die}, the differences are obvious, where the artifacts for intermediate and sparse cases can be clearly observed especially on the side of the {\em Die} with four pips.  This makes sense because a complex SLF, such as one due to specularity or transparency, requires more samples to accurately represent its variation.  For a simpler SLF, using fewer observations will not degrade the reconstruction quality much. As an extreme case, one observation is sufficient for recovering a Lambertian surface.
	
	\subsection{Scalability}
	\label{sec:results_scalability}
	
	Scalable video coding (SVC) \cite{svc} is a desirable feature for video streaming in practical applications, since it enables the videos to be encoded, transmitted, and displayed with various spatial or temporal resolutions or qualities under different display device or bandwidth scenarios. The proposed SLF representation method can naturally provide LF compression scalability with two kinds of modalities, namely spatial and fidelity scalability.
	
\begin{figure*}
		\centering
		\subfigure[Groundtruth]{\includegraphics[width=0.24\linewidth]{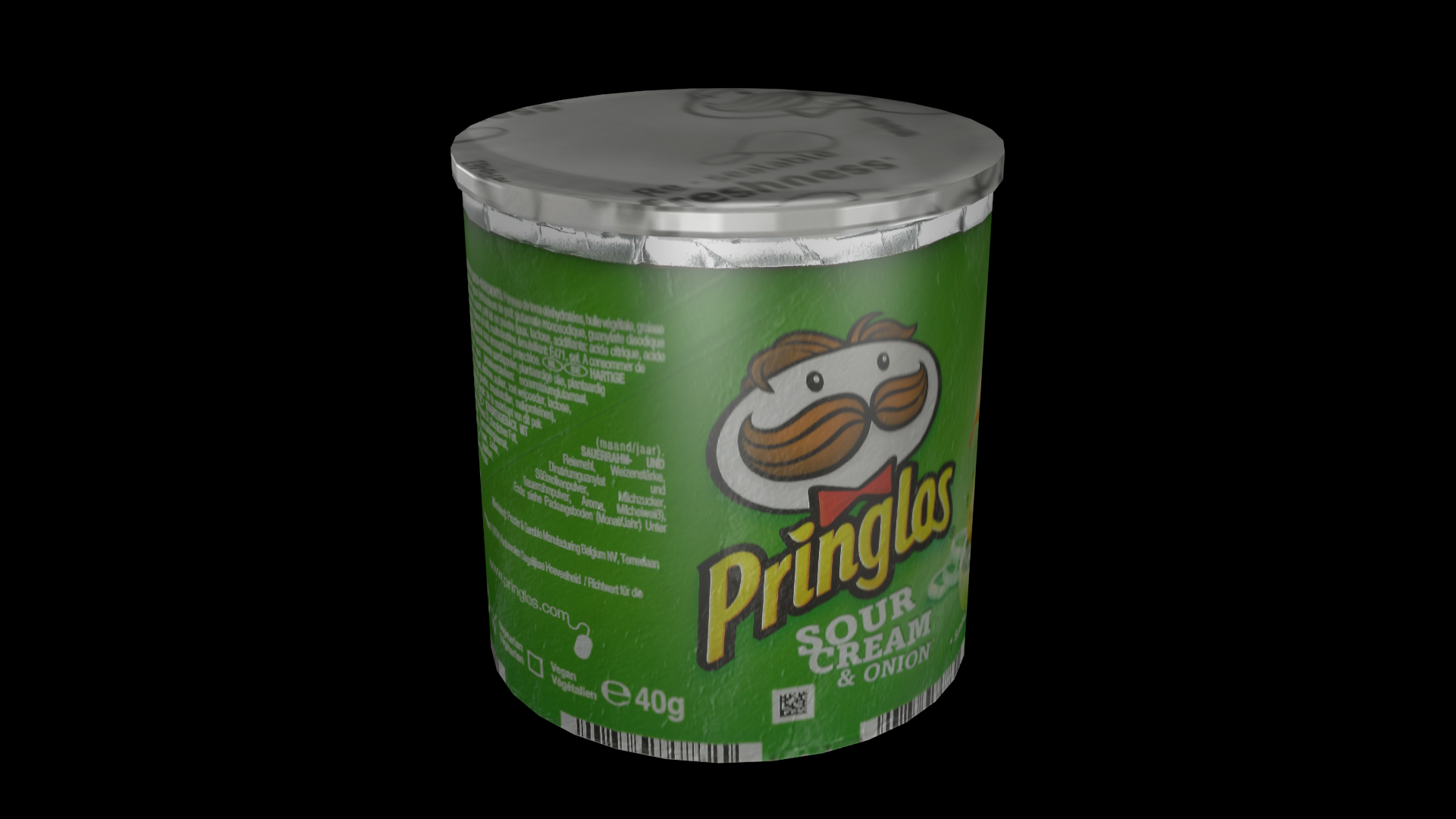}}
		\subfigure[350K points (2.48 MB)]{\includegraphics[width=0.24\linewidth]{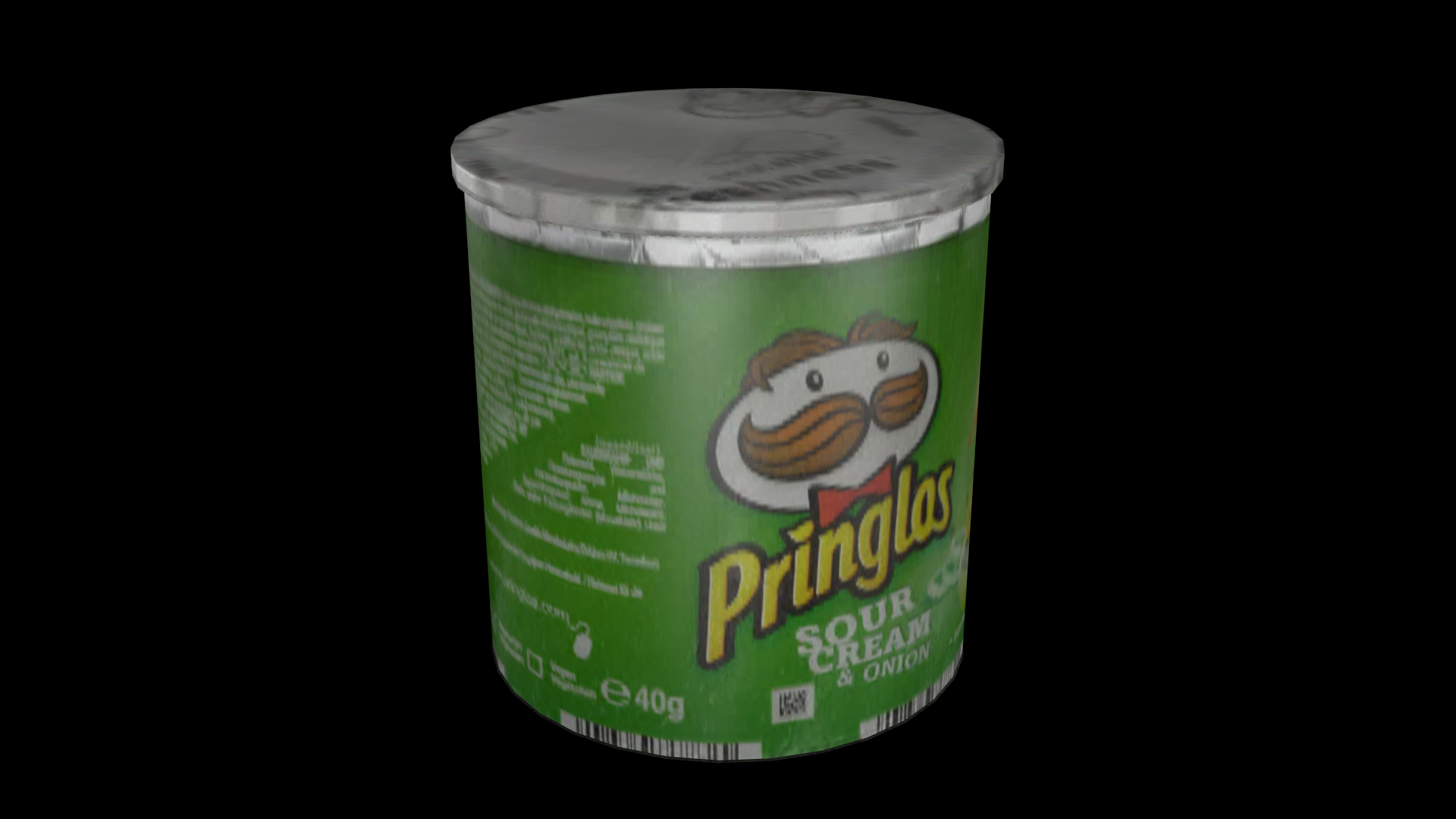}}
		\subfigure[91K points (0.75 MB)]{\includegraphics[width=0.24\linewidth]{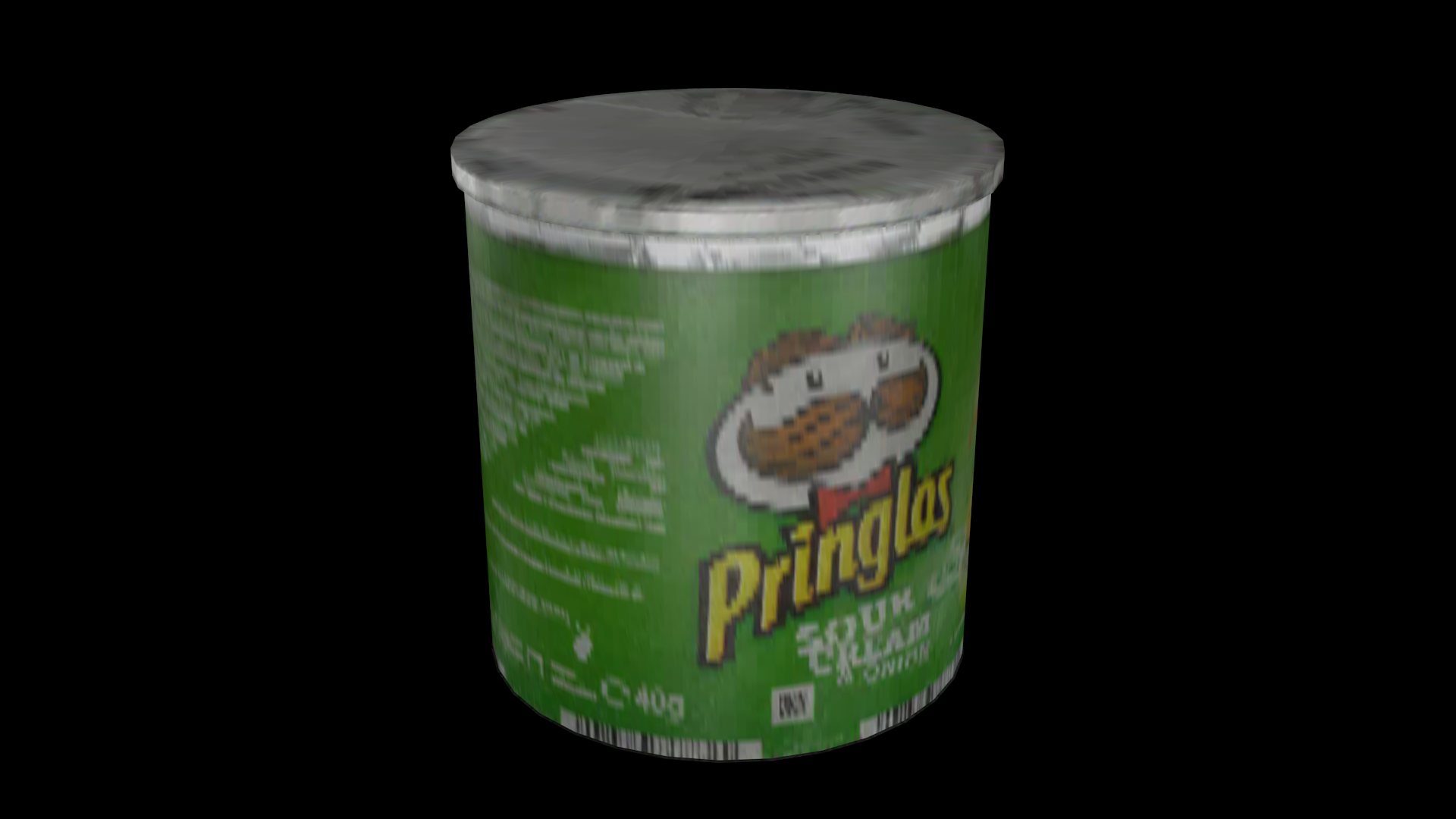}}
		\subfigure[22K points (0.33 MB)]{\includegraphics[width=0.24\linewidth]{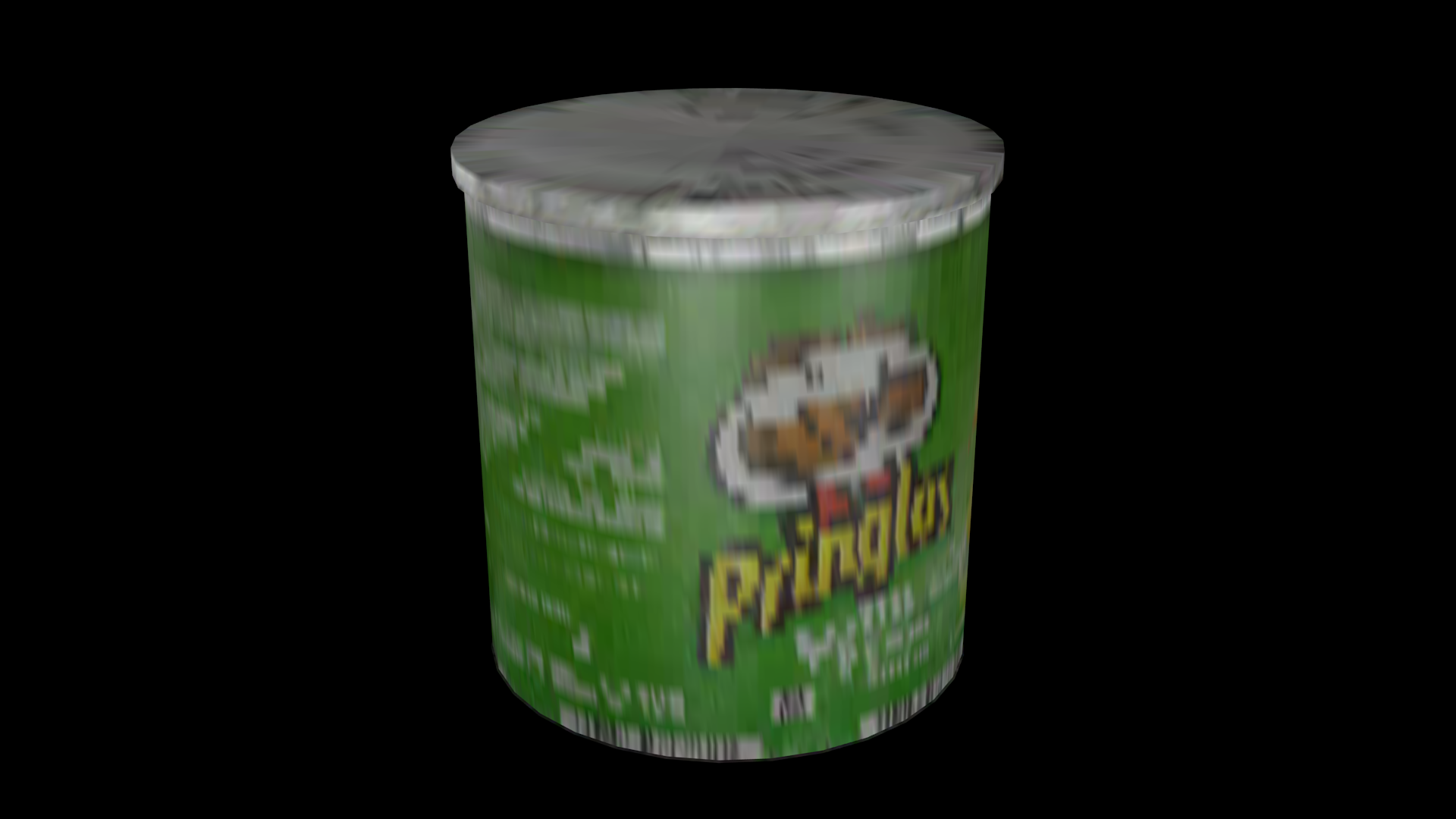}}
		\caption{Reconstruction of {\em Can} from a specific viewpoint as a function of point cloud resolution, where the number of surface points  decreases from (b) to (d). The groundtruth image is shown in (a). The numbers in brackets are the compressed SLF data size by RAHT.}
		\label{fig:spatial_scalability}
	\end{figure*}

	Spatial scalability can be achieved by sampling the point cloud in 3D space with different spatial resolution. The point cloud with lower resolutions can then be used to predict the point cloud with higher resolutions by taking advantages of the high correlation across 3D space.  We show reconstructions of {\em Can} with different point cloud resolutions in Fig.~\ref{fig:spatial_scalability}. A higher 3D resolution will simultaneously increase the 2D reconstruction resolution with a commensurate bitrate. But both of them can reconstruct the view map well for each point, as one can still observe light reflections in the lowest resolution reconstruction in Fig.~\ref{fig:spatial_scalability}(d).
	
\begin{figure*}
		\centering
		\subfigure[Groundtruth]{\includegraphics[width=0.24\linewidth]{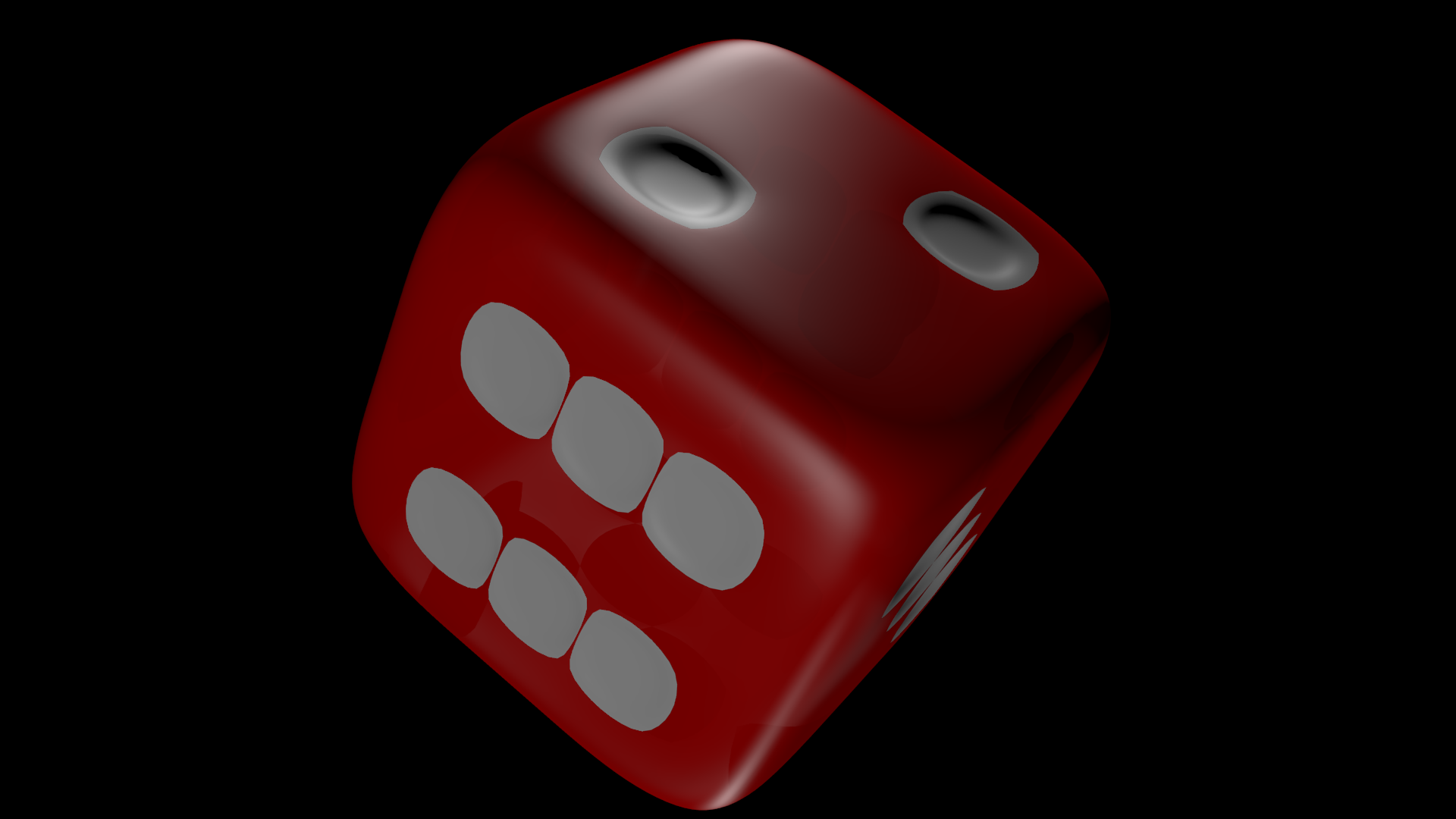}}
		\subfigure[$N=128$ (3.90 MB)]{\includegraphics[width=0.24\linewidth]{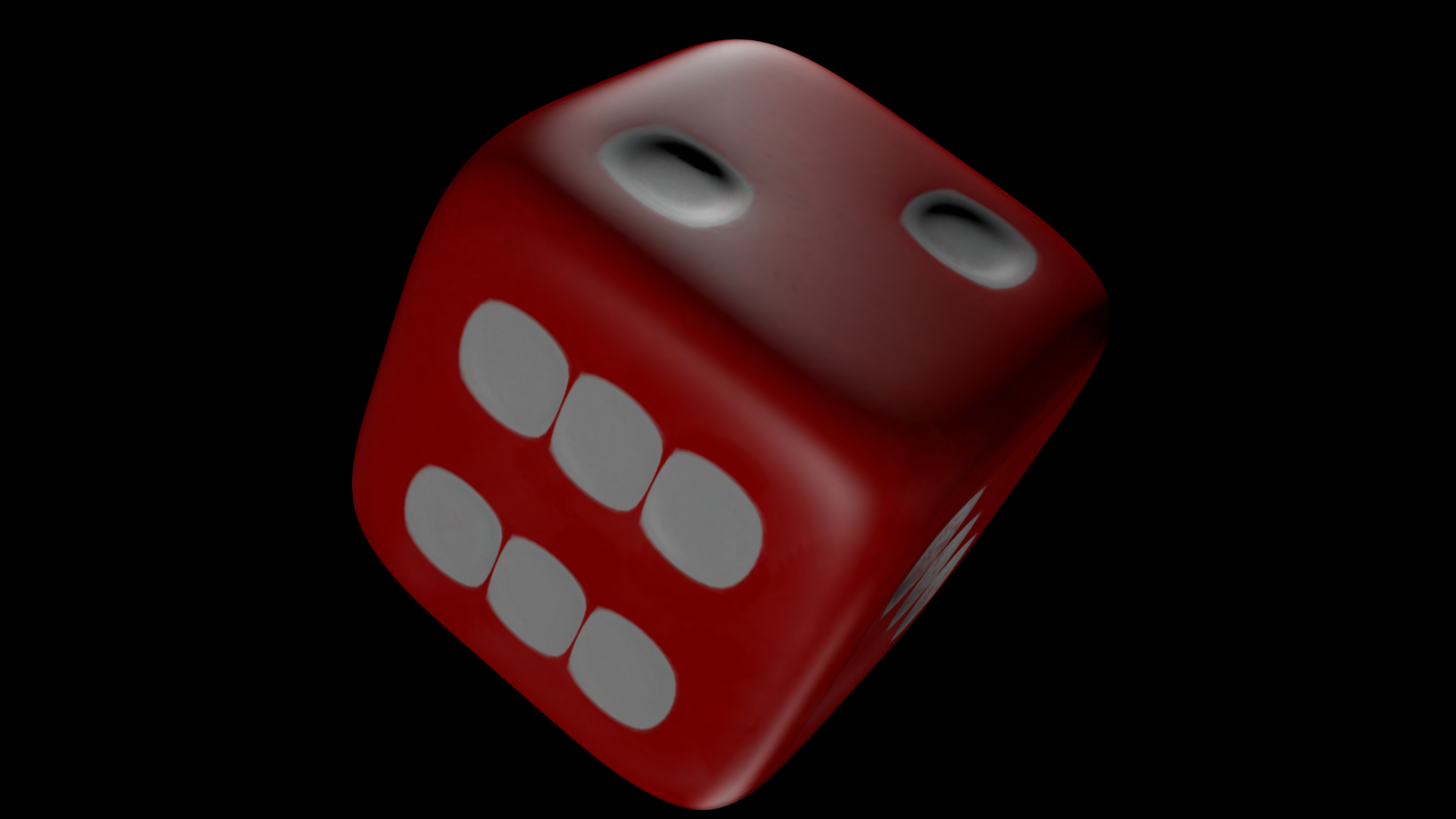}}
		\subfigure[$N=32$ (1.71 MB)]{\includegraphics[width=0.24\linewidth]{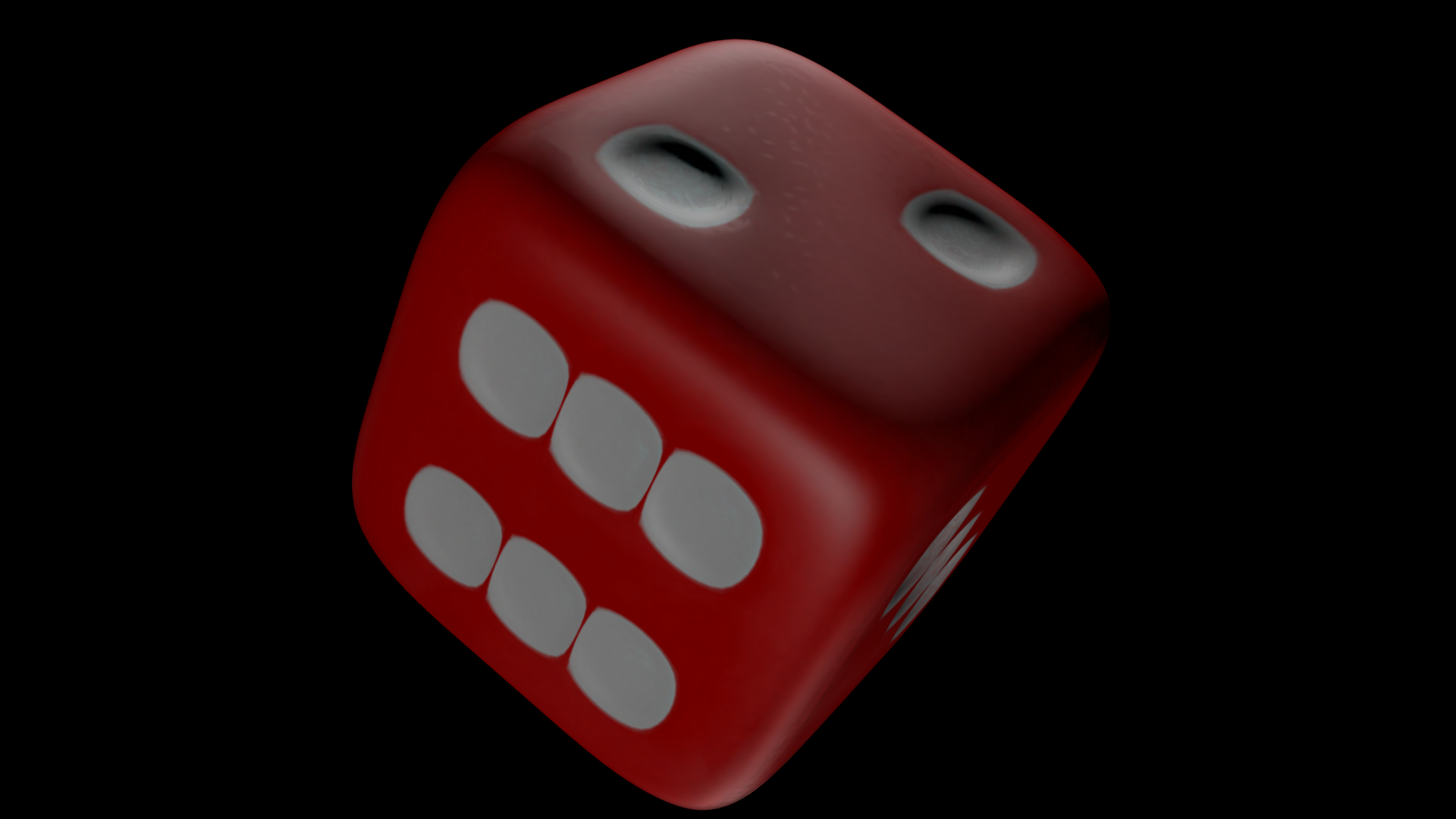}}
		\subfigure[$N=1$ (0.30 MB)]{\includegraphics[width=0.24\linewidth]{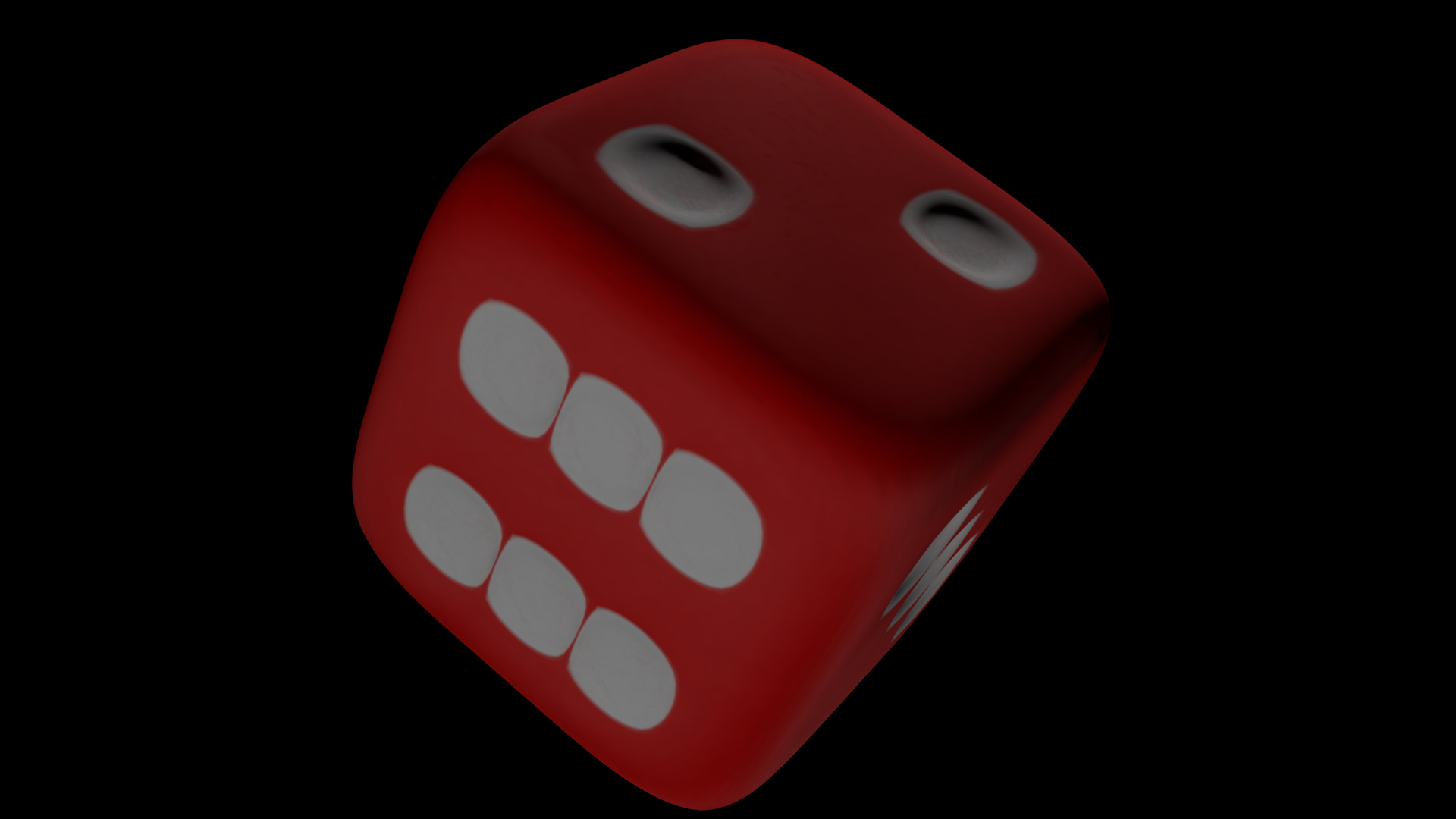}}
		\caption{Reconstruction of {\em Die} from a specific viewpoint as a function of $N$, \ie the number of basis functions, where $N$  decreases from (b) to (d). The groundtruth image is shown in (a). The numbers in brackets are the compressed SLF data size by RAHT.}
		\label{fig:fidelity_scalability_nbasis}
	\end{figure*}

\begin{figure*}
		\centering
		\subfigure[Groundtruth]{\includegraphics[width=0.24\linewidth]{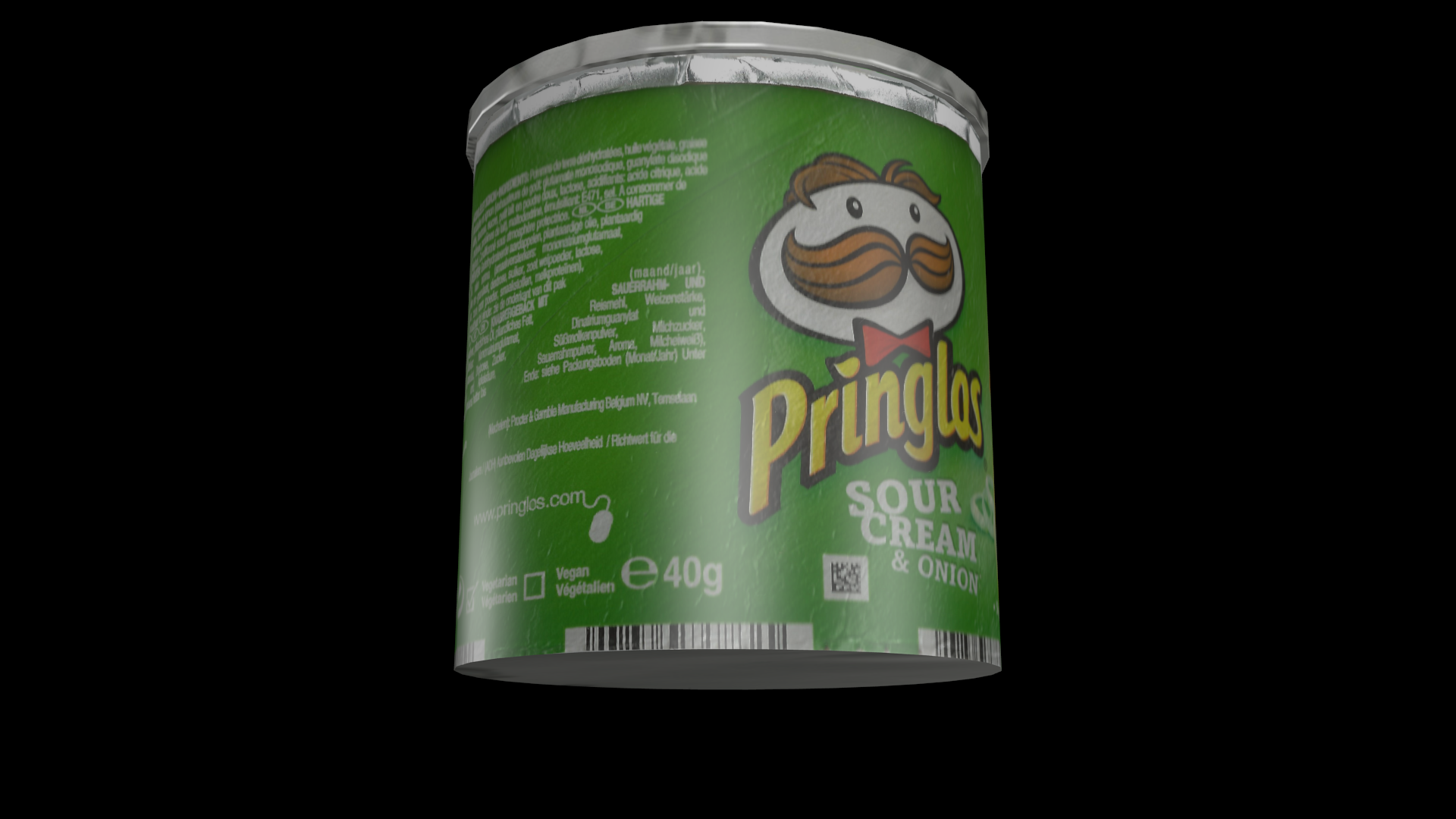}}
		\subfigure[$Q=8$ (2.48 MB)]{\includegraphics[width=0.24\linewidth]{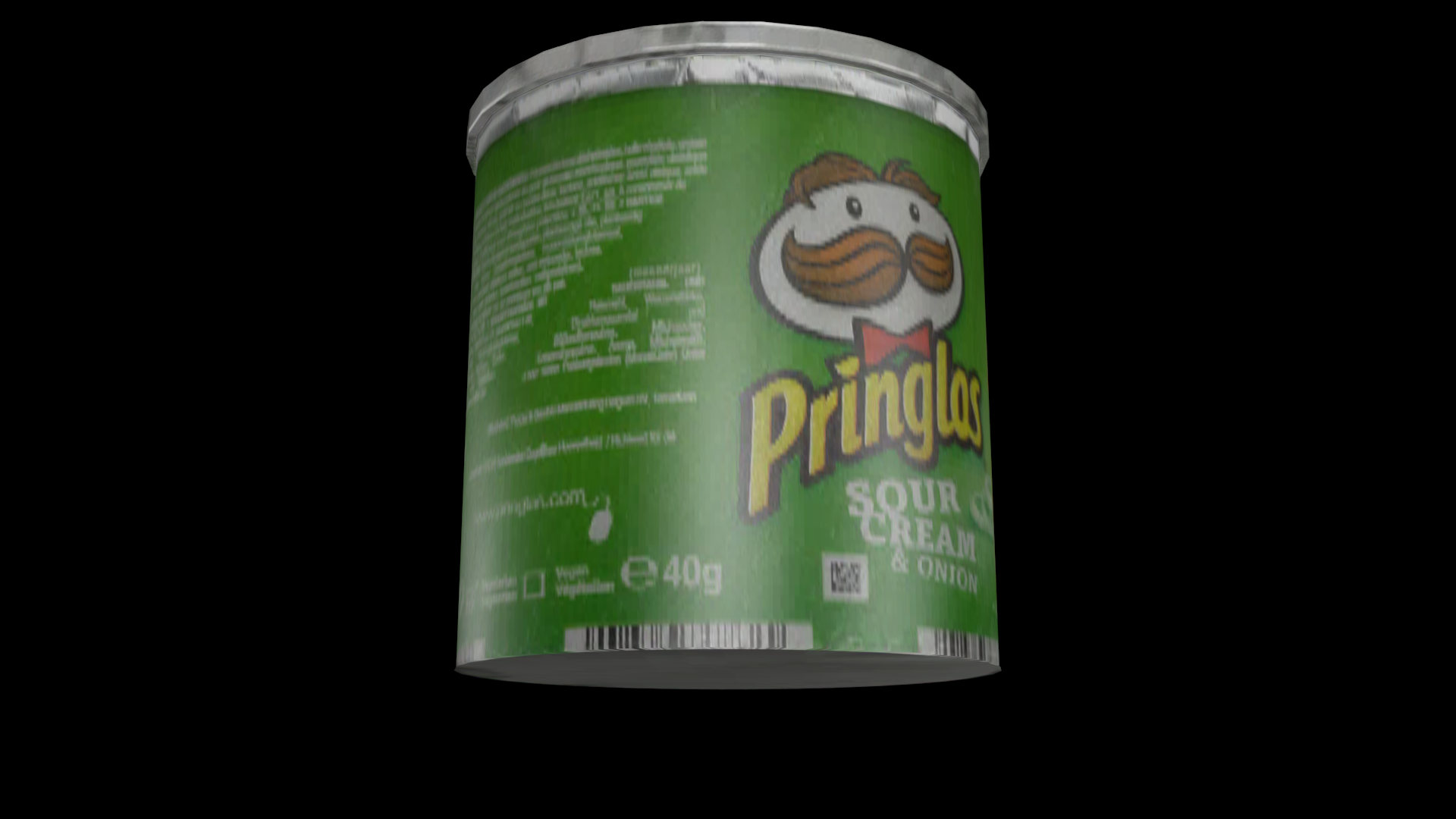}}
		\subfigure[$Q=16$ (1.08 MB)]{\includegraphics[width=0.24\linewidth]{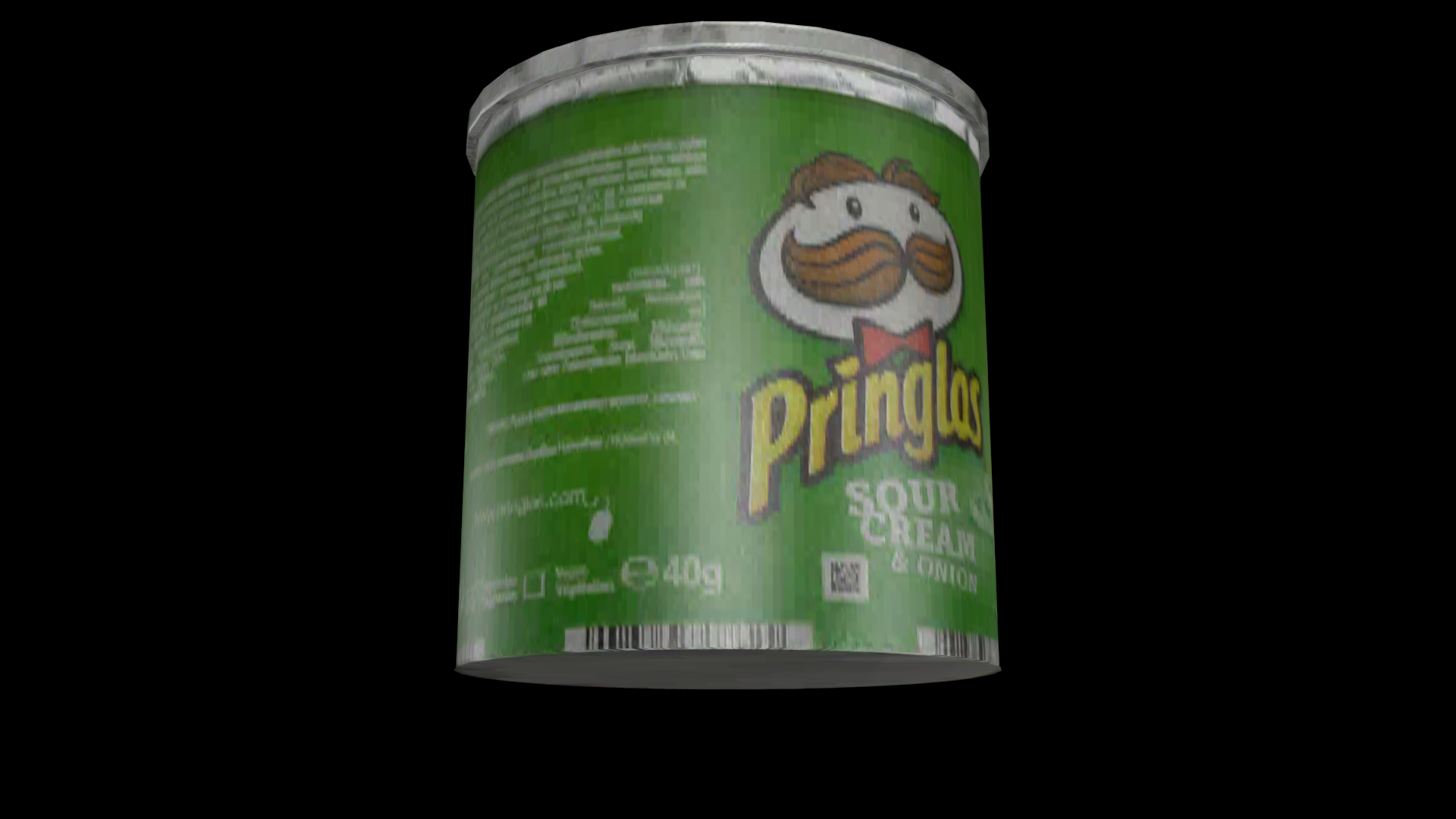}}
		\subfigure[$Q=32$ (0.44 MB)]{\includegraphics[width=0.24\linewidth]{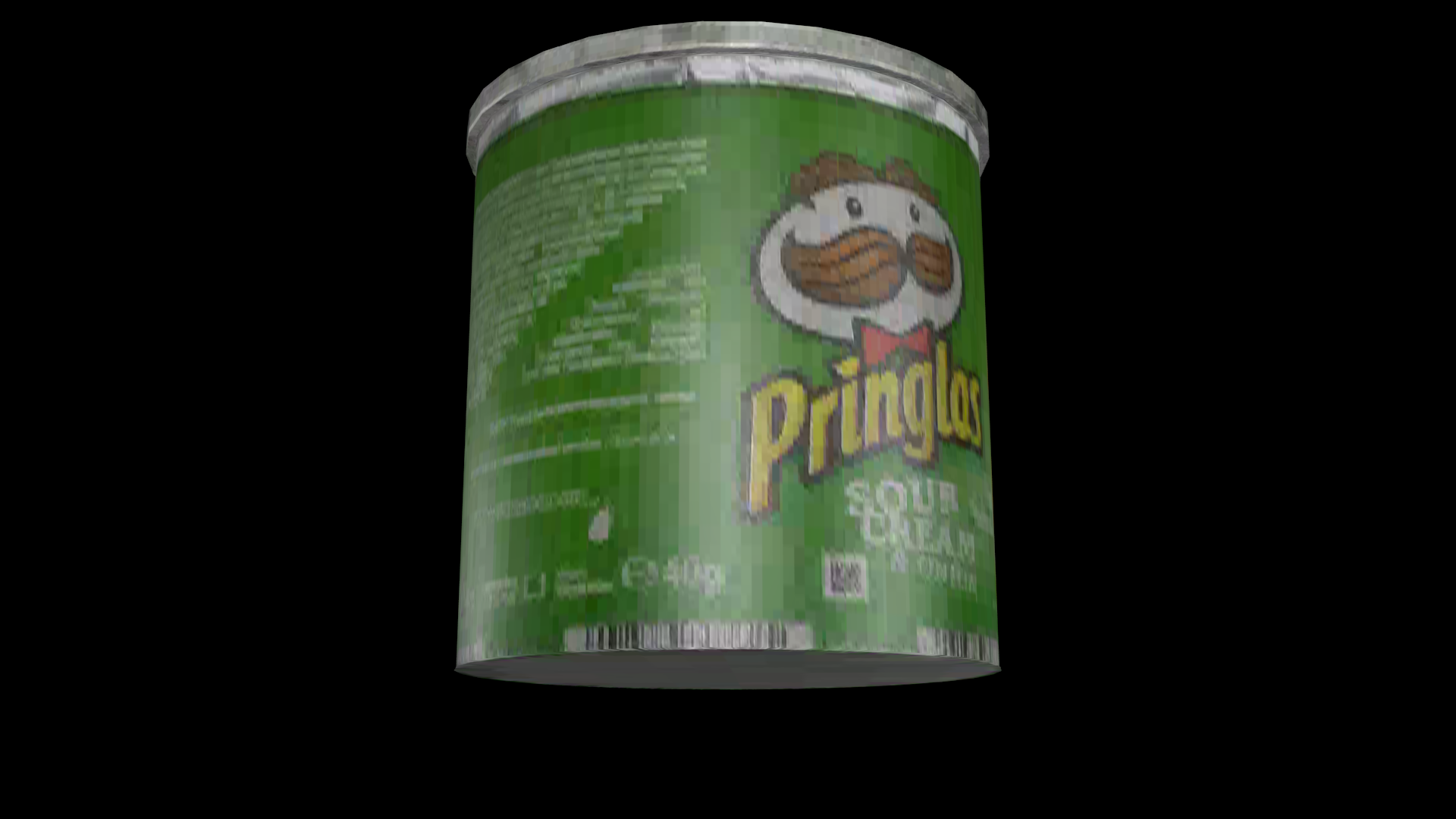}}
		\caption{Reconstruction of {\em Can} from a specific viewpoint as a function of $Q$, \ie the quantization step-size, where $Q$ increases from (b) to (d). The groundtruth image is shown in (a). The numbers in brackets are the compressed SLF data size by RAHT.}
		\label{fig:fidelity_scalability_qs}
	\end{figure*}

	Fidelity (or quality) scalability can be achieved in two ways, either by encoding different numbers of SLF coefficients, or by encoding with varying quantization step-sizes $Q$.  The former separates a video stream into several subsets by dividing the SLF coefficients into different levels. As shown in Fig.~\ref{fig:fidelity_scalability_nbasis}, the {\em Die} is reconstructed with different numbers of SLF coefficients. It can be seen that using only one coefficient (DC value) can represent only diffuse surfaces that reflect light equally in all directions. Using more high-frequency coefficients can produce higher reconstruction quality that can represent more complex surface material and lighting phenomena such as reflection and refraction.  Alternatively, fidelity scalability can be achieved by varying the quantization step-size $Q$ when compressing the SLF coefficients. As shown in Fig.~\ref{fig:fidelity_scalability_qs}, quantization of SLF coefficients can introduce blocking artifacts as in traditional image coding.

	\subsection{Reconstruction from Arbitrary Viewpoints}
	\label{sec:results_recon}
	
	Next we subjectively evaluate the ability of the proposed scheme to reconstruct scenes from arbitrary points of view.  This is one of the most significant advantages of our scheme over image-based or image+depth-based LF compression. 
	The latter cannot model global occlusions well, and hence requires images to be captured from camera arrays that are dense and thus compact (lest the camera arrays themselves fill the scene), and hence supports only a relatively narrow range of viewpoints.
	
\begin{figure}
		\centering
		{\includegraphics[width=1\linewidth]{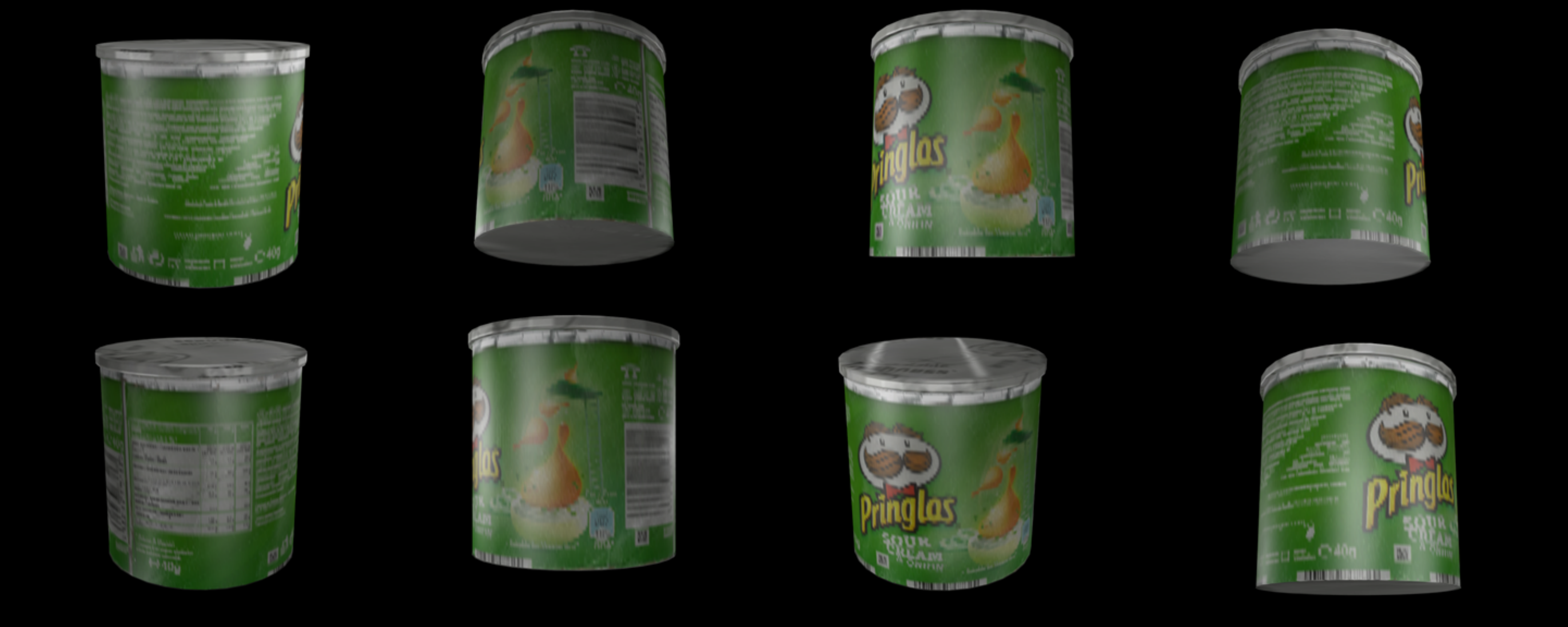}}
		\caption{Rendering {\em Can} from arbitrary viewpoints.}
		\label{fig:recon_arb_can}
	\end{figure}

\begin{figure}
		\centering
		{\includegraphics[width=1\linewidth]{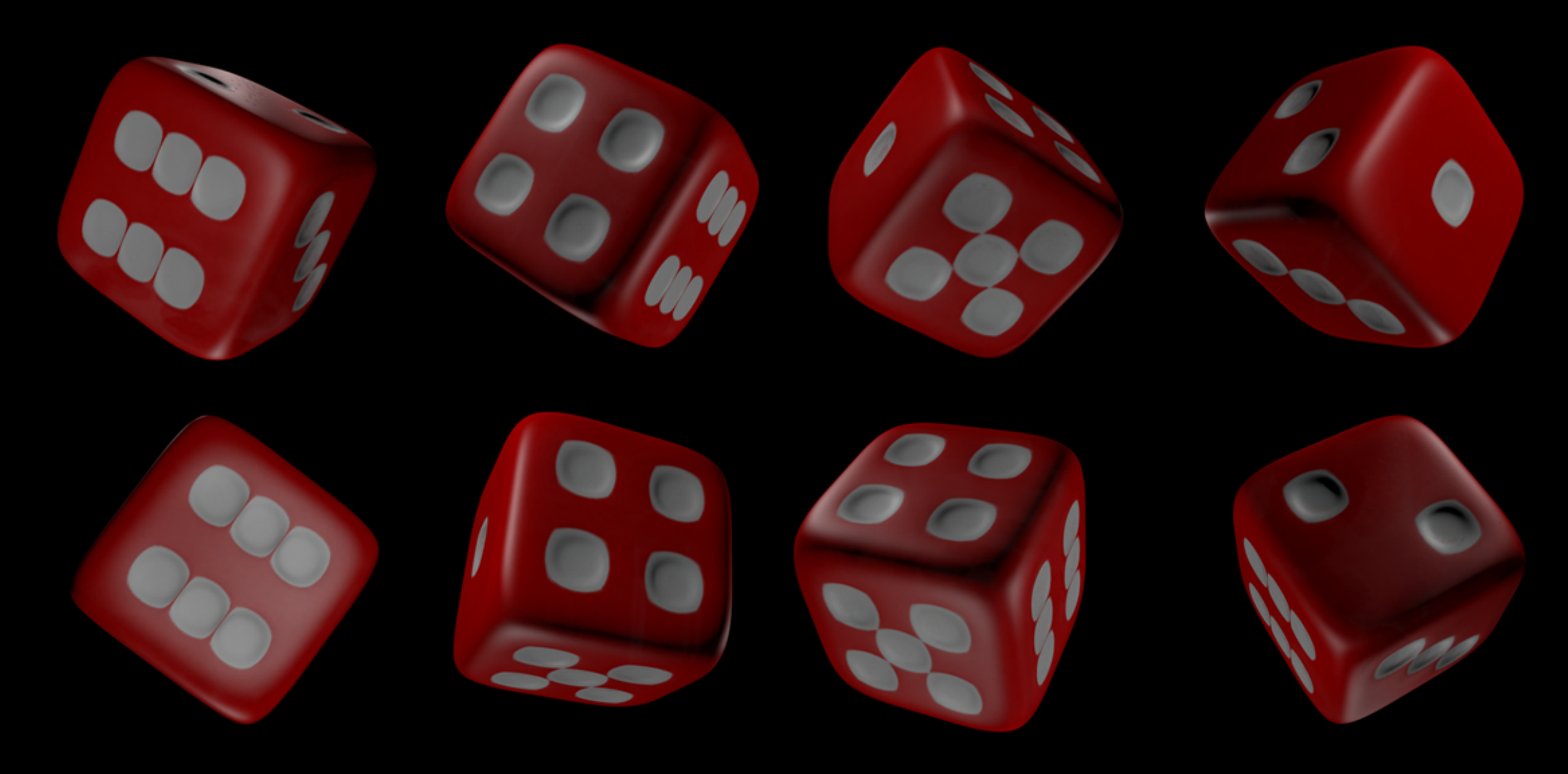}}
		\caption{Rendering {\em Die} from arbitrary viewpoints.}
		\label{fig:recon_arb_die}
	\end{figure}

\begin{figure}
		\centering
		{\includegraphics[width=1\linewidth]{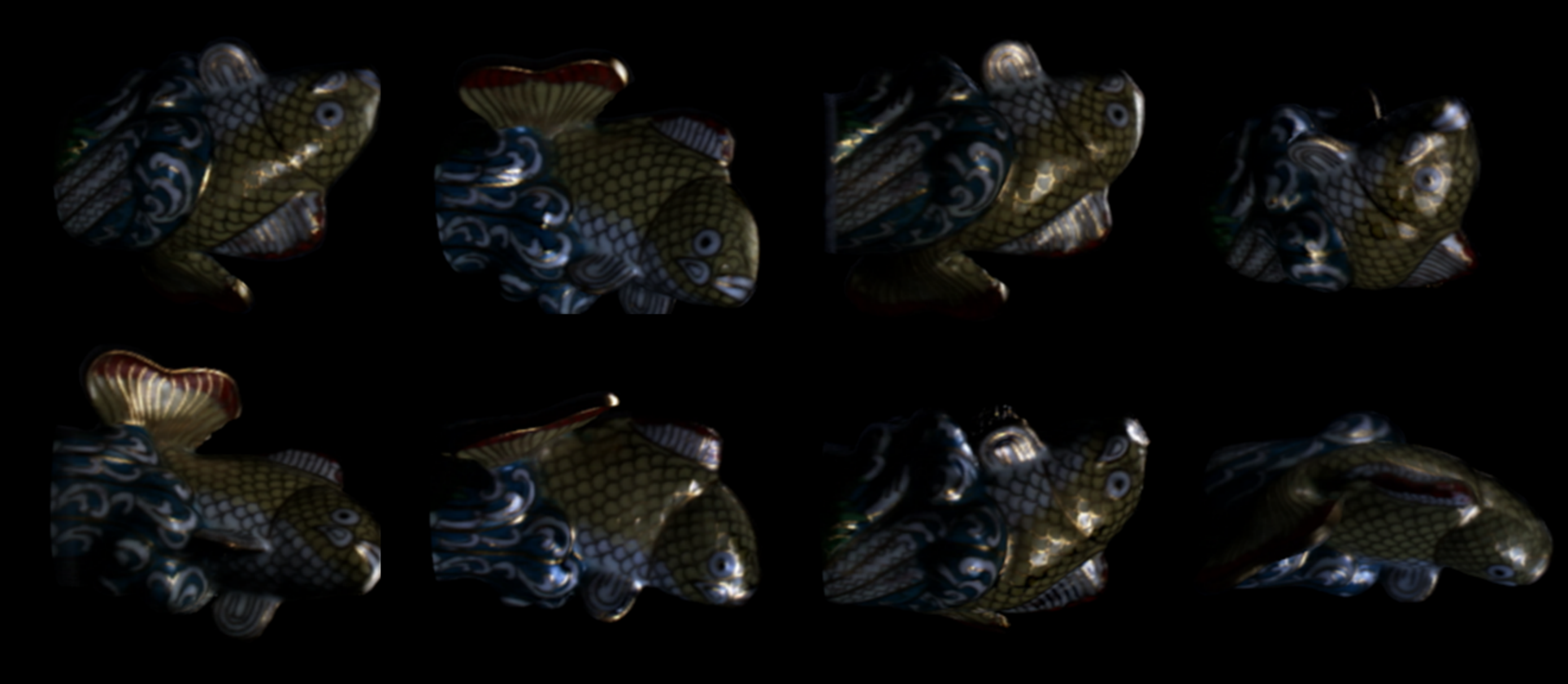}}
		\caption{Rendering {\em Fish} from arbitrary viewpoints.}
		\label{fig:recon_arb_fish}
	\end{figure}

\begin{figure}
		\centering
		{\includegraphics[width=1\linewidth]{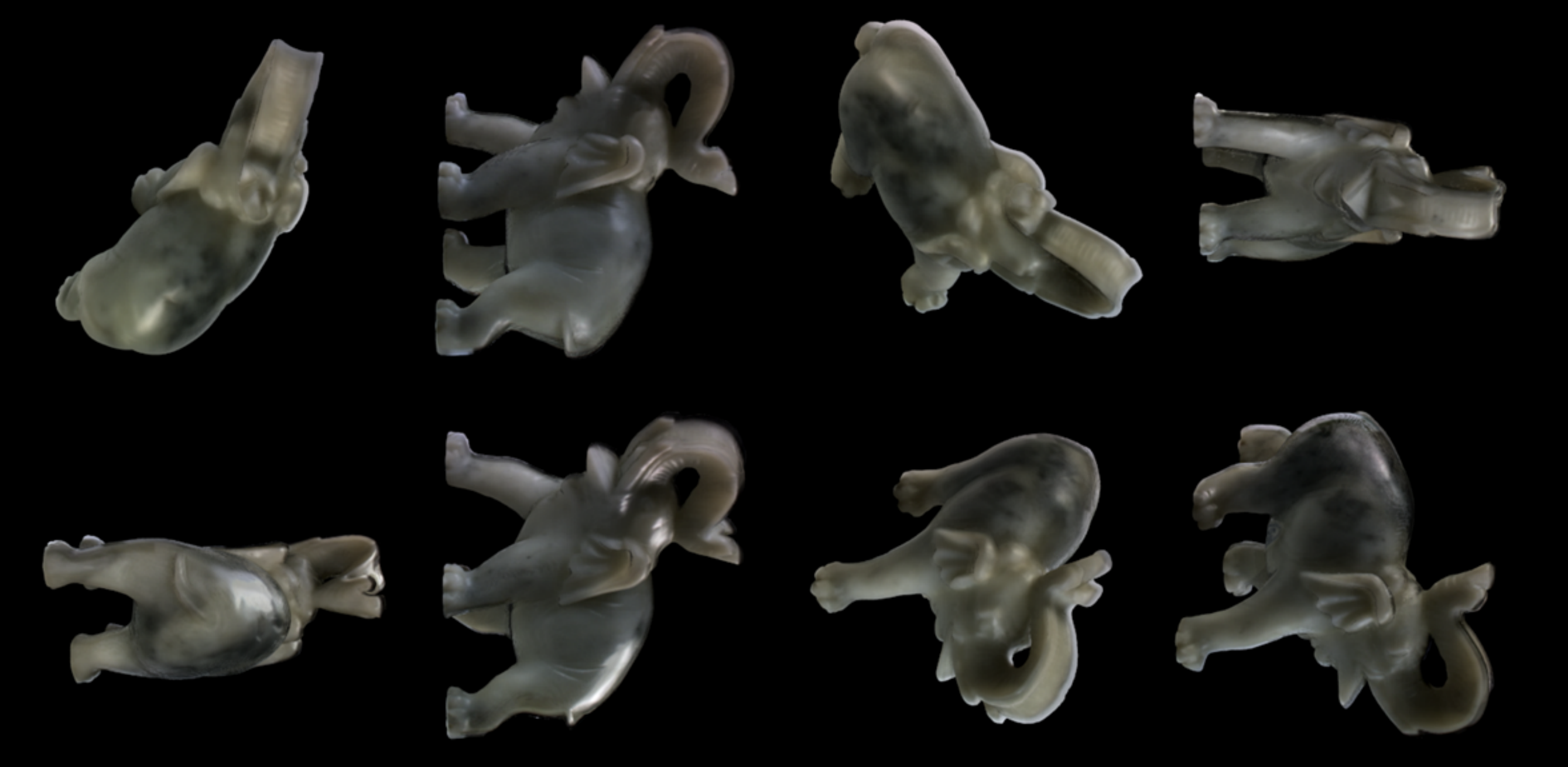}}
		\caption{Rendering {\em Elephant} from arbitrary viewpoints.}
		\label{fig:recon_arb_elephant}
	\end{figure}

	Figs.~\ref{fig:recon_arb_can} and \ref{fig:recon_arb_die} show reconstructions of two synthetic datasets {\em Can} and {\em Die} from several viewpoints. Figs.~\ref{fig:recon_arb_fish} and \ref{fig:recon_arb_elephant} show reconstructions of two real datasets {\em Elephant} and {\em Fish}. From the results, it can be seen that the proposed scheme is adaptive and robust to different surface materials and light conditions. Light reflection effects on the object surface such as specularity can be easily observed.  The real datasets further demonstrate that the proposed method can better tackle the occlusion issue, such as the {\em Elephant} legs and trunk.
	
	\subsection{Comparison with Image based LF Compression}
	\label{sec:results_rd_compare}
	
\begin{figure}
		\centering
		\includegraphics[width=1\linewidth]{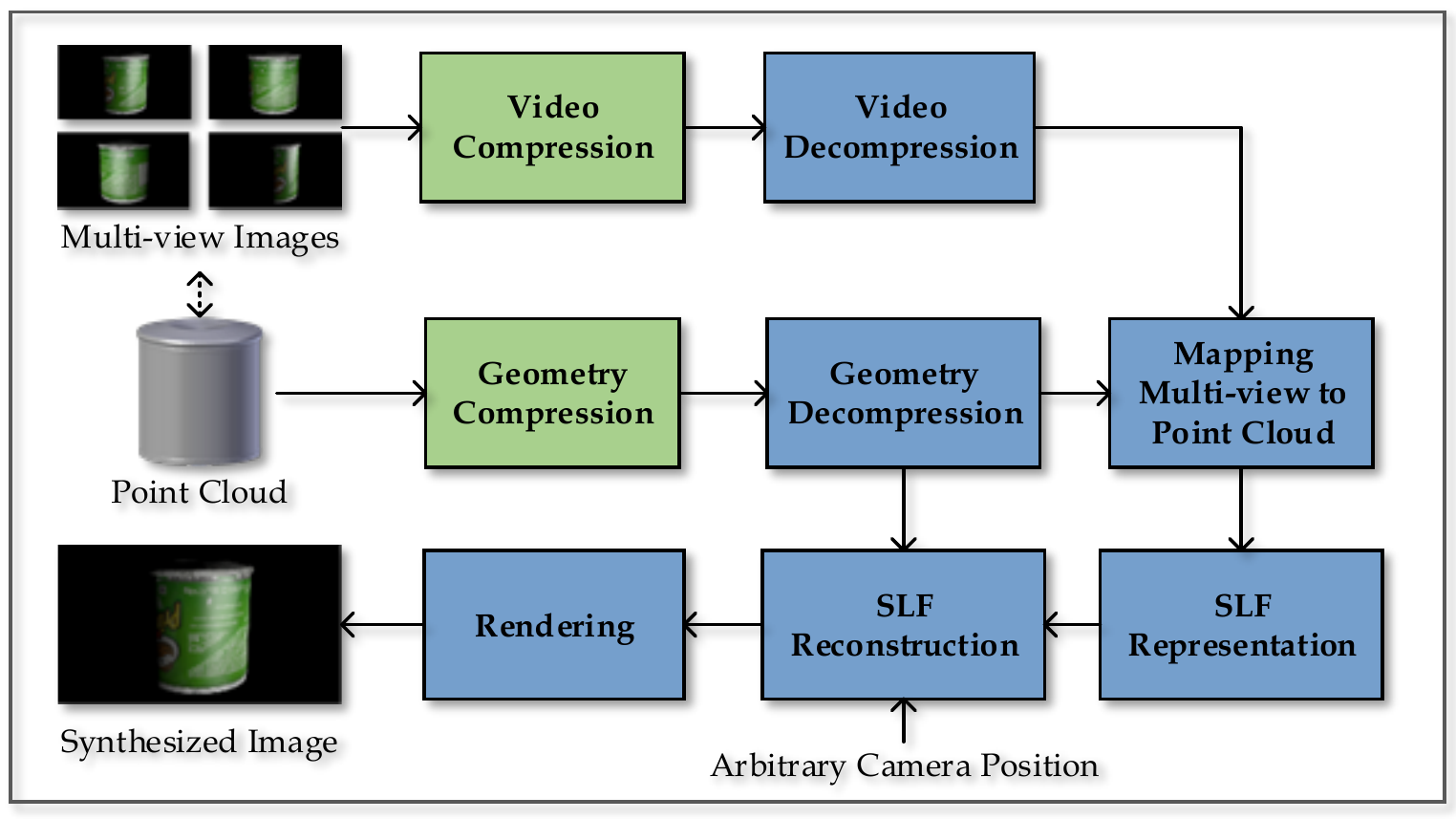}\\
		\caption{The framework of IGC for fair comparison with the proposed scheme. Green and blue boxes indicate the processes on encoder and decoder sides, respectively.}\label{fig:framework_image}
	\end{figure}

	A fair, quantitative comparison between our scheme and image based LF compression is not straightforward, since our SLF scheme can reconstruct viewpoints that are far from the original camera positions, while an image based LF cannot. However, we can fairly compare our SLF compression scheme with an alternative reference scheme, which we call {\em image-plus-geometry compression} (IGC), as illustrated in Fig.~\ref{fig:framework_image}. At the encoder side, instead of compressing and transmitting the coefficients of the SLF representation, IGC compresses and transmits the captured source images, using a video codec. Like SLF compression, IGC also compresses and transmits the geometry. At the decoder side, instead of directly decoding the SLF representation and then using the representation to render arbitrary points of view, IGC decodes the source images, constructs the SLF representation from the decoded images, and finally uses the SLF representation to render arbitrary points of view.  Thus the major difference between the SLF compression framework and IGC is that in SLF compression, the SLF construction is performed at the encoder and the SLF representation is compressed, while in IGC, the source images are compressed and the SLF construction is performed at the decoder, which significantly increases the complexity of the decoder. The running time of decoder side between the proposed scheme and the IGC is compared in Table~\ref{tab:time}. One can see that the proposed scheme either by RAHT or by TMC is much faster than IGC, and RAHT is much faster than TMC. 
	
	\begin{table}
		\centering
		\caption{Decoder running time comparison.}
		
		\begin{tabular}{|c|c|c|}
			\hline
			\textbf{Running Time (s)} & \textbf{IGC} & \textbf{Proposed}  \bigstrut\\
			\hline
			{Geometry Decompression} & 0.12  & 0.12  \bigstrut\\
			\hline
			{SLF Coef. Decompression (RAHT / TMC)} & -     & 5.19 / 23.90  \bigstrut\\
			\hline
			{Images Decompression} & 1.92  & -  \bigstrut\\
			\hline
			{SLF representation} & 35.13 & -  \bigstrut\\
			\hline
			{Rendering} & 0.51  & 0.51  \bigstrut\\
			\hline
			\textbf{In Total (RAHT / TMC)} & \textbf{37.68} & \textbf{5.82 / 24.53}  \bigstrut\\
			\hline
		\end{tabular}%
		\label{tab:time}%
	\end{table}%
	
\begin{figure*}
		\centering
		\subfigure[Dense case]{\includegraphics[width=0.32\linewidth]{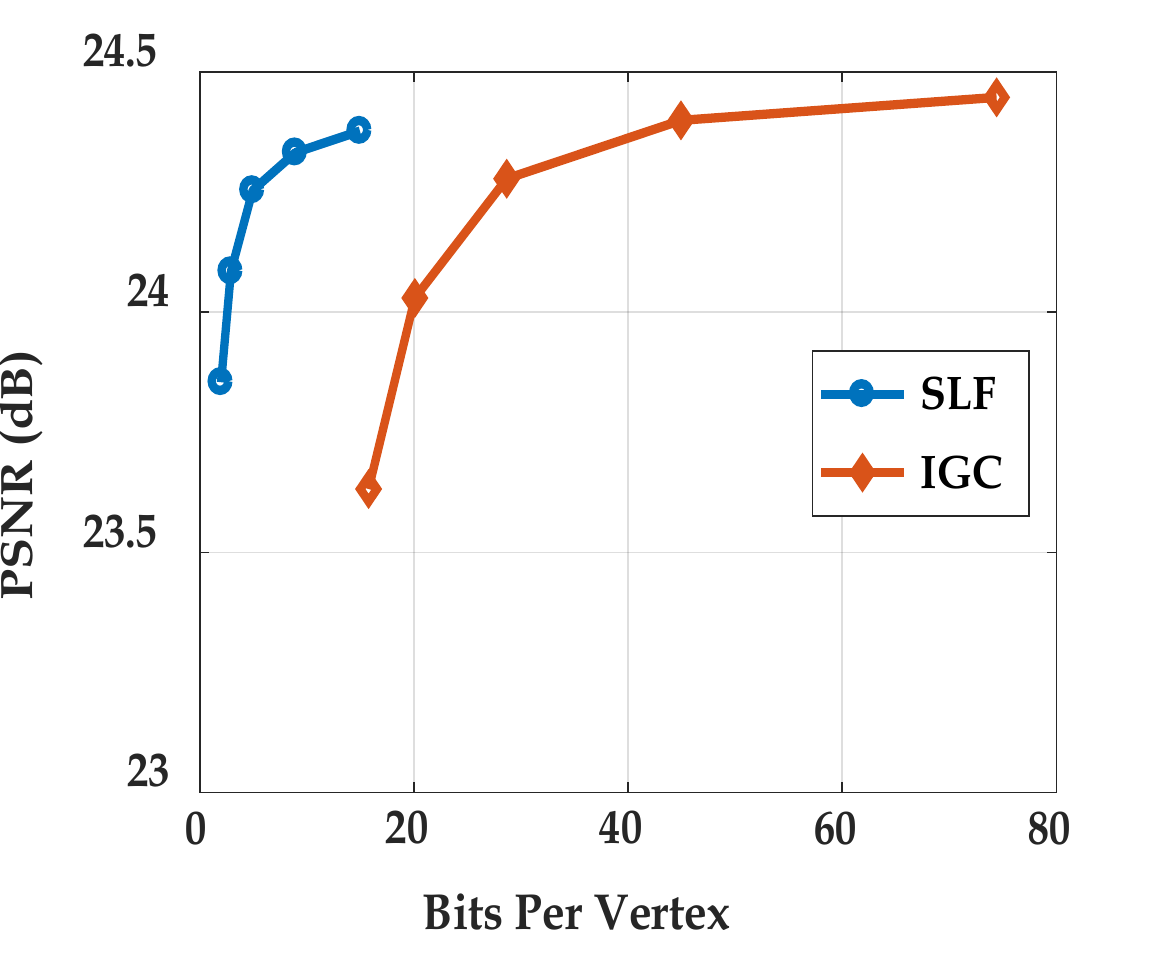}}
		\subfigure[Intermediate case]{\includegraphics[width=0.32\linewidth]{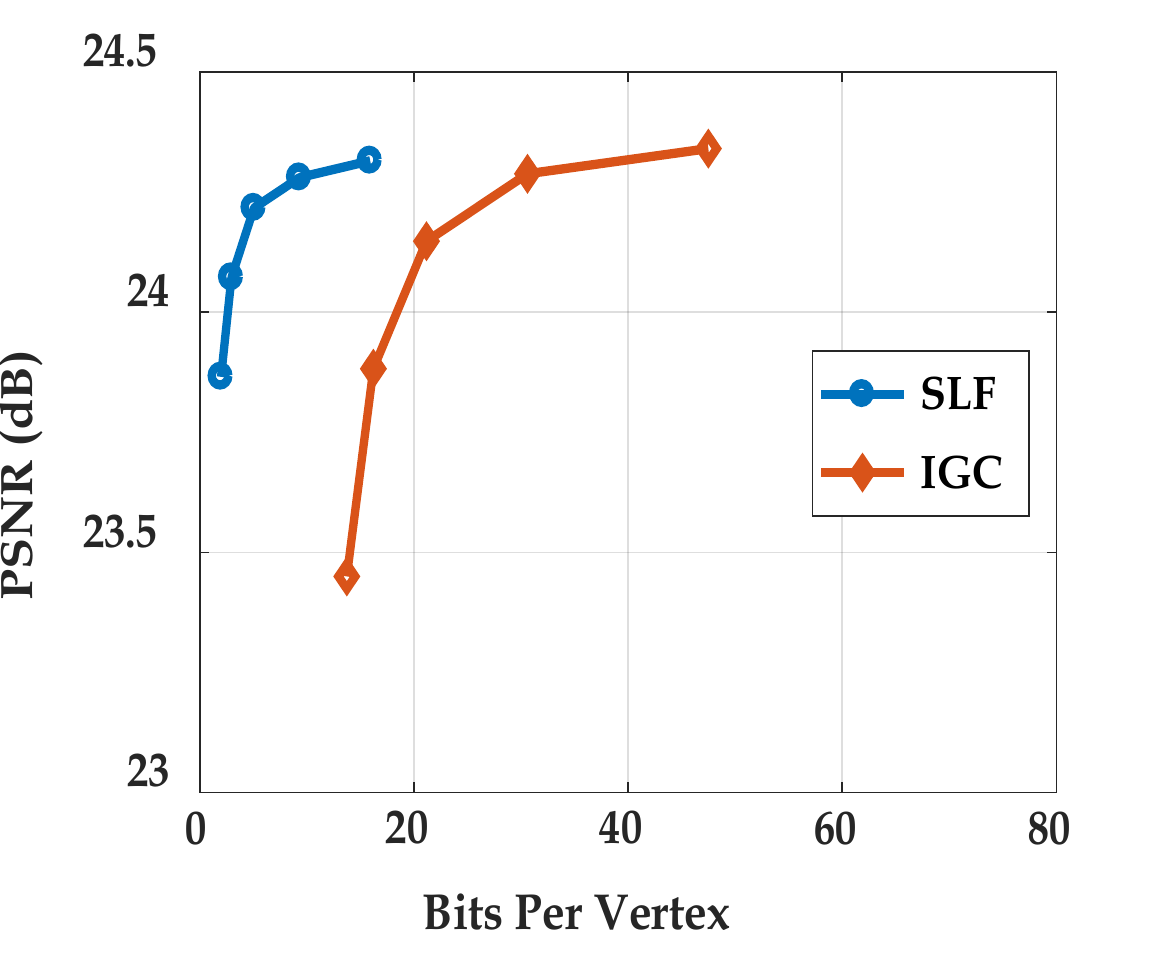}}
		\subfigure[Sparse case]{\includegraphics[width=0.32\linewidth]{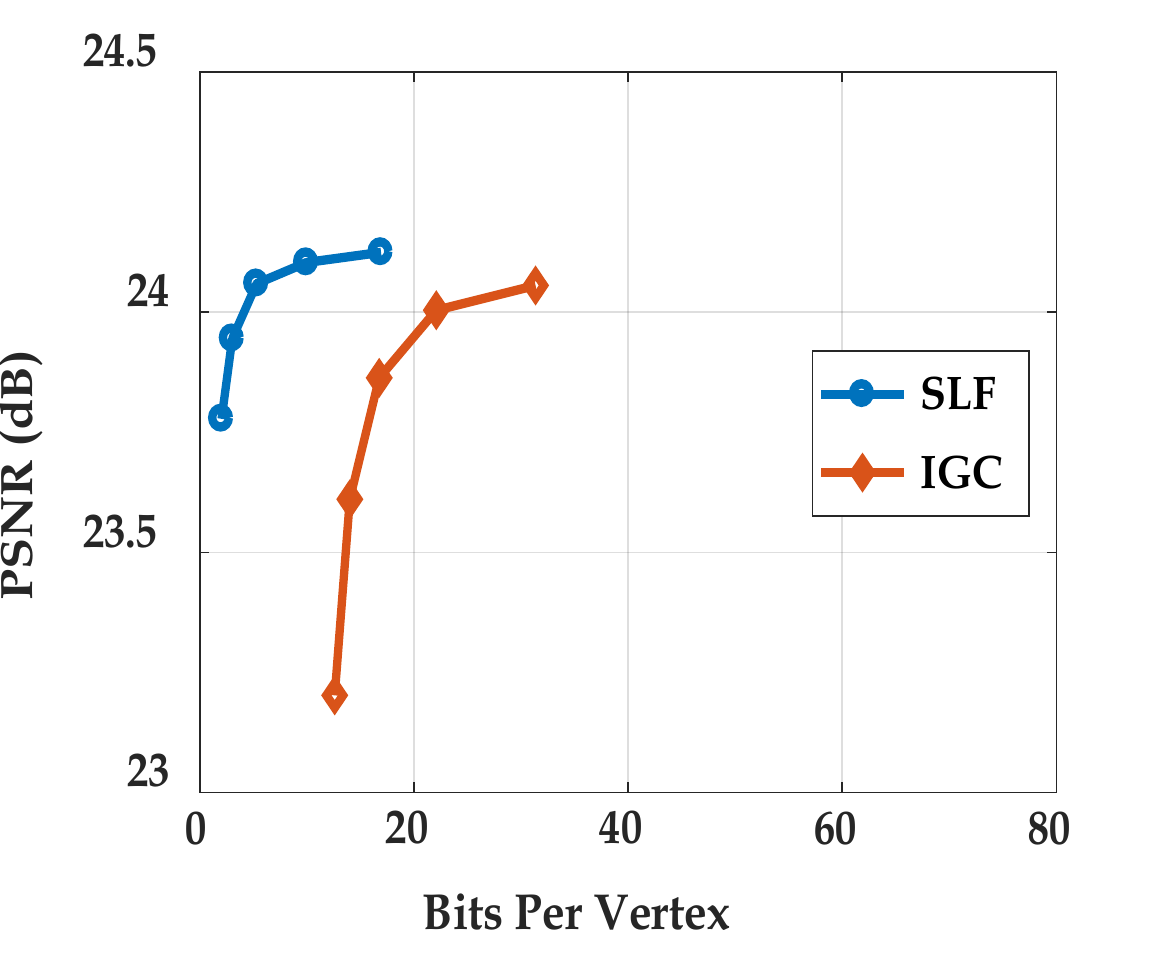}}
		\caption{RD performance comparisons between the proposed method (SLF) and IGC under different camera densities for {\em Fish}.}
		\label{fig:rdcompare_fish}
	\end{figure*}

\begin{figure*}
		\centering
		\subfigure[Dense case]{\includegraphics[width=0.32\linewidth]{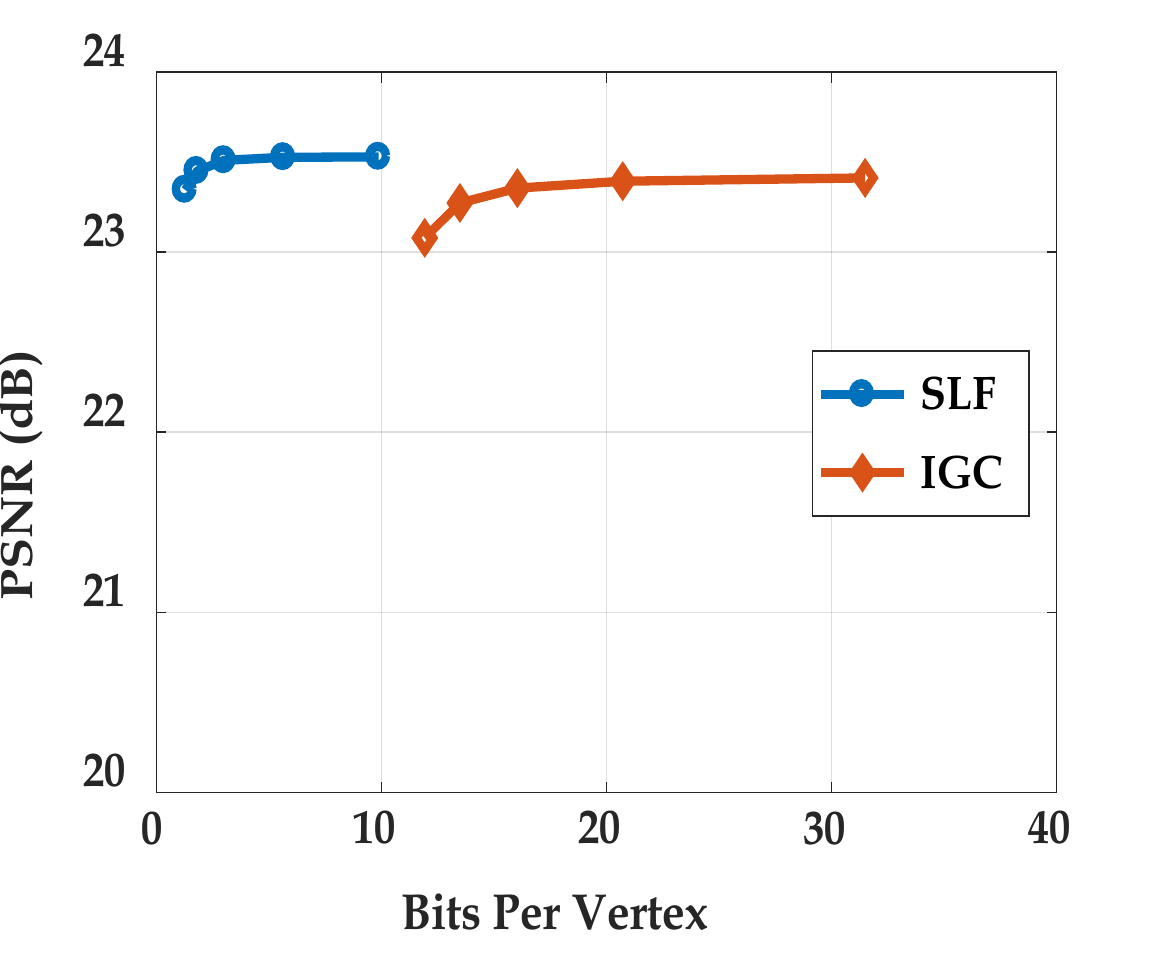}}
		\subfigure[Intermediate case]{\includegraphics[width=0.32\linewidth]{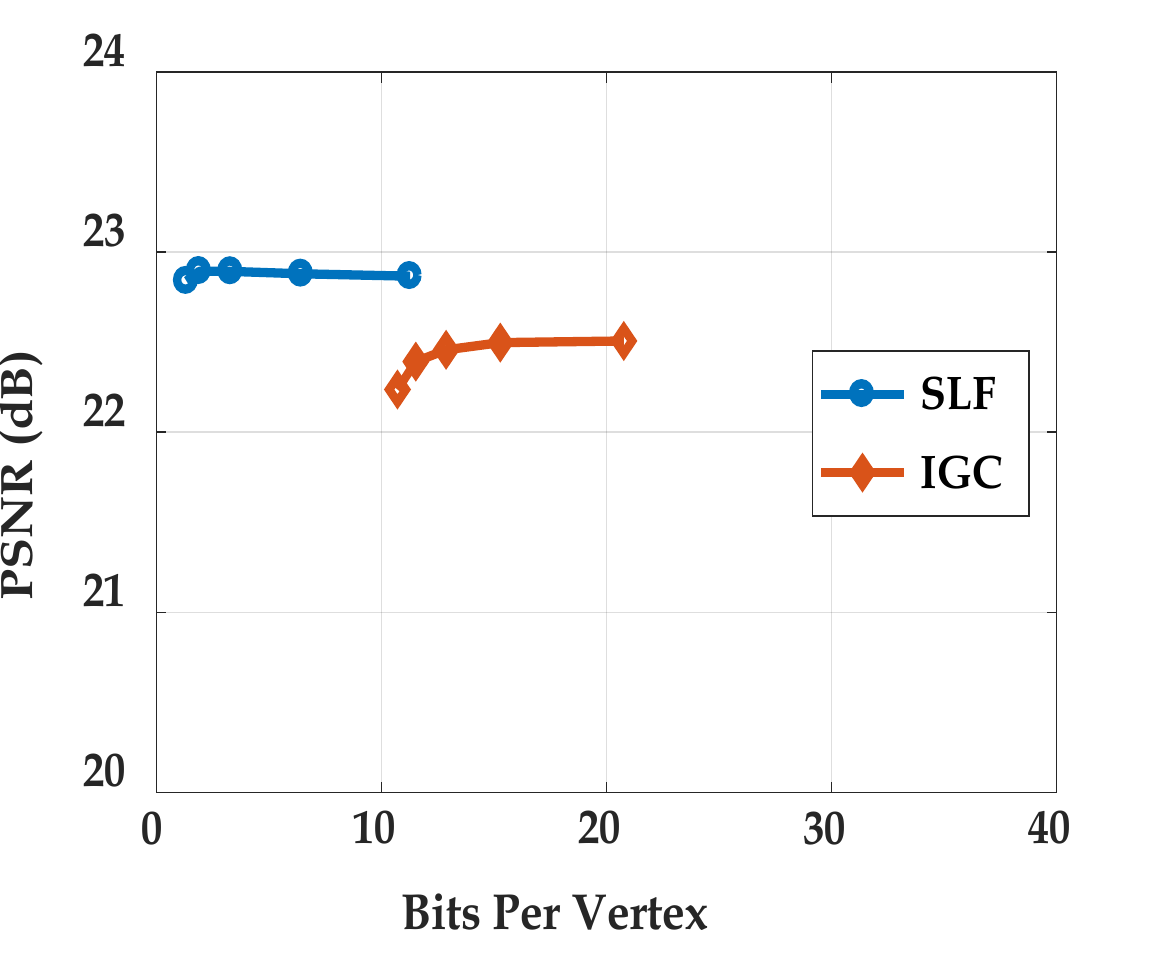}}
		\subfigure[Sparse case]{\includegraphics[width=0.32\linewidth]{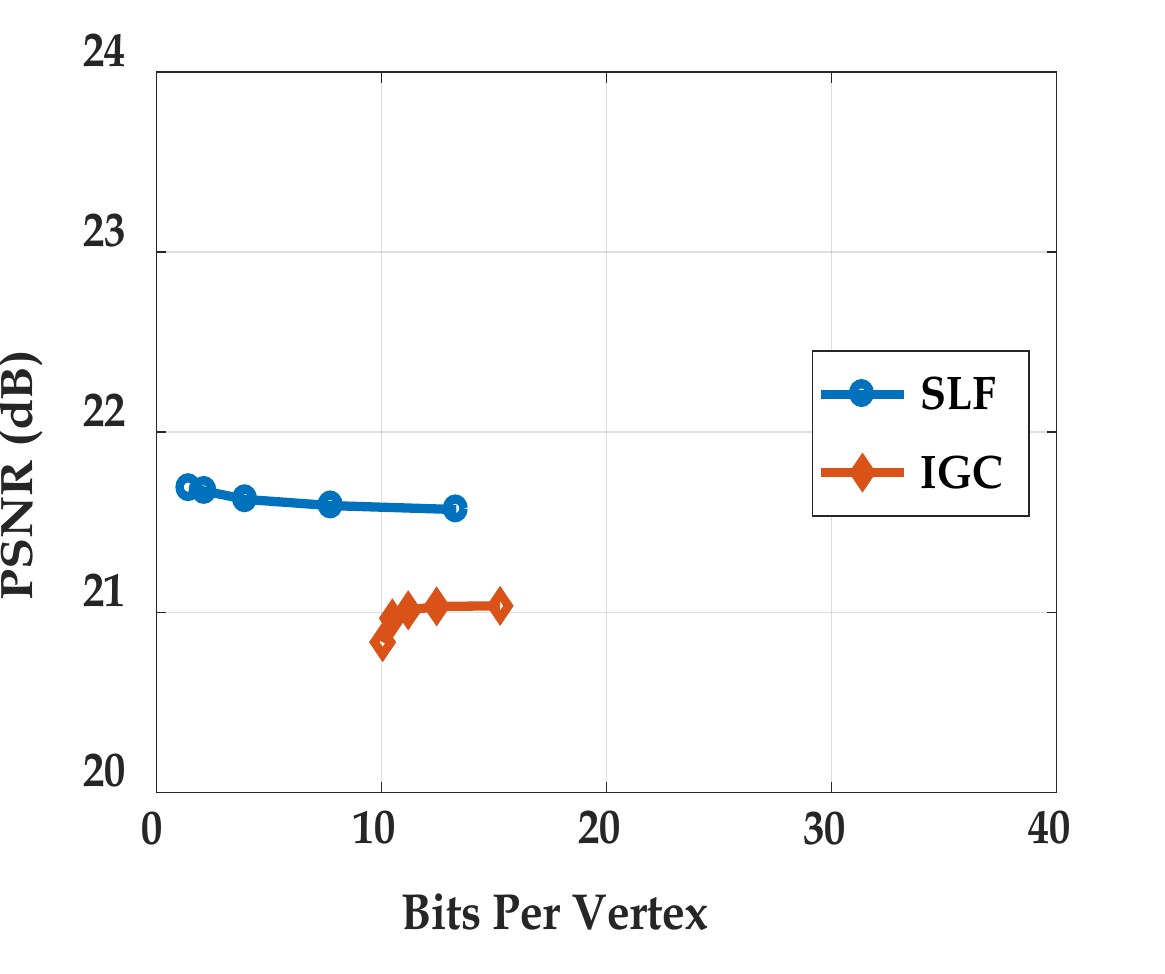}}
		\caption{RD performance comparisons between the proposed method (SLF) and IGC under different camera densities for {\em Elephant}.}
		\label{fig:rdcompare_elephant}
	\end{figure*}

	We compare the RD performance of IGC with that of the proposed SLF compression scheme to gain some insight. To compress the input images for IGC, we use reference software HM-16.2 \cite{hm} of HEVC \cite{hevc}, which is the state-of-the-art video coding standard, with common test conditions of low delay. Experimental results of the RD performance comparison are shown in Figs.~\ref{fig:rdcompare_fish} and~\ref{fig:rdcompare_elephant}, where the {\em Fish} and {\em Elephant} are  compressed under three different input camera densities, respectively. We can draw several conclusions. First, the proposed SLF compression can achieve superior RD performance than IGC. Second, the bitrates can be significantly decreased but with marginal PSNR losses (even with some gains) for the {\em Elephant} dataset. Since the point cloud compression standard is still  developing by MPEG, we can expect the performance of SLF compression can be further improved with even lower complexity.
	
	\section{Conclusions}
	\label{sec:conclusion}
	
	In this work, we propose a new light field (LF) representation and compression framework based on the surface light field (SLF).  Instead of directly processing multi-view images, we map the multi-view images to a point cloud and compress the geometry and view-dependent colors (\ie view maps) jointly. The advantage of the proposed SLF representation over multi-view image-based LFs is that occlusions are more accurately modeled, thereby reducing the camera density needed for capture, and making the view maps easier to compress.  We are able to achieve a compact, robust, and scalable representation by approximating the view maps as a linear combination of B-Spline wavelet functions, and compressing the corresponding coefficients using point cloud codecs to remove the spatial redundancy between neighboring points. The proposed scheme enables efficient free-viewpoint rendering, while recovering complex surface radiances and textures with good quality. Experimental results indicate that the proposed method achieves superior rate-distortion performance with lower decoder complexity compared to an image-plus-geometry compression (IGC) scheme.
	

	\bibliographystyle{IEEEtran}
	\bibliography{IEEEabrv,strings}
	\label{sec:ref}

\end{document}